\documentclass[letterpaper, 11pt]{article}
\usepackage{graphicx}
\usepackage{amsmath}
\usepackage{amsfonts}
\usepackage{hyperref}
\usepackage{amsthm}
 \usepackage{amscd}
\usepackage{mathrsfs}
\usepackage{caption}
\usepackage{subcaption}
\usepackage{float}
\usepackage[top=1in,left=1in,right=1in,bottom=1in]{geometry}

\newtheorem{theorem}{Theorem}[section]

\newtheorem{example}{Example}[section]
\counterwithin{figure}{section}
\hypersetup{
	colorlinks = true,
	linkcolor  = red,
	citecolor  = blue,
	filecolor  = cyan,
	urlcolor   = magenta,}

\newcommand{\ds}{\displaystyle}

\numberwithin{equation}{section}

\title{Soliton solutions associated with a class of third-order ordinary linear differential operators}
\author{Tuncay Aktosun\\
Department of Mathematics\\
University of Texas at Arlington\\
Arlington, TX 76019-0408, USA\\
\\
Abdon E. Choque-Rivero\\
Instituto de F\'isica y Matem\'aticas\\
Universidad Michoacana de San Nicolás de Hidalgo\\
Morelia, Michoac\'an, Mexico\\
\\
Ivan Toledo\\
Department of Mathematics\\
University of Texas at San Antonio\\
San Antonio, TX 78249, USA\\
\\
Mehmet Unlu\\
Department of Mathematics\\
Recep Tayyip Erdogan University\\
53100 Rize, Turkey}

\date{}

\begin{document}

\maketitle

\begin{abstract}
Explicit solutions to the related integrable nonlinear evolution equations are constructed by solving the inverse scattering problem
in the reflectionless case for the third-order differential equation $d^3\psi/dx^3+Q\,d\psi/dx+P\psi=k^3\psi,$ where
$Q$ and $P$ are the potentials in the Schwartz class and $k^3$ is the spectral parameter. The input data set used to solve
the relevant inverse problem consists of the bound-state poles of a transmission coefficient and the corresponding bound-state dependency constants.
Using the time-evolved dependency constants, explicit solutions to the related integrable evolution equations are obtained. In the special cases of the
Sawada--Kotera equation and the modified bad Boussinesq equation, the method presented here explains the physical origin of the constants
appearing in the relevant
$\mathbf N$-soliton solutions algebraically constructed, but without any physical insight, by the bilinear method of Hirota.
\end{abstract}



\newpage

\section{Introduction}
\label{section1}

Consider the third-order linear ordinary differential equation
\begin{equation}
\label{1.1}
\psi'''+Q(x,t)\,\psi'+P(x,t)\,\psi=k^3\,\psi, \qquad x\in\mathbb R,
\end{equation}
where $x$ is the independent variable representing the spacial coordinate and taking values
on the real line $\mathbb R,$
the prime denotes the $x$-derivative, $t$ is a parameter representing the time coordinate, $k^3$ is the
spectral parameter, and the coefficients $Q$ and $P$ are some complex-valued functions of $x$ and $t$
vanishing rapidly as $x\to\pm\infty$ for each fixed $t\in\mathbb R.$ For simplicity, we assume that $Q$
and $P$ belong to the Schwartz class $S(\mathbb R)$ in the $x$-variable, which is the class of infinitely differentiable
functions $\theta(x)$ in such a way that $x^j d^l\theta(x)/dx^l$ vanishes as $x\to\pm\infty$ for any pair of nonnegative 
integers $j$ and $l.$ We refer to $Q$ and $P$ as the potentials. Our results hold under
weaker assumptions on $Q$ and $P,$ but for simplicity we assume that the potentials $Q$ and $P$
belong to the Schwartz class in $x.$
We can write \eqref{1.1} as $L\psi=\lambda\psi$ by letting $\lambda:=k^3$ and defining the linear differential
operator $L$ as
 \begin{equation}
 \label{1.2}
L:=D^3+QD+P,
\end{equation}
with $D$ representing
the differential operator $d/dx.$ 

The direct scattering problem for \eqref{1.1} consists of the determination of the corresponding scattering coefficients and
the bound-state information when the potentials $Q$ and $P$ are known for all $x\in\mathbb R$ and at any fixed value of $t.$
On the other hand, the inverse scattering problem for \eqref{1.1} consists of the determination of
the potentials $Q$ and $P$
 by using the input data set consisting
of the scattering coefficients and the bound-state information.
Unless the theory of the direct and inverse scattering scattering for \eqref{1.1} is used to obtain solutions to 
some integrable nonlinear evolution equations, the dependence of the potentials
$Q$ and $P$ on the time variable $t$ can be arbitrary.
On the other hand, when we consider various integrable evolution equations associated with
the linear differential operator $L$ given in \eqref{1.2}, the time dependence of 
the potentials cannot be arbitrary and is determined by a linear differential operator
$A,$ where the pair of operators $L$ and $A$ is known as the Lax pair.
In that case, the specific operator $A$ in the Lax pair governs 
the time evolution of solutions to the corresponding integrable evolution equation.

There are various integrable evolution equations associated with $L$ appearing
in \eqref{1.2}, and each such equation corresponds to a particular choice of the potentials $Q$ and $P$
and a particular time dependence for those potentials.
In our paper, we consider three such cases. In the first case, the potentials 
$Q$ and $P$ belong to the Schwartz class in $x$ and their time dependence is governed by
the operator $A$ given in \eqref{1.3}. The second case is a special subcase of the first case,
where the operator $L$ in \eqref{1.2} becomes either $L_1$ or $L_2$
defined in \eqref{1.11} and the corresponding operator $A$ becomes
$A_1$ or $A_2,$ given in \eqref{1.12} and \eqref{1.13}, respectively.
In the third case, we again consider the operator
$L$ given in \eqref{1.2} for some special choices of
the potentials $Q$ and $P,$ but associated
with a new operator $A$ which is different from
the operator $A$ given in \eqref{1.3}. This third case
corresponds to a modified version of the integrable bad Boussinesq equation
with the Lax pair $(L,A)$ given in \eqref{6.2} and \eqref{6.3}.
Even though we do not consider it in our paper, there is also a special case
of the operator $L$ in \eqref{1.2}, and that fourth special case is given
in \eqref{1.18} with the corresponding $A$ specified in
\eqref{1.19}. 
%In this fourth case, even though the operator $L$ in \eqref{1.18}
%is a special case of the operator $L$ appearing in \eqref{1.2}, the corresponding operator
%$A$ given in
%\eqref{1.19} is not a special case of the operator $A$ appearing in \eqref{1.3}.

In order to consider the first aforementioned case, we introduce the linear differential operator $A$ given by
 \begin{equation}
 \label{1.3}
A:=9\,D^5+15\,Q D^3+\big(15\,P+15\,Q_x\big)D^2+\big(15\,P_x+10\,Q_{xx}+5\,Q^2\big)D
      +\big(10\,P_{xx}+10\,Q \,P\big),
\end{equation}
where the subscripts $x$ and $t$ denote the appropriate partial derivatives. We 
determine the integrable system of nonlinear partial differentials corresponding to the Lax pair
$L$ and $A$ given in \eqref{1.2} and \eqref{1.3}, respectively, by imposing the Lax compatibility condition
\cite{L1968}
 \begin{equation}
 \label{1.4}
L_t+LA-AL=0,
\end{equation}
where $L_t$ is the differential operator given by
 \begin{equation*}
%\label{1.5}
L_t=Q_tD+P_t.
\end{equation*}
It can be verified directly that the operator equality \eqref{1.4} holds
provided that the potentials $Q$ and $P$ satisfy the coupled system of two fifth-order nonlinear partial differential equations
given by
 \begin{equation}
  \label{1.6}
 \begin{cases}
 Q_t+Q_{xxxxx}+5\,Q_x\,Q_{xx}+5\,Q\, Q_{xxx}+5\,Q^2\,Q_x
           +15\,Q_{xx}\,P+15\,Q_x\, P_x-30\,P \,P_x=0,\\
\noalign{\medskip}
 \begin{aligned}
P_t+P_{xxxxx}+5\,Q\, P_{xxx}+15\,Q_x \,P_{xx}+20\,Q_{xx}\,P_x+5\,Q^2\,P_x+&10\,Q_{xxx}\,P-15\,P \,P_{xx}\\
           &+10\,Q\, Q_x P-15\,(P_x)^2=0,
\end{aligned}        
\end{cases}
\end{equation}
where we have $x\in\mathbb R$ and the $t$-domain can be chosen
either as $\mathbb R$ or $t>0.$
Thus, the system \eqref{1.6} is integrable in the sense of the inverse scattering transform method \cite{GGKM1967}, and its
initial-value problem can be solved with the help of the direct and inverse scattering theory for \eqref{1.1} with the
appropriate time evolution of the scattering data governed by the linear operator $A$ in \eqref{1.3}. 

For simplicity, we choose our notation by suppressing the appearance of
$t$ in the arguments of the potentials and all other quantities associated with \eqref{1.1}, \eqref{1.6}, and any other integrable systems
related to \eqref{1.1}.
For example, we mainly use
$Q(x)$ instead of $Q(x,t)$ and write $P(x)$ instead of $P(x,t).$

Let us now consider the second aforementioned case. The uncoupled version of \eqref{1.6} is obtained when the sum of the last three terms on the left-hand side of the first equality
of \eqref{1.6} vanishes, i.e. when we have 
 \begin{equation}
  \label{1.7}
15\,Q_{xx} P+15\,Q_x\, P_x-30\,P\, P_x=0.
\end{equation}
Integrating \eqref{1.7} and using the fact that the potentials $Q$ and $P$ vanish as $x\to\pm\infty,$ we get
 \begin{equation}
  \label{1.8}
Q_x P-P^2=0.
\end{equation}
From \eqref{1.8}, we observe that \eqref{1.6} has exactly two special subcases given by 
 \begin{equation}
  \label{1.9}
P(x)\equiv 0, \quad P(x)\equiv Q_x(x).
\end{equation}
In the first special subcase $P(x)\equiv 0,$
the integrable system \eqref{1.6} reduces to
the single integrable evolution equation
\begin{equation}
 \label{1.10}
Q_t+Q_{xxxxx}+5\,Q_x\,Q_{xx}+5\,Q\,Q_{xxx}+5\,Q^2\,Q_x=0, \qquad x,\,t\in\mathbb R,
\end{equation}
which is known as the Sawada--Kotera equation \cite{H1989,K1980,SK1974}. In the second special subcase where we have
$P(x)\equiv Q_x(x),$ 
the integrable system \eqref{1.6} again reduces to the single integrable evolution equation given
in \eqref{1.10}.
In applications, the quantity $Q$ in the Sawada--Kotera
equation is usually required to be real valued.
Using each of the two special subcases in the equalities given in \eqref{1.2} and \eqref{1.3},
we observe that \eqref{1.10} is associated with two distinct Lax pairs $(L_1,A_1)$ and
$(L_2,A_2),$ where we have defined
 \begin{equation}
 \label{1.11}
L_1:=D^3+QD,\quad
L_2:=D^3+QD+Q_x,
\end{equation}
 \begin{equation}
 \label{1.12}
A_1:=9\,D^5+15\,Q D^3+15\,Q_x\,D^2+\big(10\,Q_{xx}+5\,Q^2\big)D,
\end{equation}
 \begin{equation}
 \label{1.13}
A_2:=9\,D^5+15\,Q D^3+30\,Q_x\,D^2+\big(25\,Q_{xx}+5\,Q^2\big)D
      +\big(10\,Q_{xxx}+10\,Q \,Q_x\big).
\end{equation}
At first, it seems awkward that the Sawada--Kotera equation \eqref{1.10}
is associated with two different third-order linear differential equations.
However, there is no contradiction here, and the clarification
for this is as follows. When $Q(x)$ is real valued
and $P(x)\equiv 0$ in \eqref{1.1}, 
with the help of \eqref{2.19} we see that
the adjoint equation \eqref{2.18} corresponding to \eqref{1.1} is given by
\begin{equation}\label{1.14}
    \overline \psi'''(k,x) + Q(x)\, \overline \psi'(k,x) + Q_x(x)\, \overline \psi(k,x) = k^3\,\overline \psi(k,x), \qquad x\in \mathbb R.
\end{equation}
This shows that the inverse scattering transform method on \eqref{1.10} can be applied in two different ways:
The first way is through the use of
the direct and inverse scattering theory for \eqref{1.1} when $Q(x)$ is real valued and 
$P(x)\equiv 0.$ The second way is through the use of
the direct and inverse scattering theory for \eqref{1.14} when $Q(x)$ is real valued.
The first way relies on the transmission coefficients
$T_{\text{\rm{l}}}(k)$ and $T_{\text{\rm{r}}}(k)$ for \eqref{1.1}.
The second way relies on  the transmission coefficients
$\overline T_{\text{\rm{l}}}(k)$ and $\overline T_{\text{\rm{r}}}(k)$ for \eqref{1.14}.
The transmission coefficients for the adjoint equation \eqref{1.14} are related to the
transmission coefficients for 
\eqref{1.1} as stated in \eqref{2.20} and \eqref{2.21}.
In this case, it turns out that we have
\begin{equation*}
%\label{1.15}
\overline T_{\text{\rm{l}}}(k)\equiv T_{\text{\rm{l}}}(k),\qquad k\in\overline{\Omega_1},
\end{equation*}
\begin{equation*}
%\label{1.16}
\overline T_{\text{\rm{r}}}(k)\equiv T_{\text{\rm{r}}}(k),\qquad k\in\overline{\Omega_3},
\end{equation*}
but the bound-state dependency constants for \eqref{1.1} and those for \eqref{1.14} do not
coincide. We elaborate on this issue in Example~\ref{example5.3} of Section~\ref{section5}.

We remark that there is another fifth-order integrable evolution equation resembling \eqref{1.10}, usually known as 
the Kaup--Kupershmidt equation \cite{K1980,K1984}, given by
\begin{equation}
 \label{1.17}
Q_t+Q_{xxxxx}+\ds\frac{25}{2}\,Q_x\,Q_{xx}+5\,Q\,Q_{xxx}+5\,Q^2\,Q_x=0, \qquad x,\,t\in\mathbb R.
\end{equation}
This corresponds to the fourth aforementioned case.
We obtain \eqref{1.17} by using $P=Q_x/2$ in the integrable system \eqref{1.6}.
Even though the coefficients in an integrable evolution equation in general can be changed
by scaling both the dependent and independent variables, it is impossible
to transform \eqref{1.17} to \eqref{1.10} or vice versa through any scaling.
This indicates that
the Kaup--Kupershmidt equation and the Sawada--Kotera equation are different from each other.
The Lax pair $(L,A)$ associated with the Kaup--Kupershmidt equation \eqref{1.17}
is obtained by using $P=Q_x/2$ in \eqref{1.2} and \eqref{1.3}, respectively, and we have
\begin{equation}
\label{1.18}
L=D^3+QD+\ds\frac{1}{2}\,Q_x,
\end{equation}
 \begin{equation}
 \label{1.19}
A=9\,D^5+15\,Q D^3+\ds\frac{45}{2}\,Q_x\,D^2+\left(\ds\frac{35}{2}\,Q_{xx}+5\,Q^2\right)D+
5\,Q_{xxx}+5\,Q\,Q_x.
\end{equation}
As seen from \eqref{1.18}, the Kaup--Kupershmidt equation can be written as
\begin{equation}
\label{1.20}
\psi'''+Q(x)\,\psi'+\ds\frac{1}{2}\,Q_x(x)\,\psi=k^3\,\psi, \qquad x\in\mathbb R.
\end{equation}
%Because \eqref{1.18} is a special case of \eqref{1.2}, even though \eqref{1.19} is not a special case of
%\eqref{1.3},
The direct and inverse scattering theory for \eqref{1.1} can be applied to determine solutions to \eqref{1.17} and
\eqref{1.20}.
However, in our paper we do not analyze \eqref{1.17}. 
We only remark that, when $Q$ is real valued, the adjoint equation
corresponding to \eqref{1.20} is \eqref{1.20} itself. This can be verified by comparing
\eqref{1.20} and \eqref{2.18} with the help of \eqref{2.19}.
The coincidence of \eqref{1.20} and its adjoint equation
 does not mean that the differential operator $L$ appearing in \eqref{1.18} is selfadjoint.

Like the Korteweg--de Vries equation \cite{GGKM1967}, the Sawada--Kotera and Kaup--Kupershmidt equations are used to describe
the propagation of surface water waves in long, narrow, shallow canals but by taking into consideration of steeper waves of shorter wavelength.
We also mention the paper \cite{WZ2022}, where the direct and inverse scattering problems are analyzed 
for a modified version of the Sawada--Kotera equation by considering a related Riemann--Hilbert problem.

In this paper we are interested in exploring explicit solutions to the integrable system \eqref{1.6} and its 
special case \eqref{1.10}. In the
literature some specific explicit solutions to \eqref{1.10} are obtained usually by using Hirota's bilinear method \cite{H1989}
or by using the dressing method
of Zakharov and Shabat \cite{P2001}, where both these methods avoid the analysis of the direct and inverse scattering theory associated with
\eqref{1.1}. Such explicit solutions, if they do not contain any singularities, correspond to soliton solutions.
Our goal is to obtain those explicit solutions, without using an ansatz but only by relying
on the analysis of the direct and inverse scattering theory for \eqref{1.1}. In general, soliton solutions for
integrable systems correspond to reflectionless scattering data sets, which consist of transmission coefficients
and time-evolved bound-state information. The bound-state information can be provided by
specifying the bound-state poles of the transmission coefficients and the bound-state normalization constants.
Alternatively, the bound-state information can be given by
specifying the bound-state poles of the transmission coefficients and
the bound-state dependency constants.
Hence, in this paper we describe soliton solutions to \eqref{1.6} and its special case \eqref{1.10},
without using any ansatzes but by using the transmission coefficients and the time-evolved
bound-state data for \eqref{1.1}, where the bound-state data set consists of the $k$-values specifying
the bound-state poles of the transmission coefficients and the
bound-state dependency constants at those $k$-values. The differential operator
$L$ appearing in \eqref{1.2} is in general not selfadjoint, even though one of its special cases
expressed in \eqref{6.2} is selfadjoint.
In case a bound state for \eqref{1.1} has any multiplicities,
it is understood that for that bound state we have as many dependency constants as the
multiplicity of that bound state. We remark that the bound-state
information for \eqref{1.1} can also be described
in terms of the values of the spectral parameter
$\lambda$ instead of the values of the
parameter $k,$ by recalling that
we have $\lambda:=k^3.$ In case the linear differential operator $L$ in \eqref{1.2}
can be expressed as a selfadjoint operator, which happens in the special case given in \eqref{6.2}, 
one can readily verify that the corresponding eigenvalues expressed in $\lambda$
are all real.

The most relevant references related to the research presented in this paper are the 1980 paper \cite{K1980} by Kaup, the 1982 paper
\cite{DTT1982} by Deift, Tomei, and Trubowitz, the 1989 paper \cite{H1989} by Hirota, the 2001 paper \cite{P2001} by Parker, the 2002 paper \cite{K2002} by Kaup, and the Ph.D.
thesis \cite{T2024} of the third author of the present paper. 
Kaup initiated \cite{K1980} the analysis of the direct and inverse scattering for \eqref{1.1} with the goal of studying solutions to the integrable nonlinear evolution equation
\eqref{1.10}. However, he was unable to formulate a proper scattering matrix and a proper Riemann--Hilbert problem
associated with \eqref{1.1}. We refer to a Riemann--Hilbert problem as properly formulated if the plus and minus regions in the complex
plane are separated by an infinite straight line passing through the origin. Such a formulation allows the use of a Fourier transformation
along that line. For example, in the analysis of the inverse scattering problem for the Schr\"odinger equation on the full line, we have a proper formulation of the Riemann--Hilbert
problem with the real axis separating the complex plane into two regions. Such a formulation allows the use of a Fourier transformation, which yields
the Marchenko integral equation \cite{AK2001,CS1989,DT1979,F1967,L1987,M2011} playing a key role in the solution to the corresponding inverse scattering problem.
The lack of a proper formulation of a Riemann--Hilbert problem for \eqref{1.1} prevented Kaup from establishing \cite{K2002} a linear integral
equation, which would be the analog of the Marchenko integral equation associated with the inverse scattering problem for the full-line
Schr\"odinger equation. Deift, Tomei, and Trubowitz studied \cite{DTT1982} the direct and inverse scattering problems for a special case of \eqref{1.1}, namely for the third-order
equation \eqref{6.4} listed in Section~\ref{section6}. 
%Even though the linear operator $L$ given in \eqref{1.2} associated with \eqref{1.1} is in general not selfadjoint,
The linear operator $L$ given in \eqref{6.2} of Section~\ref{section6} associated with \eqref{6.4} is selfadjoint. Deift, Tomei, and Trubowitz 
introduced \cite{DTT1982} a proper Riemann--Hilbert problem
on the complex $k$-plane, enabling them to
solve the corresponding inverse scattering problem for \eqref{6.4}. However, in order to formulate their Riemann--Hilbert problem properly, 
in their paper they used 
the severe assumption that the two transmission coefficients
for \eqref{6.4} are identically equal to $1$ for all $k$-values. They also used the assumption
that the two secondary reflection coefficients
are identically zero. That latter assumption is not as severe as the former assumption, and in fact
it helps to discard solutions to \eqref{6.5} that blow up at finite time. 
The former assumption prevented Deift, Tomei, and Trubowitz from obtaining
explicit solutions to the modified bad Boussinesq equation \eqref{6.5}. This is because such solutions are usually obtained 
in the reflectionless case by using the bound-state poles of the transmission coefficients.
Hirota used his bilinear method to obtain the $\mathbf N$-soliton solution \cite{H1973}
to the bad Boussinesq equation and the $\mathbf N$-soliton solution \cite{H1989} to
the Sawada--Kotera equation. Those ad hoc solutions were introduced algebraically 
by Hirota without any connection to the direct and inverse scattering theory
associated with the corresponding linear third-order equations.
Parker \cite{P2001} tried to explain the derivation of Hirota's ad hoc $\mathbf N$-soliton solutions to
the Sawada--Kotera equation by using a modification of the dressing method \cite {Z1990} of
Zakharov and Shabat. In his 2002 review paper \cite{K2002} Kaup summarized the efforts on the inverse scattering transform
associated with the third-order equation \eqref{1.1} in the special case corresponding to the Sawada--Kotera
equation \eqref{1.10}, but without mentioning the 
relevant paper \cite{DTT1982} by Deift, Tomei, and Trubowitz.
Inspired by \cite{DTT1982}, Toledo formulated \cite{T2024} a proper Riemann--Hilbert problem associated with the inverse scattering problem
for \eqref{1.1} by using the assumption that the two secondary reflection coefficients 
are identically zero, but without using the assumption that the two transmission coefficients
are identically equal to $1$ for all $k$-values.

Our paper uses the method of \cite{T2024} in the reflectionless case, and it is organized as follows. 
In Section~\ref{section2} we present a brief summary of the direct scattering problem for \eqref{1.1}
by introducing the basic solutions to \eqref{1.1} and the scattering coefficients for \eqref{1.1} and by providing the relevant properties
of those solutions and scattering coefficients. This is first done without using any particular time dependence of the
potentials $Q$ and $P.$ At the end of Section~\ref{section2}, in case the operator $A$ given in \eqref{1.3}
is used to govern the time evolution of the potentials $Q$ and $P,$
we describe the corresponding time evolution of the basic solutions and the scattering coefficients for \eqref{1.1}.
In Section~\ref{section3} we provide a brief
description of the bound states associated with \eqref{1.1} and the corresponding bound-state dependency constants. 
At the end of Section~\ref{section3}, we consider \eqref{1.1} in the reflectionless case and when 
the potentials $Q$ and $P$ are in the Schwartz class but with arbitrary time dependence. In that case, we show that
the left and right transmission coefficients must be the reciprocals of each other.
This means that, in the reflectionless case, the bound-state poles of the left
transmission coefficient and the bound-state poles of the right transmission coefficient
are symmetrically located with respect to the origin of the complex $k$-plane
when the poles and zeros of a transmission coefficient are symmetrically located.
In Section~\ref{section4}
we consider \eqref{1.1} in the reflectionless case, and we formulate a proper Riemann--Hilbert problem
in the complex $k$-plane by introducing the relevant plus and
minus regions and the plus and minus functions in those regions. We also describe how the potentials $Q$ and $P$ are recovered
from the solution to the aforementioned Riemann--Hilbert problem. In Section~\ref{section5} we show how 
the relevant solutions to \eqref{1.1} are explicitly
constructed by solving the corresponding Riemann--Hilbert problem using the input data set consisting of the bound-state poles of a transmission
coefficient and the corresponding bound-state dependency constants. With the input data set containing the time-evolved dependency constants, the
constructed potentials $Q$ and $P$ yield explicit solutions to the integrable system \eqref{1.6}. By using the appropriate restrictions on the locations of the
bound-state $k$-values and the appropriate restrictions on the dependency constants, we obtain real-valued solutions to the
Sawada--Kotera equation \eqref{1.10}. 
In fact, we show that Hirota's $\mathbf N$-soliton solution to the Sawada--Kotera equation can be obtained from a particular $2\mathbf N$-soliton solution to \eqref{1.6}
by using some appropriate restrictions on our input data set.
Hence, our method using the input data set consisting of the bound-state poles of a transmission coefficient and
the corresponding time-evolved bound-state dependency constants
explains the physical origin of the
$\mathbf N$-soliton solution to the Sawada--Kotera equation algebraically constructed \cite{H1989} by the bilinear method of Hirota.
Since our general technique is based on the
solution to the inverse scattering problem for \eqref{1.1} in the reflectionless case without requiring any specific
time dependence for the potentials $Q$ and $P,$
it is applicable to other integrable evolution equations
associated with \eqref{1.1}. 
In applying our method to a specific integrable system, we simply use the appropriate
time evolution particular to that integrable system.
This is further illustrated in 
Section~\ref{section6}, where we
show how our method yields explicit solutions to a modified version the bad Boussinesq equation, again explaining
the physical origin of the $\mathbf N$-soliton solution obtained by Hirota's method \cite{H1973}.

\section{The direct scattering problem}
\label{section2}

In this section we present a summary of the direct scattering theory for the third-order equation \eqref{1.1}. 
We suppress the appearance of the parameter $t$ in the arguments
of the quantities related to \eqref{1.1}. We choose the notation used in the recent Ph.D. thesis \cite{T2024}
of the third author.
We recall that the dependence of
the potentials $Q$ and $P$ on the time variable
$t$ is arbitrary, unless we apply the direct and inverse scattering theory
for \eqref{1.1} to solve a specific integrable system of nonlinear evolution
equations. At the end of this section, we specify the particular time evolution
of the solutions and the scattering coefficients in the special case when
the time evolution of the potentials $Q$ and $P$ is governed
by the particular operator $A$ given in \eqref{1.3}.

\begin{figure}[!ht]
     \centering
         \includegraphics[width=1.8in]{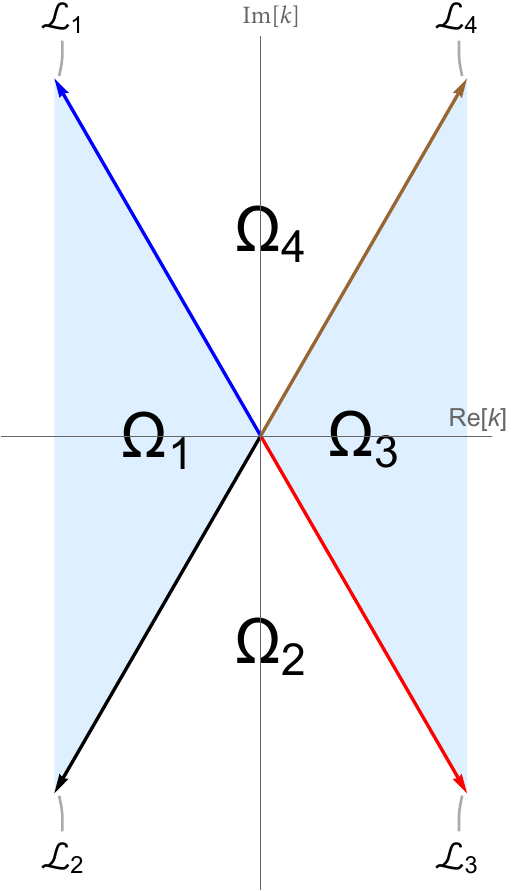}
\caption{The directed half lines
$\mathcal L_1,$ $\mathcal L_2,$ $\mathcal L_3,$ $\mathcal L_4$
and the open sectors $\Omega_1,$ $\Omega_2,$ $\Omega_3,$ $\Omega_4$ in the complex $k$-plane.}
\label{figure2.1}
\end{figure}

 It is convenient to use
the four directed half lines $\mathcal L_1,$ $\mathcal L_2,$ $\mathcal L_3,$ $\mathcal L_4$ to divide the complex $k$-plane
$\mathbb C$ into the four sectors $\Omega_1,$ $\Omega_2,$ $\Omega_3,$ $\Omega_4$ as shown in Figure~\ref{figure2.1}.
We use $z$ to denote the special complex number $e^{2\pi i/3},$ which is equivalently defined as
\begin{equation}\label{2.1}
z:=-\ds\frac{1}{2}+i\,\ds\frac{\sqrt{3}}{2}.
\end{equation}
The directed half lines $\mathcal L_1,$ $\mathcal L_2,$ $\mathcal L_3,$ $\mathcal L_4$ are defined by the 
respective parametrizations given by
\begin{equation}\label{2.2}
\mathcal L_1:=\{k\in\mathbb C: k=zs \text{\rm{ for }}  s\in[0,+\infty)\},
\end{equation}
\begin{equation}\label{2.3}
\mathcal L_2:=\{k\in\mathbb C: k=z^2s \text{\rm{ for }}  s\in[0,+\infty)\},
\end{equation}
\begin{equation}\label{2.4}
\mathcal L_3:=\{k\in\mathbb C: k=-zs \text{\rm{ for }}  s\in[0,+\infty)\},
\end{equation}
\begin{equation}\label{2.5}
\mathcal L_4:=\{k\in\mathbb C: k=-z^2s \text{\rm{ for }}  s\in[0,+\infty)\}.
\end{equation}
As seen from Figure~\ref{figure2.1}, the four open sectors $\Omega_1,$ $\Omega_2,$ $\Omega_3,$ $\Omega_4$ in $\mathbb C$ are respectively defined by using
the parametrizations given by
\begin{equation}\label{2.6}
\Omega_1:=\left\{k\in\mathbb C: \ds\frac{2\pi}{3}<\arg[k]<\ds\frac{4\pi}{3}\right\},
\end{equation}
\begin{equation}\label{2.7}
\Omega_2:=\left\{k\in\mathbb C: -\ds\frac{2\pi}{3}<\arg[k]<-\ds\frac{\pi}{3}\right\},
\end{equation}
\begin{equation}\label{2.8}
\Omega_3:=\left\{k\in\mathbb C: -\ds\frac{\pi}{3}<\arg[k]<\ds\frac{\pi}{3}\right\},
\end{equation}
\begin{equation}\label{2.9}
\Omega_4:=\left\{k\in\mathbb C: \ds\frac{\pi}{3}<\arg[k]<\ds\frac{2\pi}{3}\right\},
\end{equation}
where $\arg[k]$ represents the argument function taking values in
the interval $(-2\pi/3,4\pi/3).$
We use $\overline{\Omega_1},$ $\overline{\Omega_2},$ $\overline{\Omega_3},$ $\overline{\Omega_4}$ to denote the closures
of $\Omega_1,$ $\Omega_2,$ $\Omega_3,$ $\Omega_4,$ respectively. Thus, the closed sector $\overline{\Omega_1}$ is obtained
from $\Omega_1$ by including its respective upper and lower boundaries $\mathcal L_1$ and $\mathcal L_2,$ the closed sector
$\overline{\Omega_2}$ is obtained from $\Omega_2$ by including its respective left and right boundaries $\mathcal L_2$ and
$\mathcal L_3,$ the closed sector $\overline{\Omega_3}$ is obtained from $\Omega_3$ by including its respective lower and upper
boundaries $\mathcal L_3$ and $\mathcal L_4,$ and the closed sector $\overline{\Omega_4}$ is obtained from $\Omega_4$
by including its respective left and right boundaries $\mathcal L_1$ and $\mathcal L_4.$  

The left Jost solution $f(k,x)$ to \eqref{1.1} is defined as the solution satisfying the spacial asymptotics 
\begin{equation}\label{2.10}
\begin{cases}
f(k,x)=e^{kx}\left[1+o(1)\right], \qquad x\to+\infty,\\
\noalign{\medskip}
f'(k,x)=k\,e^{kx}\left[1+o(1)\right], \qquad x\to+\infty,\\
\noalign{\medskip}
f''(k,x)=k^2\,e^{kx}\left[1+o(1)\right],  \qquad x\to+\infty.
\end{cases}
\end{equation}
The right Jost solution $g(k,x)$ to \eqref{1.1} is defined as the solution satisfying the spacial asymptotics 
\begin{equation}\label{2.11}
\begin{cases}
g(k,x)=e^{kx}\left[1+o(1)\right], \qquad x\to-\infty,\\
\noalign{\medskip}
g'(k,x)=k\,e^{kx}\left[1+o(1)\right], \qquad x\to-\infty,\\
\noalign{\medskip}
g''(k,x)=k^2\,e^{kx}\left[1+o(1)\right],  \qquad x\to-\infty.
\end{cases}
\end{equation}

The basic properties of the Jost solutions $f(k,x)$ and $g(k,x)$ to \eqref{1.1} are listed in the following theorem. In the theorem, the
dependence of the potentials on the time variable $t$ is arbitrary. We refer the reader
to \cite{T2024} for a proof. Even though we assume that the two potentials $Q(x)$ and $P(x)$ are restricted to the
Schwartz class, the results stated in the theorem are valid under milder restrictions
on the two potentials.

\begin{theorem}\label{theorem2.1}
Assume that the potentials $Q(x)$ and $P(x)$ in \eqref{1.1} belong to the Schwartz class $\mathcal{S(\mathbb R)}.$
Let $\Omega_1$ and $\Omega_3$ be the sectors defined in \eqref{2.6} and
\eqref{2.8}, respectively, and let
$\overline{\Omega_1}$ and $\overline{\Omega_3}$
denote the corresponding closures, respectively.
We have the following:
\begin{enumerate}
    \item[\text{\rm{(a)}}] 
    The left Jost solution $f(k,x)$ to \eqref{1.1} satisfying \eqref{2.10} exists and is uniquely determined.
    
    \item[\text{\rm{(b)}}] For each fixed $x\in \mathbb R,$ the quantity $f(k,x)$ is analytic in $k\in \Omega_1,$ is continuous
    in $k\in\overline{\Omega_1},$ and has the large $k$-asymptotics  
    \begin{equation}\label{2.12}
    f(k,x)=e^{kx}\left[1+O\left(\ds\frac{1}{k}\right)\right], \qquad k\to \infty \text{\rm{ in }} \overline{\Omega_1}.
    \end{equation}
    
     \item[\text{\rm{(c)}}] For each fixed $k\in \overline{\Omega_1},$ the quantity $f(k,x)$ is continuous in $x\in \mathbb R.$
     
    \item[\text{\rm{(d)}}] The right Jost solution $g(k,x)$ to \eqref{1.1} satisfying \eqref{2.11} exists and is uniquely determined.
    
    \item[\text{\rm{(e)}}] For each fixed $x\in \mathbb R,$ the quantity $g(k,x)$ is analytic in $k\in \Omega_3,$ is continuous in
    $k\in\overline{\Omega_3},$ and has the large $k$-asymptotics given by
    \begin{equation}\label{2.13}
    g(k,x)=e^{kx}\left[1+O\left(\ds\frac{1}{k}\right)\right],\qquad k\to \infty \text{\rm{ in }} \overline{\Omega_3}.
    \end{equation}
    
     \item[\text{\rm{(f)}}] For each fixed $k\in \overline{\Omega_3},$ the quantity $g(k,x)$ is continuous in $x\in \mathbb R.$
\end{enumerate}
\end{theorem}

In terms of the left Jost solution $f(k,x),$ we introduce the left scattering coefficients $T_{\text{\rm{l}}}(k),$
$L(k),$ and $M(k)$ via the spacial asymptotics as $x\to -\infty$ given by
\begin{equation}\label{2.14}
f(k,x)=\begin{cases}
e^{kx}\,T_{\text{\rm{l}}}(k)^{-1}[1+o(1)]+e^{zkx}L(k)\,T_{\text{\rm{l}}}(k)^{-1}[1+o(1)],\qquad k\in \mathcal L_1,\\
\noalign{\medskip}
e^{kx}\,T_{\text{\rm{l}}}(k)^{-1}[1+o(1)], \qquad k\in \Omega_1,\\
\noalign{\medskip}
e^{kx}\,T_{\text{\rm{l}}}(k)^{-1}[1+o(1)]+e^{z^2 kx}M(k)\,T_{\text{\rm{l}}}(k)^{-1}[1+o(1)], \qquad k\in \mathcal L_2,
\end{cases}
\end{equation}
where we recall that $z$ is the special constant appearing in \eqref{2.1}
and that $\mathcal L_1$ and $\mathcal L_2$ are the directed half lines defined in \eqref{2.2} and \eqref{2.3}, respectively.
We refer to $L(k)$ as the left primary reflection coefficient, $M(k)$ as the left secondary reflection coefficient, and $T_{\text{\rm{l}}}(k)$
as the left transmission coefficient. We remark that the $k$-domains of $L(k),$ $M(k),$ and $T_{\text{\rm{l}}}(k)$ are $\mathcal L_1,$
$\mathcal L_2,$ and $\overline{\Omega_1},$ respectively.
In a similar manner, in terms of the right Jost solution $g(k,x),$
we introduce the right scattering coefficients $T_{\text{\rm{r}}}(k),$
$R(k),$ and $N(k)$ via the spacial asymptotics as $x\to +\infty$ given by
\begin{equation}\label{2.15}
g(k,x)=\begin{cases}
e^{kx}\,T_{\text{\rm{r}}}(k)^{-1}[1+o(1)]+e^{zkx}R(k)\,T_{\text{\rm{r}}}(k)^{-1}[1+o(1)], \qquad k\in \mathcal L_3,\\
\noalign{\medskip}
e^{kx}\,T_{\text{\rm{r}}}(k)^{-1}[1+o(1)], \qquad k\in \Omega_3,\\
\noalign{\medskip}
e^{kx}\,T_{\text{\rm{r}}}(k)^{-1}[1+o(1)]+e^{z^2 kx}N(k)\,T_{\text{\rm{r}}}(k)^{-1}[1+o(1)],\qquad k\in \mathcal L_4,
\end{cases}
\end{equation}
where we recall that $\mathcal L_3$ and $\mathcal L_4$ are the directed half lines defined in \eqref{2.4} and \eqref{2.5}, respectively.
We refer to $R(k)$ as the right primary reflection coefficient, $N(k)$ as the right secondary reflection coefficient, and $T_{\text{\rm{r}}}(k)$
as the right transmission coefficient. We remark that the $k$-domains of $R(k),$ $N(k),$ and $T_{\text{\rm{r}}}(k)$ are $\mathcal L_3,$
$\mathcal L_4,$ and $\overline{\Omega_3},$ respectively.

\begin{figure}[!ht]
     \centering
         \includegraphics[width=2.15in]{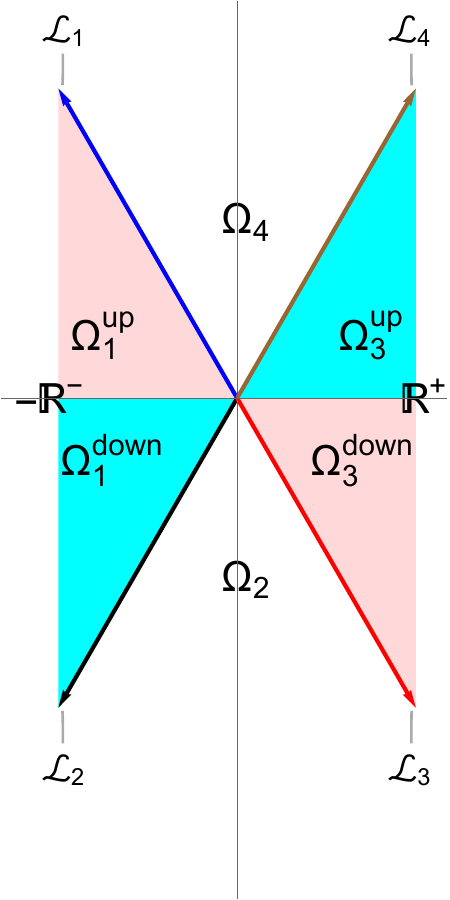}      \hskip .25in
         \includegraphics[width=2.25in]{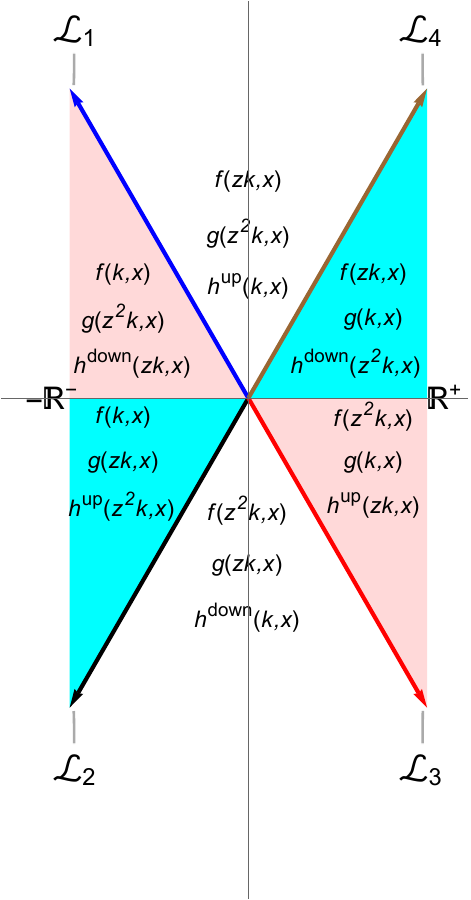} 
\caption{The complex $k$-plane is divided into the six sectors $\Omega^\text{\rm{up}}_1,$ $\Omega^\text{\rm{down}}_1,$ 
$\Omega_2,$ $\Omega^\text{\rm{down}}_3,$ $\Omega^\text{\rm{up}}_3,$ and $\Omega_4$ as shown on the left plot, and the 
$k$-domains of three basic solutions to \eqref{1.1} in each of the six regions, respectively, are shown on the right plot.}
\label{figure2.2}
\end{figure}

We introduce the solution $h^{\text{\rm{down}}}(k,x)$ to \eqref{1.1} with the $k$-domain $\overline{\Omega_2}$ and 
the solution $h^{\text{\rm{up}}}(k,x)$
to \eqref{1.1} with the $k$-domain $\overline{\Omega_4}$ via
  \begin{equation}\label{2.16}
        h^\text{\rm{down}}(k,x) := \left[\overline f(-z^2k^\ast,x)^\ast\, ;\, \overline g(-z k^\ast,x)^\ast \right],\qquad x\in\mathbb R,
    \end{equation}
    \begin{equation}\label{2.17}
    h^{\text{\rm{up}}}(k,x):=\left[\overline f(-zk^\ast,x)^\ast\, ;\,\overline g(-z^2 k^\ast,x)^\ast \right],\qquad x\in\mathbb R,
    \end{equation}
    where the asterisk denotes complex conjugation and we recall that
     $\overline{\Omega_2}$ and $\overline{\Omega_4}$
    are the closures of the sectors
    $\Omega_2$ and
     $\Omega_4$ defined in \eqref{2.7} and \eqref{2.9}, respectively.
We remark that, on the right-hand sides of \eqref{2.16} and \eqref{2.17}, we have the $2$-Wronskians of the Jost solutions to the adjoint
equation
\begin{equation}\label{2.18}
    \overline \psi'''(k,x) + \overline Q(x)\, \overline \psi'(k,x) + \overline P(x)\, \overline \psi(k,x) = k^3\,\overline \psi(k,x), \qquad x\in \mathbb R,
\end{equation}
with
\begin{equation}
\label{2.19}
    \overline Q(x):= Q(x)^\ast, \qquad \overline P(x):= Q'(x)^\ast - P(x)^\ast, \qquad x\in \mathbb R.
\end{equation}
It is known \cite{T2024} that the left transmission coefficient $\overline T_{\text{\rm{l}}}(k)$ 
for the adjoint equation \eqref{2.18} is related to the right
transmission coefficient 
$T_{\text{\rm{r}}}(k)$ for \eqref{1.1} as
\begin{equation}
\label{2.20}
\overline T_{\text{\rm{l}}}(k)=T_{\text{\rm{r}}}(-k^\ast)^\ast, \qquad k\in\overline{\Omega_1}.
\end{equation}
Similarly,
the right transmission coefficient $\overline T_{\text{\rm{r}}}(k)$ 
for the adjoint equation \eqref{2.18} is related to the left
transmission coefficient 
$T_{\text{\rm{l}}}(k)$ for \eqref{1.1} as
\begin{equation}
\label{2.21}
\overline T_{\text{\rm{r}}}(k)=T_{\text{\rm{l}}}(-k^\ast)^\ast, \qquad k\in\overline{\Omega_3}.
\end{equation}
We emphasize that the asterisk denotes complex conjugation and the overbar in our
notation does not denote complex conjugation but is used to denote
the quantities related to the adjoint equation \eqref{2.18}. We note that the $2$-Wronskian $\left[F(x);G(x)\right]$ of two
functions $F(x)$ and $G(x)$ is defined as
\begin{equation*}
%\label{2.22}
    \left[F(x) \,; G(x) \right] = F(x)\, G'(x) - F'(x)\, G(x).
\end{equation*}

The quantity $\overline f(k,x)$ appearing in \eqref{2.16} and \eqref{2.17} is the left Jost solution to \eqref{2.18} satisfying the asymptotics similar to \eqref{2.10}, i.e. 
\begin{equation*}
%\label{2.23}
\begin{cases}
\overline f(k,x)=e^{kx}\left[1+o(1)\right], \qquad x\to+\infty,\\
\noalign{\medskip}
\overline f'(k,x)=k\,e^{kx}\left[1+o(1)\right], \qquad x\to+\infty,\\
\noalign{\medskip}
\overline f''(k,x)=k^2\,e^{kx}\left[1+o(1)\right],  \qquad x\to+\infty,
\end{cases}
\end{equation*}
and $\overline g(k,x)$ appearing in \eqref{2.16} and \eqref{2.17} is the right Jost solution to \eqref{2.18} satisfying the asymptotics similar to \eqref{2.11}, i.e. 
\begin{equation*}
%\label{2.24}
\begin{cases}
\overline g(k,x)=e^{kx}\left[1+o(1)\right], \qquad x\to-\infty,\\
\noalign{\medskip}
\overline g'(k,x)=k\,e^{kx}\left[1+o(1)\right], \qquad x\to-\infty,\\
\noalign{\medskip}
\overline g''(k,x)=k^2\,e^{kx}\left[1+o(1)\right],  \qquad x\to-\infty.
\end{cases}
\end{equation*}

The spacial asymptotics of  $h^\text{\rm{down}}(k,x)$ in its $k$-domain $\overline{\Omega_2}$ are known \cite{T2024}.
As $x\to+\infty$ we have
\begin{equation}\label{2.25}
h^\text{\rm{down}}(k,x)=\begin{cases}
\begin{aligned}
z(1-&z) k e^{kx}\,T_{\text{\rm{l}}}(z^2k)^{-1}[1+o(1)]\\
&-z(1-z)k e^{z^2kx}M(k)\,T_{\text{\rm{l}}}(k)^{-1}[1+o(1)], \qquad k\in \mathcal L_2,
\end{aligned}\\
\noalign{\medskip}
z(1-z) k e^{kx}\,T_{\text{\rm{l}}}(z^2k)^{-1}[1+o(1)], \qquad k\in \Omega_2,\\
\noalign{\medskip}
z(1-z) k e^{kx}\,T_{\text{\rm{l}}}(z^2k)^{-1}[1+o(1)],\qquad k\in \mathcal L_3,
\end{cases}
\end{equation}
where we recall that $M(k)$ and $T_{\text{\rm{l}}}(k)$ are the left scattering coefficients appearing in \eqref{2.14}.
Similarly, as $x\to-\infty$ we have
\begin{equation}\label{2.26}
h^\text{\rm{down}}(k,x)=\begin{cases}
z(1-z) k e^{kx}\,T_{\text{\rm{r}}}(zk)^{-1}[1+o(1)], \qquad k\in \mathcal L_2,\\
\noalign{\medskip}
z(1-z) k e^{kx}\,T_{\text{\rm{r}}}(zk)^{-1}[1+o(1)],  \qquad k\in \Omega_2,\\
\noalign{\medskip}
\begin{aligned} z(1-&z) ke^{kx}\,T_{\text{\rm{r}}}(zk)^{-1}[1+o(1)]\\
&-z(1-z)k e^{zkx}R(k)\,T_{\text{\rm{r}}}(k)^{-1}[1+o(1)],\qquad k\in \mathcal L_3,\end{aligned}
\end{cases}
\end{equation}
where we recall that $R(k)$ and $T_{\text{\rm{r}}}(k)$ are the right scattering coefficients appearing in \eqref{2.15}.
In a similar manner, the spacial asymptotics of  $h^\text{\rm{up}}(k,x)$ in its $k$-domain $\overline{\Omega_4}$ are known \cite{T2024}.
As $x\to+\infty$ we have
\begin{equation}\label{2.27}
h^\text{\rm{up}}(k,x)=\begin{cases}
\begin{aligned}
-z(1-&z) k e^{kx}\,T_{\text{\rm{l}}}(zk)^{-1}[1+o(1)]\\
&+z(1-z) k e^{zkx}L(k)\,T_{\text{\rm{l}}}(k)^{-1}[1+o(1)], \qquad k\in \mathcal L_1,
\end{aligned}\\
\noalign{\medskip}
-z(1-z) k e^{kx}\,T_{\text{\rm{l}}}(z k)^{-1}[1+o(1)],\qquad k\in \Omega_4,\\
\noalign{\medskip}
-z(1-z) k e^{kx}\,T_{\text{\rm{l}}}(z k)^{-1}[1+o(1)],\qquad k\in \mathcal L_4,
\end{cases}
\end{equation}
where we recall that $L(k)$ is the left primary reflection coefficient appearing in \eqref{2.14}.
Similarly, as $x\to-\infty$ we have
\begin{equation}\label{2.28}
h^\text{\rm{up}}(k,x)=\begin{cases}
-z(1-z) k e^{kx}\,T_{\text{\rm{r}}}(z^2 k)^{-1}[1+o(1)], \qquad k\in \mathcal L_1,\\
\noalign{\medskip}
-z(1-z) k e^{kx}\,T_{\text{\rm{r}}}(z^2 k)^{-1}[1+o(1)], \qquad k\in \Omega_4,\\
\noalign{\medskip}
\begin{aligned} -z(1-&z) k e^{kx}\,T_{\text{\rm{r}}}(z^2 k)^{-1}[1+o(1)]\\
&+z(1-z)ke^{z^2 kx}N(k)\,T_{\text{\rm{r}}}(k)^{-1}[1+o(1)],\qquad k\in \mathcal L_4,\end{aligned}
\end{cases}
\end{equation}
where we recall that $N(k)$ is the right secondary reflection coefficient appearing in \eqref{2.15}.

Since \eqref{1.1} is a linear homogeneous ordinary differential equation, any constant multiple of a solution to \eqref{1.1} is
also a solution. As seen from \eqref{2.25}--\eqref{2.28}, it is convenient to let
\begin{equation}\label{2.29}
m(k,x):=\ds\frac{ T_{\text{\rm{r}}}(zk)\, h^\text{\rm{down}}(k,x)}{z(1-z) k}, \qquad k\in\overline{\Omega_2}, \quad x\in\mathbb R,
\end{equation}
\begin{equation}\label{2.30}
n(k,x):=\ds\frac{ T_{\text{\rm{r}}}(z^2k)\, h^\text{\rm{up}}(k,x)}{-z(1-z) k}, \qquad  k\in\overline{\Omega_4}, \quad x\in\mathbb R,
\end{equation}
so that $m(k,x)$ and $n(k,x)$ remain solutions to \eqref{1.1} with
the respective $k$-domains $\overline{\Omega_2}$ and
$\overline{\Omega_4},$ but with simpler spacial asymptotics.
As $x\to+\infty,$ with the help of \eqref{2.25} and \eqref{2.29} we obtain
\begin{equation}\label{2.31}
m(k,x)=\begin{cases}
\begin{aligned}
e^{kx}&\,T_{\text{\rm{l}}}(z^2k)^{-1}\,T_{\text{\rm{r}}}(zk)
[1+o(1)]\\
&-e^{z^2kx}M(k)\,T_{\text{\rm{l}}}(k)^{-1} \,T_{\text{\rm{r}}}(zk)
[1+o(1)], \qquad k\in \mathcal L_2,
\end{aligned}\\
\noalign{\medskip}
e^{kx}\,T_{\text{\rm{l}}}(z^2k)^{-1}\,T_{\text{\rm{r}}}(zk)[1+o(1)], \qquad k\in \Omega_2,\\
\noalign{\medskip}
e^{kx}\,T_{\text{\rm{l}}}(z^2k)^{-1}\,T_{\text{\rm{r}}}(zk)[1+o(1)],\qquad k\in \mathcal L_3.
\end{cases}
\end{equation}
As $x\to-\infty,$ with the help of \eqref{2.26} and \eqref{2.29}, we have
\begin{equation}\label{2.32}
m(k,x)=\begin{cases}
e^{kx}[1+o(1)], \qquad k\in \mathcal L_2,\\
\noalign{\medskip}
e^{kx}[1+o(1)],  \qquad k\in \Omega_2,\\
\noalign{\medskip}
e^{kx}[1+o(1)]-e^{zkx}R(k)\,T_{\text{\rm{r}}}(k)^{-1}\,T_{\text{\rm{r}}}(zk)[1+o(1)],\qquad k\in \mathcal L_3.
\end{cases}
\end{equation}
As $x\to+\infty$ we get
\begin{equation}\label{2.33}
n(k,x)=\begin{cases}
\begin{aligned}
e^{kx}&\,T_{\text{\rm{l}}}(zk)^{-1}\,T_{\text{\rm{r}}}(z^2k)[1+o(1)]\\
&-e^{zkx}L(k)\,T_{\text{\rm{l}}}(k)^{-1}\,T_{\text{\rm{r}}}(z^2k)[1+o(1)], \qquad k\in \mathcal L_1,
\end{aligned}\\
\noalign{\medskip}
e^{kx}\,T_{\text{\rm{l}}}(z k)^{-1}\,T_{\text{\rm{r}}}(z^2k)[1+o(1)],\qquad k\in \Omega_4,\\
\noalign{\medskip}
e^{kx}\,T_{\text{\rm{l}}}(z k)^{-1}\,T_{\text{\rm{r}}}(z^2k)[1+o(1)],\qquad k\in \mathcal L_4.
\end{cases}
\end{equation}
Similarly, as $x\to-\infty$ we have
\begin{equation}\label{2.34}
n(k,x)=\begin{cases}
 e^{kx}[1+o(1)], \qquad k\in \mathcal L_1,\\
\noalign{\medskip}
e^{kx}[1+o(1)], \qquad k\in \Omega_4,\\
\noalign{\medskip}
e^{kx}[1+o(1)]
-e^{z^2 kx}N(k)\,T_{\text{\rm{r}}}(k)^{-1}\,T_{\text{\rm{r}}}(z^2k)[1+o(1)],\qquad k\in \mathcal L_4.
\end{cases}
\end{equation}
For each fixed $x\in\mathbb R,$ we have \cite{T2024} the large $k$-asymptotics
\begin{equation}\label{2.35}
\begin{cases}
m(k,x)=e^{kx}\left[1+O\left(\ds\frac{1}{k}\right)\right], \qquad  k\to\infty  \text{\rm{ in }} \overline{\Omega_2},\\
\noalign{\medskip}
n(k,x)=e^{kx}\left[1+O\left(\ds\frac{1}{k}\right)\right], \qquad  k\to\infty  \text{\rm{ in }} \overline{\Omega_4}.
\end{cases}
\end{equation}

In the reflectionless case, the spacial asymptotics of $m(k,x)$ and $n(k,x)$ are even simpler, and from \eqref{2.31}--\eqref{2.34} we obtain
\begin{equation*}
%\label{2.36}
m(k,x)=\begin{cases}
e^{kx}\,T_{\text{\rm{l}}}(z^2k)^{-1}\,T_{\text{\rm{r}}}(zk)
[1+o(1)],\qquad x\to+\infty, \quad k\in \overline{\Omega_2},\\
\noalign{\medskip}
e^{kx}[1+o(1)],\qquad  x\to-\infty, \quad k\in \overline{\Omega_2},
\end{cases}
\end{equation*}
\begin{equation}\label{2.37}
n(k,x)=\begin{cases}
e^{kx}\,T_{\text{\rm{l}}}(zk)^{-1}\,T_{\text{\rm{r}}}(z^2k)
[1+o(1)],\qquad x\to+\infty, \quad k\in \overline{\Omega_4},\\
\noalign{\medskip}
e^{kx}[1+o(1)],\qquad  x\to-\infty, \quad k\in \overline{\Omega_4}.
\end{cases}
\end{equation}

We refer to $f(k,x),$ $g(k,x),$ $h^\text{\rm{down}}(k,x),$ and $h^\text{\rm{up}}(k,x)$ as the four basic solutions to \eqref{1.1}.
With the help of \eqref{2.1}, we can show that $\psi(zk,x)$ and $\psi(z^2k,x)$ are solutions
to \eqref{1.1} whenever $\psi(k,x)$ is a solution to \eqref{1.1}. In fact, the $k$-domain of $\psi(zk,x)$ is obtained from the $k$-domain
of $\psi(k,x)$ by a clockwise rotation of $2\pi/3$ radians around the origin of the complex $k$-plane $\mathbb C.$ Similarly, the $k$-domain
of $\psi(z^2k,x)$ is obtained from the $k$-domain of $\psi(k,x)$ by a clockwise rotation of $4\pi/3$ radians around the origin of the complex
$k$-plane $\mathbb C.$ Thus, at every $k$-value in the complex plane, with the help of the four basic solutions $f(k,x),$ $g(k,x),$
$h^\text{\rm{down}}(k,x),$ and $h^\text{\rm{up}}(k,x),$ we can form three linearly independent solutions to \eqref{1.1}.

In the right plot of Figure~\ref{figure2.2}, we present three solutions to \eqref{1.1} at each $k$-value in the complex $k$-plane. As seen
from Figure~\ref{figure2.2}, this is done by displaying a specific set of three solutions to \eqref{1.1} in each of the six closed
sectors $\overline{\Omega_1^\text{\rm{up}}},$ $\overline{\Omega_1^\text{\rm{down}}},$ $\overline{\Omega_2},$ $\overline{\Omega_3^\text{\rm{down}}},$
$\overline{\Omega_3^\text{\rm{up}}},$ and $\overline{\Omega_4},$ respectively. We remark that $\Omega_1^\text{\rm{up}},$ $\Omega_1^\text{\rm{down}},$
$\Omega_3^\text{\rm{down}},$ and $\Omega_3^\text{\rm{up}}$ are the respective open sectors parametrized as
\begin{equation}\label{2.38}
\Omega_1^\text{\rm{up}}:=\left\{k\in\mathbb C: \ds\frac{2\pi}{3}<\arg[k]<\pi\right\},
\end{equation}
\begin{equation}\label{2.39}
\Omega_1^\text{\rm{down}}:=\left\{k\in\mathbb C: \pi<\arg[k]<\ds\frac{4\pi}{3}\right\},
\end{equation}
\begin{equation*}
%\label{2.40}
\Omega_3^\text{\rm{down}}:=\left\{k\in\mathbb C: -\ds\frac{\pi}{3}<\arg[k]<0\right\},
\end{equation*}
\begin{equation*}
%\label{2.41}
\Omega_3^\text{\rm{up}}:=\left\{k\in\mathbb C: 0<\arg[k]<\ds\frac{\pi}{3}\right\}.
\end{equation*}
The closed sector $\overline{\Omega_1^\text{\rm{up}}}$ is the closure of $\Omega_1^\text{\rm{up}},$ and it is obtained from $\Omega_1^\text{\rm{up}}$
by including its upper boundary $\mathcal L_1$ and its lower boundary $-\mathbb R^-.$ We use $-\mathbb R^-$ to denote the directed half line
given by
\begin{equation*}
%\label{2.40}
-\mathbb R^-:=\{k=-s: s\in[0,+\infty)\}.
\end{equation*}
Similarly, the closed sector $\overline{\Omega_1^\text{\rm{down}}}$ is the closure of $\Omega_1^\text{\rm{down}},$ and it is obtained from
$\Omega_1^\text{\rm{down}}$ by including its upper boundary $-\mathbb R^-$ and its lower boundary $\mathcal L_2.$ The closed sector
$\overline{\Omega_3^\text{\rm{down}}}$ is the closure of $\Omega_3^\text{\rm{down}},$ and it is obtained from $\Omega_3^\text{\rm{down}}$ by including
its lower boundary $\mathcal L_3$ and its upper boundary $\mathbb R^+,$ where $\mathbb R^+$ is the directed half line corresponding to the
real interval $[0,+\infty)$ in the complex $k$-plane. Similarly,
$\overline{\Omega_3^\text{\rm{up}}}$ is the closure of $\Omega_3^\text{\rm{up}},$ and it is obtained from $\Omega_3^\text{\rm{up}}$ by including its lower
boundary $\mathbb R^+$ and its upper boundary $\mathcal L_4.$

We define the $3$-Wronskian of three functions $F(x),$ $G(x),$ and $H(x)$ as
\begin{equation*}
%\label{2.41}
\left[F(x);G(x);H(x)\right]:=\begin{vmatrix}
F(x) & G(x) & H(x)\\
F'(x) & G'(x) & H'(x)\\
F''(x) & G''(x) & H''(x)
\end{vmatrix},
\end{equation*}
where we have the determinant of the relevant $3\times 3$ matrix on the right-hand side. 
The $3$-Wronskian of any three solutions to \eqref{1.1} at any particular $k$-value is zero if and only if those three
solutions are linearly dependent. Furthermore, because of the absence of $\psi''$ in \eqref{1.1}, 
the $3$-Wronskian of those three solutions is independent of $x$ and its value can be evaluated at any particular
$x$-value. For example, using their asymptotics as $x\to \pm\infty,$ we evaluate the $3$-Wronskian of $f(k,x),$ $g(zk,x),$ and
$h^\text{\rm{up}}(z^2k,x)$ in $\overline{\Omega_1^\text{\rm{down}}}$ as
\begin{equation}\label{2.42}
\left[f(k,x);g(zk,x);h^\text{\rm{up}}(z^2k,x)\right]=-9 z^2 k^4 \,T_{\text{\rm{l}}}(k)^{-1}\,T_{\text{\rm{r}}}(zk)^{-1},
\qquad k\in\overline{\Omega_1^\text{\rm{down}}}.
\end{equation}
Similarly, 
using their asymptotics as $x\to \pm\infty,$ we evaluate the $3$-Wronskian of $f(k,x),$ $g(z^2k,x),$ and
$h^\text{\rm{down}}(zk,x)$ in $\overline{\Omega_1^\text{\rm{up}}}$ as
\begin{equation}\label{2.43}
\left[f(k,x);g(z^2k,x);h^\text{\rm{down}}(zk,x)\right]=-9 z k^4 \,T_{\text{\rm{l}}}(k)^{-1}\,T_{\text{\rm{r}}}(zk)^{-1},
\qquad k\in\overline{\Omega_1^\text{\rm{up}}}.
\end{equation}
Using \eqref{2.30} and \eqref{2.42} we have
\begin{equation}\label{2.44}
\left[f(k,x);g(zk,x);n(z^2k,x)\right]=-3 z(1-z)k^3 \,T_{\text{\rm{l}}}(k)^{-1},
\qquad k\in\overline{\Omega_1^\text{\rm{down}}},
\end{equation}
and using \eqref{2.29} and \eqref{2.43} we get
\begin{equation}\label{2.45}
\left[f(k,x);g(z^2k,x);m(zk,x)\right]=3(1-z^2)k^3 \,T_{\text{\rm{l}}}(k)^{-1},
\qquad k\in\overline{\Omega_1^\text{\rm{up}}}.
\end{equation}
Comparing \eqref{2.42} and \eqref{2.43} with \eqref{2.44} and \eqref{2.45}, we see another advantage of using 
$m(k,x)$ and $n(k,x)$ instead of $h^\text{\rm{down}}(k,x)$ and $h^\text{\rm{up}}(k,x),$ respectively.
This is because, by using \eqref{2.44} and \eqref{2.45}, we can directly relate the bound-state poles 
the left transmission coefficient $T_{\text{\rm{l}}}(k)$ to the linear dependence of the three relevant solutions
to \eqref{1.1} in the appropriate $k$-domain.

If the potentials $Q$ and $P$ appearing in \eqref{1.1} depend on the time parameter $t,$ then the Jost solutions $f(k,x)$ and $g(k,x)$
to \eqref{1.1} also depend on $t.$ In general, the time dependence
of the potentials can be arbitrary. However, if those potentials satisfy a certain
integrable system, then the time evolution of the potentials must be compatible 
with the operator $A$ in the corresponding Lax pair $(L,A)$ for that integrable system.
Hence, in order to solve the integrable system \eqref{1.6} and its special case \eqref{1.10}, 
we need to use the time dependence governed by the linear operator $A$ specified in \eqref{1.3}.
In that case, the time evolution of $f(k,x)$ and $g(k,x)$ is determined with the help of the linear
operator $A$ given in \eqref{1.3}. In fact, the time evolution of any solution $\psi(k,x)$ to \eqref{1.1} is determined by using the requirement
\cite{A2005,A2009} that the quantity $\partial\psi(k,x)/\partial t-A\psi(k,x)$ is a solution to \eqref{1.1}. This yields the time evolution
for the Jost solutions $f(k,x)$ and $g(k,x),$ and we have
\begin{equation}\label{2.46}
\ds\frac{\partial f(k,x)}{\partial t}-A\,f(k,x)=-9k^5 f(k,x), \qquad
 k\in\overline{\Omega_1}, \quad x\in\mathbb R,
\end{equation}
\begin{equation}\label{2.47}
\ds\frac{\partial g(k,x)}{\partial t}-A\,g(k,x)=-9k^5 g(k,x), \qquad  k\in\overline{\Omega_3}, \quad  x\in\mathbb R.
\end{equation}
We use \eqref{2.14} in the asymptotics of \eqref{2.46} as $x\to-\infty,$ and similarly we use \eqref{2.15} in the asymptotics of \eqref{2.47} as $x\to+\infty.$
By equating the corresponding coefficients of $e^{kx},$ $e^{zkx},$ and $e^{z^2kx},$ respectively,
on both sides of the two aforementioned asymptotic equalities,
%
%
%
%
%
%With the help of \eqref{2.14}, \eqref{2.15}, \eqref{2.46}, and \eqref{2.47}, 
we obtain the time evolution of the scattering coefficients for \eqref{1.1}
as
\begin{equation}\label{2.48}
\begin{cases}
T_{\text{\rm{l}}}(k)\mapsto T_{\text{\rm{l}}}(k),\qquad k\in\overline{\Omega_1},\\
\noalign{\medskip}
T_{\text{\rm{r}}}(k)\mapsto T_{\text{\rm{l}}}(k),\qquad k\in\overline{\Omega_3},\\
\noalign{\medskip}
L(k)\mapsto L(k)\, \exp\left(9(z^2-1)k^5 t\right),\qquad k\in\mathcal L_1,\\
\noalign{\medskip}
M(k)\mapsto M(k)\, \exp\left(9(z-1)k^5 t\right),\qquad k\in\mathcal L_2,\\
\noalign{\medskip}
R(k)\mapsto R(k)\, \exp\left(9(z^2-1)k^5 t\right),\qquad k\in\mathcal L_3,\\
\noalign{\medskip}
N(k)\mapsto N(k)\, \exp\left(9(z-1)k^5 t\right),\qquad k\in\mathcal L_4.
\end{cases}
\end{equation}
As indicated in \eqref{2.48}, the transmission coefficients do not change in time.
On the other hand, the reflection coefficients at any time $t$ are obtained from 
the corresponding reflection coefficients at $t=0$ by using the indicated multiplicative exponential factors in \eqref{2.48}.
For example, the value $L(k)$ at any time $t$ is obtained by
multiplying the initial value of $L(k)$ at $t=0$ with the exponential factor
$e^{9(z^2-1)k^5t}.$

\section{The bound states and bound-state dependency constants}
\label{section3}

A bound-state solution to \eqref{1.1} is defined as a nontrivial solution which is square integrable in $x\in\mathbb R.$ If the bound state
occurs at the $k$-value $k_j$ in the complex $k$-plane, then the number of linearly independent square-integrable solutions to \eqref{1.1} at $k=k_j,$ i.e.
to the equation
\begin{equation}
\label{3.1}
\psi'''+Q(x)\,\psi'+P(x)\,\psi=k_j^3\,\psi, \qquad x\in\mathbb R,
\end{equation}
determines the
multiplicity of the bound state at $k=k_j.$ If there is only one linearly independent bound-state solution at $k=k_j,$ then the bound state at
$k=k_j$ is simple. 
We recall that we suppress the $t$-dependence of the potentials $Q$ and $P$ and any quantities related
to \eqref{3.1} in our notation.

Suppose that the complex constant $k_j$ is located in the open sector $\Omega_1^\text{\rm{down}}$ and that it
corresponds to a bound state. Then, we have
$T_{\text{\rm{l}}}(k_j)^{-1}=0.$ From \eqref{2.44} we see that the three solutions
$f(k_j,x),$ $g(zk_j,x),$ and $n(z^2k_j,x)$ to
\eqref{3.1} become linearly dependent.
This allows us
to express $f(k_j,x)$ as a linear combination of $g(zk_j,x)$ and $n(z^2k_j,x)$ as
\begin{equation}\label{3.2}
f(k_j,x)=D(k_j)\,g(zk_j,x)+W(k_j)\,n(z^2k_j,x), \qquad x\in\mathbb R,
\end{equation}
for some complex constants $D(k_j)$ and $W(k_j).$

Let us divide the open sector $\Omega_1^\text{\rm{down}}$ specified in \eqref{2.39} into two parts, the first of which is the sector
with $\arg[k]\in (\pi,7\pi/6)$ and the second sector is described by using $\arg[k]\in [7\pi/6,4\pi/3).$ If the bound state at $k=k_j$
occurs when $\arg[k_j]\in (\pi,7\pi/6),$ then from \eqref{3.2} we get $D(k_j)=0$ and hence \eqref{3.2} yields
\begin{equation}\label{3.3}
f(k_j,x)=W(k_j) \, n(z^2k_j,x),\qquad \arg[k_j]\in \left(\pi,\ds\frac{7\pi}{6}\right), \quad x\in\mathbb R,
\end{equation}
where $W(k_j)\neq 0$ and the complex constant $W(k_j)$ corresponds to the dependency constant at the bound state with $k=k_j.$
On the other hand, if we have $\arg[k_j]\in [7\pi/6,4\pi/3),$ then from \eqref{3.2} we get $W(k_j)=0$ and hence \eqref{3.2} yields
\begin{equation}\label{3.4}
f(k_j,x)=D(k_j) \, g(zk_j,x),\qquad \arg[k_j]\in \left[\ds\frac{7\pi}{6},\ds\frac{4\pi}{3}\right), \quad x\in\mathbb R,
\end{equation}
where $D(k_j)\neq 0$ and the complex constant $D(k_j)$ corresponds to the dependency constant at the bound state with $k=k_j.$

By using the following argument, we show how each of \eqref{3.3} and \eqref{3.4} follows from \eqref{3.2}.
When $k_j\in\Omega_1^\text{\rm{down}},$ the real and imaginary parts of the complex constant $k_j$ are both negative, i.e. we have
\begin{equation}\label{3.5}
\text{Re}[k_j]<0, \quad \text{Im}[k_j]<0,\qquad k_j\in\Omega_1^\text{\rm{down}}.
\end{equation}
Using \eqref{3.5} in the first line of \eqref{2.10} we see that $f(k_j,x)$ decays exponentially as $x\to +\infty.$ Since
$T_{\text{\rm{l}}}(k_j)^{-1}=0,$ from the second line of the right-hand side of \eqref{2.14} it follows that $f(k_j,x)$ also vanishes
as $x\to -\infty.$ Since those decays are exponential, $f(k_j,x)$ corresponds to a bound-state solution to \eqref{3.1}. With the help of the first line of
the right-hand side of \eqref{2.11}, we get 
\begin{equation}\label{3.6}
g(zk_j,x)=\text{exp}\left(\bigg(-\ds\frac{1}{2}+i\,\ds\frac{\sqrt{3}}{2}\bigg)\bigg(\text{Re}[k_j]+i\,\text{Im}[k_j]\bigg)x\right)\left[1+o(1)\right], \qquad x\to -\infty,
\end{equation}
which yields
\begin{equation}\label{3.7}
\left| g(zk_j,x) \right|=\left(\text{exp}\bigg(-\ds\frac{1}{2}\,\text{Re}[k_j]-\ds\frac{\sqrt{3}}{2}\,\text{Im}[k_j]\bigg)x\right)\left[1+o(1)\right], \qquad x\to -\infty.
\end{equation}
Using \eqref{3.5} on the right-hand side of \eqref{3.7}, we see that $g(zk_j,x)$ decays exponentially as $x\to -\infty.$ On the other hand,
from the second line of the right-hand side of \eqref{2.15} we observe that
\begin{equation}\label{3.8}
\left| g(zk_j,x)\right|=\left| T_{\text{\rm{r}}}(zk_j)^{-1}\right|\left(\text{exp}\bigg(-\ds\frac{1}{2}\,\text{Re}[k_j]-\ds\frac{\sqrt{3}}{2}\,\text{Im}[k_j]\bigg)x\right)\left[1+o(1)\right], \qquad x\to +\infty.
\end{equation}
When $\arg[k_j]\in (\pi,7\pi/6),$ we see from \eqref{3.8} that $\left| g(zk_j,x)\right|$ becomes unbounded as $x\to +\infty.$
On the other hand, when $\arg[k_j]\in [7\pi/6,4\pi/3),$ we see from \eqref{3.8} that $g(zk_j,x)$ decays exponentially as $x\to +\infty.$
Hence, for \eqref{3.2} to hold we must have $D(k_j)=0$ when $\arg[k_j]\in (\pi,7\pi/6).$ In a similar way, since $T_{\text{\rm{l}}}(k_j)^{-1}=0,$
from the first line of the right-hand side of \eqref{2.37} it follows that $n(z^2k_j,x)$ vanishes exponentially as $x\to +\infty.$ From the second line of
the right-hand side of \eqref{2.37}, we get
\begin{equation}\label{3.9}
n(z^2k_j,x)=\text{exp}\left(\bigg(-\ds\frac{1}{2}-i\,\ds\frac{\sqrt{3}}{2}\bigg)\bigg(\text{Re}[k_j]+i\,\text{Im}[k_j]\bigg)x\right)\left[1+o(1)\right], \qquad x\to +\infty,
\end{equation}
which yields
\begin{equation}\label{3.10}
\left| n(z^2k_j,x)\right|=\left(\text{exp}\bigg(-\ds\frac{1}{2}\,\text{Re}[k_j]+\ds\frac{\sqrt{3}}{2}\,\text{Im}[k_j]\bigg)x\right)\left[1+o(1)\right], \qquad x\to +\infty.
\end{equation}
From \eqref{3.10} we see that $n(z^2k_j,x)$ decays exponentially as $x\to +\infty$ when $\arg[k_j]\in(\pi,7\pi/6).$ On the other hand, 
when $\arg[k_j]=7\pi/6,$ it follows from \eqref{3.10} that $n(z^2k_j,x)$ remains bounded as $x\to +\infty$ but does not vanish as $x\to +\infty.$
When $\arg[k_j]\in(7\pi/6,4\pi/3),$ from \eqref{3.10} it follows that $\left| n(z^2k_j,x)\right|$ becomes unbounded as
$x\to +\infty$ and hence $n(z^2k_j,x)$ does not vanish as $x\to +\infty.$ Thus, for \eqref{3.2} to hold we must have $W(k_j)=0$ when
$\arg[k_j]\in[7\pi/6,4\pi/3).$ This concludes the justification of \eqref{3.3} and \eqref{3.4}.

Next, we determine the time evolution of the dependency constants $W(k_j)$ and $D(k_j)$ appearing in \eqref{3.3} and \eqref{3.4},
respectively. We first determine the time evolution of the dependency constant $W(k_j).$ In analogy with \eqref{2.46} and \eqref{2.47},
we have the time evolution of the solution $n(k,x)$ given by 
\begin{equation}\label{3.11}
\ds\frac{\partial}{\partial t}n(k,x)-A\,n(k,x)=-9k^5 \,n(k,x), \qquad k\in\overline{\Omega_1^\text{\rm{up}}}, \quad x\in\mathbb R.
\end{equation}
From \eqref{2.46} and \eqref{3.11} we have
\begin{equation}\label{3.12}
\ds\frac{\partial f(k_j,x)}{\partial t}-A\,f(k_j,x)=-9k_j^5 \,f(k_j,x), \qquad x\in\mathbb R,
\end{equation}
\begin{equation}\label{3.13}
\ds\frac{\partial}{\partial t}n(z^2k_j,x)-A\,n(z^2k_j,x)=-9zk_j^5 \,n(z^2k_j,x), \qquad x\in\mathbb R,
\end{equation}
where we have used $z^{10}=z$ in \eqref{3.13}. Using \eqref{3.3} in \eqref{3.12} we obtain 
\begin{equation}\label{3.14}
\ds\frac{\partial}{\partial t}\left(W(k_j)\,n(z^2k_j,x)\right)-A\left(W(k_j)\,n(z^2k_j,x)\right)=-9k_j^5 \,W(k_j)\,n(z^2k_j,x), \qquad x\in\mathbb R.
\end{equation}
Taking the $t$-derivative of the first term on the left-hand side of \eqref{3.14}, we get
\begin{equation}\label{3.15}
\begin{split}
\ds\frac{\partial W(k_j)}{\partial t}\,n(z^2k_j,x)+W(k_j)\,&\ds\frac{\partial n(z^2k_j,x)}{\partial t}-W(k_j)\,A\,n(z^2k_j,x)
\\
&=-9k_j^5 \,W(k_j)\,n(z^2k_j,x), \qquad x\in\mathbb R.
\end{split}
\end{equation}
Multiplying both sides of \eqref{3.13} by $W(k_j)$ and subtracting the resulting equality from \eqref{3.15}, we obtain   
\begin{equation}\label{3.16}
\ds\frac{\partial W(k_j)}{\partial t}\,n(z^2k_j,x)=(-9k_j^5+9zk_j^5) \,W(k_j)\,n(z^2k_j,x), \qquad x\in\mathbb R,
\end{equation}
which simplifies to
\begin{equation}\label{3.17}
\ds\frac{\partial W(k_j)}{\partial t}=9(z-1)k_j^5 \,W(k_j), \qquad x\in\mathbb R.
\end{equation}
Thus, the time evolution of $W(k_j)$ is given by
\begin{equation}\label{3.18}
W(k_j)=U(k_j)\,e^{9(z-1)k_j^5t}, \qquad \arg[k_j]\in (\pi,7\pi/6),
\end{equation}
where $U(k_j)$ is the dependency constant at $t=0.$ We refer to $U(k_j)$ as the initial bound-state dependency constant at $k=k_j.$

Next, we determine the time evolution of the dependency constant $D(k_j)$ appearing in \eqref{3.4}. We again assume that
we have a simple bound state at $k=k_j$ with $\arg[k_j]\in [7\pi/6,4\pi/3).$ From \eqref{2.46} and \eqref{2.47} we have
\begin{equation}\label{3.19}
\ds\frac{\partial g(zk_j,x)}{\partial t}-A\,g(zk_j,x)=-9z^2k_j^5 \,g(zk_j,x), \qquad x\in\mathbb R,
\end{equation}
where we have used $z^5=z^2$ in \eqref{3.19}. Using \eqref{3.4} in \eqref{3.12} we obtain
\begin{equation}\label{3.20}
\ds\frac{\partial}{\partial t}\big(D(k_j)\,g(zk_j,x)\big)-A\, \big(D(k_j)\,g(zk_j,x)\big)=-9k_j^5 \,D(k_j)\,g(zk_j,x), \qquad x\in\mathbb R.
\end{equation}
By taking the $t$-derivative of the first term on the left-hand side of \eqref{3.20}, we get
\begin{equation}\label{3.21}
\ds\frac{\partial D(k_j)}{\partial t}\,g(zk_j,x)+D(k_j)\,\ds\frac{\partial g(zk_j,x)}{\partial t}-D(k_j)\,A\,g(zk_j,x)=-9k_j^5\, D(k_j)\,g(zk_j,x), \qquad x\in\mathbb R.
\end{equation}
Multiplying both sides of \eqref{3.19} by $D(k_j)$ and subtracting the resulting equality from \eqref{3.21}, we obtain
\begin{equation}\label{3.22}
\ds\frac{\partial D(k_j)}{\partial t}\,g(zk_j,x)=\left(-9k_j^5+9z^2k_j^5\right)D(k_j)\,g(zk_j,x), \qquad x\in\mathbb R,
\end{equation}
which simplifies to
\begin{equation}\label{3.23}
\ds\frac{\partial D(k_j)}{\partial t}=9(z^2-1)k_j^5 \,D(k_j), \qquad x\in\mathbb R.
\end{equation}
Thus, the time evolution of $D(k_j)$ is given by
\begin{equation}\label{3.24}
D(k_j)=E(k_j)\,e^{9(z^2-1)\,k_j^5\,t},\qquad \arg[k_j]\in [7\pi/6,4\pi/3),
\end{equation}
where $E(k_j)$ is the dependency constant at $t=0.$ We refer to $E(k_j)$ as the initial dependency constant at the bound state at $k=k_j.$

A simple bound state at $k=k_j$ with $k_j\in\Omega_1^\text{\rm{down}}$ occurs when $T_{\text{\rm{l}}}(k)^{-1}$ has a simple zero
at $k=k_j$ or equivalently when $T_{\text{\rm{l}}}(k)$ has a simple pole at $k=k_j.$ A double pole at $k=k_j$ for $T_{\text{\rm{l}}}(k)$
corresponds to a bound state at $k=k_j$ having multiplicity two. 
If the bound state at $k=k_j$ has multiplicity $m_j,$ then instead of having a single bound-state
dependency constant $W(k_j)$ in \eqref{3.3} or $D(k_j)$ in \eqref{3.4}, we have $m_j$ bound-state dependency constants.
We refer the reader to \cite{AE2019,AE2022,AEU2023} for handling bound states with multiplicities and determining the corresponding dependency
constants in the presence of bound states with multiplicities. We postpone further elaborations on the multiplicities of the bound states for \eqref{1.1}
to a future paper.

Let us remark that a bound state can also be described with the help of 
a normalization constant instead of a dependency constant. For example, let us assume
that $k_j$ in $\Omega_1^\text{\rm{down}}$ corresponds to a simple bound-state pole of $T_{\text{\rm{l}}}(k).$ As seen from \eqref{3.3},
as a bound-state wavefunction, we can use either $f(k_j,x)$ or $n(z^2k_j,x)$ when $\arg[k_j]\in\left(\pi,7\pi/6\right)$ and we can use either
$f(k_j,x)$ or $g(zk_j,x)$ when $\arg[k_j]\in\left[7\pi/6,4\pi/3\right).$ Thus, when $\arg[k_j]\in\left(\pi,7\pi/6\right)$ the bound-state data set can be
specified by using $k_j$ and $W(k_j),$ and when $\arg[k_j]\in\left[7\pi/6,4\pi/3\right)$ the bound-state data set can be specified by using
$k_j$ and $D(k_j).$ Alternatively, we can introduce the left bound-state normalization constant $c_{\text{\rm{l}}}(k_j)$ by using
\begin{equation}\label{3.25}
c_{\text{\rm{l}}}(k_j):=\ds\frac{1}{\sqrt{\ds\int_{-\infty}^\infty dx \left| f(k_j,x) \right|^2}}, \qquad k_j\in\Omega_1^\text{\rm{down}},
\end{equation}
so that the quantity $c_{\text{\rm{l}}}(k_j)\, f(k_j,x)$ is a normalized bound-state solution to \eqref{3.1}. Instead of using the
left bound-state normalization constant, we can alternatively use the right bound-state normalization constant $c_{\text{\rm{r}}}(k_j)$ by letting
\begin{equation}\label{3.26}
c_{\text{\rm{r}}}(k_j):=
\begin{cases}
\ds\frac{1}{\sqrt{\ds\int_{-\infty}^\infty dx\, \left| n(z^2k_j,x)\right|^2}}, \qquad \arg[k_j]\in\left(\pi,\ds\frac{7\pi}{6}\right),\\
\noalign{\medskip}
\ds\frac{1}{\sqrt{\ds\int_{-\infty}^\infty dx\, \left| g(zk_j,x)\right|^2}}, \qquad \arg[k_j]\in\left[\ds\frac{7\pi}{6},\ds\frac{4\pi}{3}\right),
\end{cases}
\end{equation}
so that $c_{\text{\rm{r}}}(k_j)\, n(z^2k_j,x)$ is a normalized bound-state solution to \eqref{3.1} when
$\arg[k_j]\in\left(\pi,7\pi/6\right)$ or that $c_{\text{\rm{r}}}(k_j)\, g(zk_j,x)$ is a normalized bound-state solution to \eqref{3.1} when
$\arg[k_j]\in\left[7\pi/6,4\pi/3\right)$. From \eqref{3.3} and \eqref{3.4}, for $x\in\mathbb R$ we have
\begin{equation}\label{3.27}
\left| f(k_j,x)\right|^2=
\begin{cases}
\left| W(k_j)\right|^2 \left|n(z^2k_j,x) \right|^2, \qquad \arg[k_j]\in\left(\pi,\ds\frac{7\pi}{6}\right),\\
\noalign{\medskip}
\left| D(k_j)\right|^2 \left|g(zk_j,x)\right|^2, \qquad \arg[k_j]\in\left[\ds\frac{7\pi}{6},\ds\frac{4\pi}{3}\right).
\end{cases}
\end{equation}
Integrating both sides of \eqref{3.27} on $x\in\mathbb R$ and using \eqref{3.25} and \eqref{3.26}, we obtain
\begin{equation}\label{3.28}
\ds\frac{1}{c_{\text{\rm{l}}}(k_j)^2}=
\begin{cases}
\left| W(k_j)\right|^2 \ds\frac{1}{c_{\text{\rm{r}}}(k_j)^2}, \qquad \arg[k_j]\in\left(\pi,\ds\frac{7\pi}{6}\right),\\
\noalign{\medskip}
\left| D(k_j)\right|^2 \ds\frac{1}{c_{\text{\rm{r}}}(k_j)^2}, \qquad \arg[k_j]\in\left[\ds\frac{7\pi}{6},\ds\frac{4\pi}{3}\right).
\end{cases}
\end{equation}
Since $c_{\text{\rm{l}}}(k_j)$ and $c_{\text{\rm{r}}}(k_j)$ are chosen as positive constants, from \eqref{3.28} we get 
\begin{equation*}
%\label{3.29}
\ds\frac{c_{\text{\rm{r}}}(k_j)}{c_{\text{\rm{l}}}(k_j)}=
\begin{cases}
\left| W(k_j)\right|, \qquad \arg[k_j]\in\left(\pi,\ds\frac{7\pi}{6}\right),\\
\noalign{\medskip}
\left| D(k_j)\right|, \qquad \arg[k_j]\in\left[\ds\frac{7\pi}{6},\ds\frac{4\pi}{3}\right),
\end{cases}
\end{equation*}
which is the analog of the second equality in (18) of \cite{AK2001}. It is possible to relate each of the dependency constants
$W(k_j)$ and $D(k_j)$ to the normalization constants $c_{\text{\rm{l}}}(k_j)$ and $c_{\text{\rm{r}}}(k_j)$ by using the residue of
$T_{\text{\rm{l}}}(k)$ at the bound-state pole with $k=k_j.$ This can be achieved by expanding both sides of \eqref{2.44} at $k=k_j$ and by equating the corresponding coefficient of $(k-k_j)$ on both sides of the expansion.

We remark that the sectors $\Omega_1^\text{\rm{up}}$ and $\Omega_1^\text{\rm{down}}$ defined in
\eqref{2.38} and \eqref{2.39}, respectively, are symmetrically located with 
respect to the directed half line $-\mathbb R^-.$ There is a one-to-one correspondence between
the point $k_j\in\Omega_1^\text{\rm{down}}$ and 
the point $k_j^\ast\in\Omega_1^\text{\rm{up}},$ where we recall that an asterisk denotes complex conjugation.
Thus, if there is a bound state occurring at a $k$-value in the sector $\Omega_1^\text{\rm{up}},$ without loss of generality
we can assume that it occurs at $k=k_j^\ast$ with $k_j\in\Omega_1^\text{\rm{down}}.$ In that case, let us determine the
counterparts of \eqref{3.18} and \eqref{3.24}.
By proceeding in a manner similar to the steps given in \eqref{3.2}--\eqref{3.24},
we introduce the bound-state dependency constants $W(k_j^\ast)$ and
$D(k_j^\ast)$ as the complex constants satisfying for $x\in\mathbb R$
the equalities 
\begin{equation}
\label{3.30}
f(k_j^\ast,x)=
\begin{cases}
W(k_j^\ast) \,n(zk_j^\ast,x), \qquad \arg[k_j^\ast]\in\left(\ds\frac{5\pi}{6},\pi\right),\\
\noalign{\medskip}
D(k_j^\ast)\, g(z^2k_j^\ast,x), \qquad \arg[k_j^\ast]\in\left(\ds\frac{2\pi}{3},\ds\frac{5\pi}{6}\right].
\end{cases}
\end{equation}
The time evolution of $W(k_j^\ast)$ and $D(k_j^\ast)$ each is respectively given by
\begin{equation}\label{3.31}
\begin{cases}
W(k_j^\ast)=U(k_j^\ast)\,e^{9(z^2-1)\,(k_j^\ast)^5\,t}, \qquad \arg[k_j^\ast]\in\left(\ds\frac{5\pi}{6},\pi\right),\\
\noalign{\medskip}
D(k_j^\ast)=E(k_j^\ast)\,e^{9(z-1)\,(k_j^\ast)^5\,t}, \qquad \arg[k_j^\ast]\in\left(\ds\frac{2\pi}{3},\ds\frac{5\pi}{6}\right],
\end{cases}
\end{equation} 
where $U(k_j^\ast)$ and $E(k_j^\ast)$ are the respective dependency constants at $t=0.$ We refer to them as the initial bound-state dependency constants
at $k=k_j^\ast.$

Let us now consider the special case when the potentials $Q$ and $P$ appearing in \eqref{1.1} are real valued. In that case, if
there is a simple bound state at $k=k_j$ with $k_j\in\Omega_1^\text{\rm{down}}$
and also a simple bound state at $k=k_j^\ast$ with $k_j^\ast\in\Omega_1^\text{\rm{up}},$
then we have
\begin{equation}\label{3.32}
\begin{cases}
W(k_j^\ast)=W(k_j)^\ast,\quad U(k_j^\ast)=U(k_j)^\ast, \qquad \arg[k_j^\ast]\in\left(\ds\frac{5\pi}{6},\pi\right),\\
\noalign{\medskip}
D(k_j^\ast)=D(k_j)^\ast, \quad E(k_j^\ast)=E(k_j)^\ast, \qquad \arg[k_j^\ast]\in\left(\ds\frac{2\pi}{3},\ds\frac{5\pi}{6}\right].
\end{cases}
\end{equation}
The proof of \eqref{3.32} can be given as follows.
When the potentials $Q$ and $P$ in \eqref{3.1} are real valued, 
by taking the complex conjugate of both sides of \eqref{3.1} we get
\begin{equation}
\label{3.33}
\psi'''(k_j,x)^\ast+Q(x)\,\psi'(k_j,x)^\ast+P(x)\,\psi(k_j,x)^\ast=(k_j^3)^\ast\,\psi(k_j,x)^\ast, \qquad x\in\mathbb R.
\end{equation}
On the other hand, by replacing $k_j$ by $k_j^\ast,$ from \eqref{3.1} we obtain
\begin{equation}
\label{3.34}
\psi'''(k_j^\ast,x)+Q(x)\,\psi'(k_j^\ast,x)+P(x)\,\psi(k_j^\ast,x)=(k_j^\ast)^3\,\psi(k_j^\ast,x), \qquad x\in\mathbb R.
\end{equation}
Comparing \eqref{3.33} and \eqref{3.34}, we see that 
$f(k_j,x)^\ast$ and $f(k_j^\ast,x)$ satisfy the same differential equation.
From \eqref{2.10} it follows that  $f(k_j,x)^\ast$ and $f(k_j^\ast,x)$ satisfy 
the same asymptotics as $x\to+\infty$ obtained
from \eqref{2.10} by taking the complex conjugate of the right-hand side of
\eqref{2.10}. Because of the uniqueness of the solution
to \eqref{3.33} with the aforementioned asymptotics as $x\to+\infty,$ we have
\begin{equation}
\label{3.35}
f(k_j^\ast,x)=f(k_j,x)^\ast, \qquad x\in\mathbb R.
\end{equation}
A similar argument yields
\begin{equation}
\label{3.36}
n(z k_j^\ast,x)=n(z^2 k_j,x)^\ast, \qquad x\in\mathbb R,
\end{equation}
\begin{equation}
\label{3.37}
g(z^2 k_j^\ast,x)=g(z k_j,x)^\ast, \qquad x\in\mathbb R.
\end{equation}
By taking the complex conjugate of both sides of \eqref{3.3} and \eqref{3.4}, for $x\in\mathbb R$ we get
\begin{equation}\label{3.38}
f(k_j,x)^\ast=
\begin{cases}
W(k_j)^\ast\, n(z^2k_j,x)^\ast, \qquad \arg[k_j]\in\left(\pi,\ds\frac{7\pi}{6}\right),\\
\noalign{\medskip}
D(k_j)^\ast \,g(zk_j,x)^\ast, \qquad \arg[k_j]\in\left[\ds\frac{7\pi}{6},\ds\frac{4\pi}{3}\right).
\end{cases}
\end{equation}
Comparing \eqref{3.30} and \eqref{3.38}, with the help of \eqref{3.35},
\eqref{3.36}, and \eqref{3.37}, we conclude that the first equalities in both lines of the right-hand side of \eqref{3.32} hold.
Since those first equalities hold for all $t\in\mathbb R,$ they also hold at $t=0.$ 
Hence, with the help of \eqref{3.31} we see that the second equalities in both lines of the right-hand side of \eqref{3.32} also hold.

When the potentials $Q$ and $P$ are in the Schwartz class, or more generally when
the Jost solutions $f(k,x)$ and $g(k,x)$ to \eqref{1.1} have meromorphic extensions
in $k$ from their respective $k$-domains
$\overline{\Omega_1}$ and 
$\overline{\Omega_3}$ to the entire complex $k$-plane, we have the following
results. In that case, we can use either of the
sets $\{f(k,x),f(zk,x),f(z^2k,x)\}$ and
 $\{g(k,x),g(zk,x),g(z^2k,x)\}$ as a fundamental set of solutions to \eqref{1.1}.
In particular, any solution in the first set can be expressed as a linear combination of the 
solutions in the other set.
This yields
\begin{equation}
\label{3.39}
F(k,x)=G(k,x)\,\Lambda(k),\qquad k\in\mathbb C,
\end{equation}
\begin{equation}
\label{3.40}
G(k,x)=F(k,x)\,\Sigma(k),\qquad k\in\mathbb C,
\end{equation}
for some $3\times 3$ matrices $\Lambda(k)$ and
$\Sigma(k),$
where we have defined
\begin{equation}
\label{3.41}
F(k,x):=\begin{bmatrix} f(k,x)&f(zk,x)&f(z^2k,x)\\
\noalign{\medskip}
\ds\frac{d f(k,x)}{dx}&\ds\frac{d f(zk,x)}{dx}&\ds\frac{d f(z^2k,x)}{dx}\\
\noalign{\medskip}
\ds\frac{d^2 f(k,x)}{dx}&\ds\frac{d^2 f(zk,x)}{dx}&\ds\frac{d^2 f(z^2k,x)}{dx}
\end{bmatrix},
\end{equation}
\begin{equation}
\label{3.42}
G(k,x):=\begin{bmatrix} g(k,x)&g(zk,x)&g(z^2k,x)\\
\noalign{\medskip}
\ds\frac{d g(k,x)}{dx}&\ds\frac{d g(zk,x)}{dx}&\ds\frac{d g(z^2k,x)}{dx}\\
\noalign{\medskip}
\ds\frac{d^2 g(k,x)}{dx}&\ds\frac{d^2 g(zk,x)}{dx}&\ds\frac{d^2 g(z^2k,x)}{dx}
\end{bmatrix}.
\end{equation}
With the help of \eqref{2.10}, \eqref{2.11}, \eqref{2.14}, \eqref{2.15}, \eqref{3.41}, and \eqref{3.42},
from the
asymptotics as $x\to\pm\infty$ for \eqref{3.39} and \eqref{3.40}, we determine
the matrices $\Lambda(k)$ and $\Sigma(k)$ 
in terms of the left and right scattering coefficients for \eqref{1.1}, respectively, as
\begin{equation}
\label{3.43}
\Lambda(k)=\begin{bmatrix} \ds\frac{1}{T_{\text{\rm{l}}}(k)}& \ds\frac{M(zk)}{T_{\text{\rm{l}}}(zk)}& \ds\frac{L(z^2k)}{T_{\text{\rm{l}}}(z^2k)}\\
\noalign{\medskip}
 \ds\frac{L(k)}{T_{\text{\rm{l}}}(k)}& \ds\frac{1}{T_{\text{\rm{l}}}(zk)}& \ds\frac{M(z^2k)}{T_{\text{\rm{l}}}(z^2k)}\\
\noalign{\medskip}
 \ds\frac{M(k)}{T_{\text{\rm{l}}}(k)}& \ds\frac{L(zk)}{T_{\text{\rm{l}}}(zk)}& \ds\frac{1}{T_{\text{\rm{l}}}(z^2k)}
\end{bmatrix},
\end{equation}
\begin{equation}
\label{3.44}
\Sigma(k)=\begin{bmatrix} \ds\frac{1}{T_{\text{\rm{r}}}(k)}& \ds\frac{N(zk)}{T_{\text{\rm{r}}}(zk)}& \ds\frac{R(z^2k)}{T_{\text{\rm{r}}}(z^2k)}\\
\noalign{\medskip}
 \ds\frac{R(k)}{T_{\text{\rm{r}}}(k)}& \ds\frac{1}{T_{\text{\rm{r}}}(zk)}& \ds\frac{N(z^2k)}{T_{\text{\rm{r}}}(z^2k)}\\
\noalign{\medskip}
 \ds\frac{N(k)}{T_{\text{\rm{r}}}(k)}& \ds\frac{R(zk)}{T_{\text{\rm{r}}}(zk)}& \ds\frac{1}{T_{\text{\rm{r}}}(z^2k)}
\end{bmatrix}.
\end{equation}
From \eqref{3.39} and \eqref{3.40}, we see that the matrices $\Lambda(k)$ and $\Sigma(k)$ are inverses of each other.
In other words, we have
\begin{equation}
\label{3.45}
\Sigma(k)=\Lambda(k)^{-1},\qquad k\in\mathbb C.
\end{equation}
In the reflectionless case for \eqref{1.1}, from \eqref{3.43} and \eqref{3.44} we see that
the matrices $\Lambda(k)$ and $\Sigma(k)$ are
diagonal. Hence, in the reflectionless case, from \eqref{3.45} we conclude that
the left and right transmission coefficients for \eqref{1.1} must be reciprocals of each other, i.e. we have
\begin{equation}
\label{3.46}
T_{\text{\rm{r}}}(k)= \ds\frac{1}{T_{\text{\rm{l}}}(k)},\qquad k\in\mathbb C.
\end{equation}
We remark that the result presented in \eqref{3.46} is used in the
choice of the input data sets
in the examples in Sections~\ref{section5} and
\ref{section6}.
The result in \eqref{3.46} has also another implication in the reflectionless case
when the zeros and the poles of a transmission coefficient
are symmetrically located
with respect to the origin
of the complex $k$-plane, as it is the case
in \eqref{5.35} and \eqref{5.74}.
In that case, the bound-state poles of $T_{\text{\rm{r}}}(k)$ in
the sector $\Omega_3$ and
the bound-state poles of $T_{\text{\rm{l}}}(k)$
in the sector
$\Omega_1$ are located symmetrically with respect to the origin
of the complex $k$-plane. Thus, in the determination of soliton
solutions to integrable systems associated with
\eqref{1.1} in that case, the input data set consisting of the bound-state poles of 
the left transmission coefficient $T_{\text{\rm{l}}}(k)$
can equivalently be viewed as the bound-state poles of the right transmission coefficient $T_{\text{\rm{r}}}(k).$

\section{The Riemann--Hilbert problem  in the reflectionless case}
\label{section4}

The inverse scattering problem for \eqref{1.1} consists of the recovery of the potentials
$Q$ and $P$ from an appropriate scattering data set. Such a data set consists of the scattering coefficients
for \eqref{1.1} and the bound-state information.
In this section we consider the inverse scattering problem for \eqref{1.1} in the reflectionless case, i.e. 
in the special case when the reflection coefficients $L(k),$ $M(k),$ $R(k),$ $N(k)$ appearing
in \eqref{2.14} and \eqref{2.15} are all zero in their respective domains.

In terms of the four basic solutions $f(k,x),$ $g(k,x),$
$h^\text{\rm{down}}(k,x),$ and $h^\text{\rm{up}}(k,x)$ appearing in \eqref{2.14}, \eqref{2.15}, \eqref{2.16}, and \eqref{2.17},
respectively, we introduce the two functions $\Phi_+(k,x)$ and  $\Phi_-(k,x)$ by letting
\begin{equation}\label{4.1}
    \Phi_+(k,x) := 
    \begin{cases}
        T_\text{\rm{l}}(k)\, f(k,x), \qquad k\in \overline{\Omega_1},\\
                \noalign{\medskip}
         \ds\frac{T_\text{\rm{r}}(zk)\, h^\text{\rm{down}}(k,x)}{z(1-z)k}, \qquad k\in \overline{\Omega_2},
    \end{cases}
\end{equation}
\begin{equation}\label{4.2}
    \Phi_-(k,x) := 
    \begin{cases}
        g(k,x), \qquad k\in \overline{\Omega_3},\\
        \noalign{\medskip}
         \ds\frac{T_\text{\rm{r}}(z^2k)\, h^\text{\rm{up}}(k,x)}{-z(1-z)k}, \qquad k\in \overline{\Omega_4},
    \end{cases}
\end{equation}
where we recall that $T_{\text{\rm{l}}}(k)$ and $T_{\text{\rm{r}}}(k)$ are the left and right transmission coefficients appearing in \eqref{2.14}
and \eqref{2.15}, respectively.
With the help of \eqref{2.29} and \eqref{2.30}, we can express \eqref{4.1} and \eqref{4.2}, respectively, as 
\begin{equation}\label{4.3}
    \Phi_+(k,x) := 
    \begin{cases}
        T_\text{\rm{l}}(k)\, f(k,x), \qquad k\in \overline{\Omega_1},\\
                \noalign{\medskip}
        m(k,x), \qquad k\in \overline{\Omega_2},
    \end{cases}
\end{equation}
\begin{equation}\label{4.4}
    \Phi_-(k,x) := 
    \begin{cases}
        g(k,x), \qquad k\in \overline{\Omega_3},\\
        \noalign{\medskip}
        n(k,x), \qquad k\in \overline{\Omega_4}.
    \end{cases}
\end{equation}

\begin{figure}[!ht]
     \centering
         \includegraphics[width=2.in,height=3.4in]{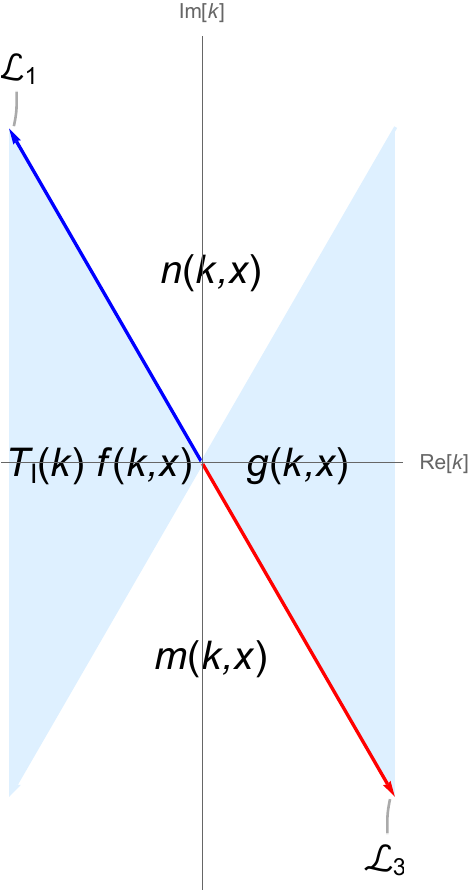}      \hskip .5in
         \includegraphics[width=2.in,height=3.4in]{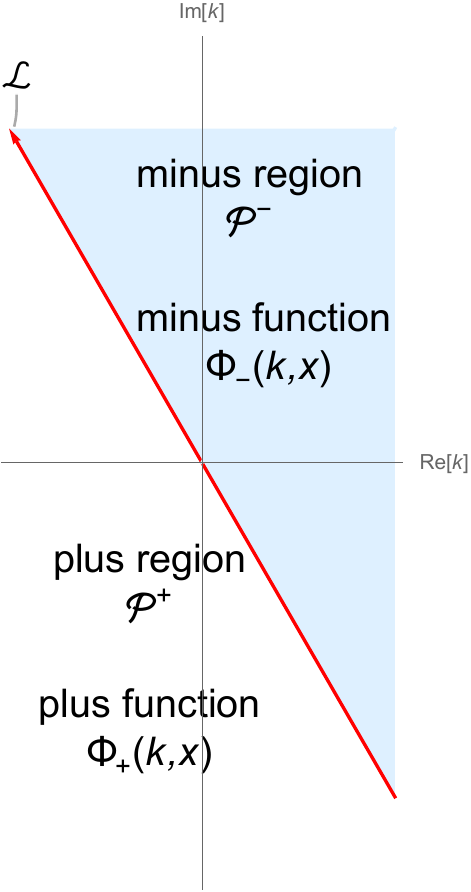} 
\caption{The $k$-domains of $T_{\text{\rm{l}}}(k) f(k,x),$ $m(k,x),$ $g(k,x),$ and $n(k,x),$ respectively, are shown on the left
plot. The right plot shows the plus and minus regions in the complex $k$-plane separated by the directed full line $\mathcal L,$
as well as the plus and minus functions in their respective $k$-domains.}
\label{figure4.1}
\end{figure}

In terms of the directed half lines $\mathcal L_1$ in \eqref{2.2} and $\mathcal L_3$ in \eqref{2.4},
we define the directed full line $\mathcal L$ by letting $\mathcal L:=\mathcal L_1\cup (-\mathcal L_3)$ so that $\mathcal L$
divides the complex $k$-plane into the open left-half complex plane $\mathcal P^+$ and the open right-half complex plane
$\mathcal P^-,$ as shown in the right plot of Figure~\ref{figure4.1}. We can parametrize $\mathcal L$  as
\begin{equation*}
%\label{4.5}
\mathcal L:=\{k\in\mathbb C: k=zs \text{\rm{ for }} s\in(-\infty,+\infty)\}.
\end{equation*}
We refer to $\mathcal P^+$ as the plus region and we use $\overline{\mathcal P^+}$ to denote its closure, i.e. we let
$\overline{\mathcal P^+}:=\mathcal P^+\cup \mathcal L.$ Similarly, we refer to $\mathcal P^-$ as the minus region and we
use $\overline{\mathcal P^-}$ to denote its closure, i.e. we let $\overline{\mathcal P^-}:=\mathcal P^-\cup \mathcal L.$
We assume that the left transmission coefficient
$T_{\text{\rm{l}}}(k)$ does not have any poles when 
$k\in\mathcal L_1\cup\mathcal L_2.$
Similarly, we assume that the right transmission coefficient
$T_{\text{\rm{r}}}(k)$ does not have any poles when 
$k\in\mathcal L_3\cup\mathcal L_4.$
It is known \cite{T2024} that the four basic solutions 
$f(k,x),$ $g(k,x),$
$h^\text{\rm{down}}(k,x),$ and $h^\text{\rm{up}}(k,x)$
are analytic in $k$ in the interior of their respective $k$-domains and are continuous in $k$
in their $k$-domains. Furthermore, when the secondary left reflection coefficient $M(k)$ appearing in \eqref{2.14}
vanishes for $k\in\mathcal L_2,$ we have
\begin{equation*}
%\label{4.6}
T_\text{\rm{l}}(k)\, f(k,x)=
\ds\frac{T_\text{\rm{r}}(zk)\, h^\text{\rm{down}}(k,x)}{z(1-z)k}, \qquad k\in \mathcal L_2.
\end{equation*}
 Similarly, when the secondary right reflection coefficient $N(k)$ appearing in \eqref{2.15}
vanishes for $k\in\mathcal L_4,$ we have
\begin{equation*}
%\label{4.7}
g(k,x)=\ds\frac{T_\text{\rm{r}}(z^2k)\, h^\text{\rm{up}}(k,x)}{-z(1-z)k}, \qquad k\in \mathcal L_4.
\end{equation*}
Thus, $\Phi_+(k,x)$ defined in \eqref{4.1} is meromorphic in $k\in\mathcal P^+,$ and it is
continuous in $k\in\overline{\mathcal P^+}$  except perhaps at the poles of the transmission coefficient
$T_{\text{\rm{l}}}(k)$ in $\Omega_1$ and at the poles of $T_{\text{\rm{r}}}(zk)$ in $\Omega_2.$
It is possible that some of those poles are offset by the zeros
of $f(k,x)$ and $h^\text{\rm{down}}(k,x),$ respectively.
Using \eqref{2.12}, \eqref{2.35}, and the known \cite{T2024}
large $k$-asymptotics of the transmission coefficients, we have
the large $k$-asymptotics of $\Phi_+(k,x)$ given by
\begin{equation}\label{4.8}
\Phi_+(k,x)=e^{kx}\left[1+O\left(\ds\frac{1}{k}\right)\right], \qquad  k\to\infty  \text{\rm{ in }} \overline{\mathcal P^+}.
\end{equation}
We refer to $\Phi_+(k,x)$ as the plus function. We remark that it is customary to use the term plus function when that function is analytic
in the plus region. We find it convenient to refer to $\Phi_+(k,x)$ as a plus function even though it is meromorphic rather than analytic
in the plus region $\mathcal P^+.$

In a similar manner, we establish that the function $\Phi_-(k,x)$ defined in \eqref{4.2} is meromorphic in
$k\in\mathcal P^-,$ and it is continuous in $k\in\overline{\mathcal P^-}$ except perhaps at the $k$-values corresponding to the poles of the transmission
coefficient of $T_{\text{\rm{r}}}(z^2k)$ in $\Omega_4.$
It is possible that some of the poles of $T_{\text{\rm{r}}}(z^2k)$ in $\Omega_4$
are offset by the zeros of $h^\text{\rm{up}}(k,x).$
Using \eqref{2.13}, \eqref{2.35}, and the known \cite{T2024}
large $k$-asymptotics of 
$T_{\text{\rm{r}}}(k)$ in its $k$-domain, we have
the large $k$-asymptotics of $\Phi_-(k,x)$ given by
\begin{equation}\label{4.9}
\Phi_-(k,x)=e^{kx}\left[1+O\left(\ds\frac{1}{k}\right)\right], \qquad  k\to\infty \text{\rm{ in }} \overline{\mathcal P^-}.
\end{equation}
We remark that we refer to $\Phi_-(k,x)$ as the minus function even though $\Phi_-(k,x)$ is meromorphic and not necessarily analytic
in the minus region $\mathcal P^-.$

The $k$-domain of $\Phi_+(k,x)$ is $\overline{\mathcal P^+}$ and the $k$-domain of $\Phi_-(k,x)$ is $\overline{\mathcal P^-}.$
The common $k$-domain of $\Phi_+(k,x)$ and $\Phi_-(k,x)$ is the directed full line $\mathcal L,$ and in fact in the reflectionless
case those two functions agree \cite{T2024} on $\mathcal L.$ Thus, we have
\begin{equation}\label{4.10}
\Phi_+(k,x)=\Phi_-(k,x), \qquad k\in\mathcal L.
\end{equation}
We view \eqref{4.10} as our Riemann--Hilbert problem as follows. As our input, we are given the transmission coefficients $T_{\text{\rm{l}}}(k)$ and
$T_{\text{\rm{r}}}(k)$ in their respective $k$-domains $\overline{\Omega_1}$ and $\overline{\Omega_3}.$ We know that the reflection coefficients 
are all zero. 
 In the reflectionless case,
the right transmission coefficient $T_\text{\rm{r}}(k)$ is related \cite{T2024} to the left transmission coefficient $T_\text{\rm{l}}(k)$ as
\begin{equation}\label{4.11}
T_\text{\rm{r}}(z^2k)=T_\text{\rm{l}}(k)\, T_\text{\rm{l}}(zk), \qquad k\in\overline{\Omega_1},
\end{equation}
and hence it is sufficient to specify only the left transmission coefficient $T_\text{\rm{l}}(k)$
in our input data set.
Our input data set also contains a dependency constant
for each pole of $T_\text{\rm{l}}(k)$ in $\Omega_1.$
In case of a pole with a multiplicity, the number of dependency constants for that pole
must match the multiplicity of the pole. However, in this paper we only deal with simple bound states.

With the given input data set, for each fixed $x\in\mathbb R,$ we solve
\eqref{4.10} to determine $\Phi_+(k,x)$ in $\overline{\mathcal P^+}$ and $\Phi_-(k,x)$ in $\overline{\mathcal P^-}$ in such a way that the large
$k$-asymptotics stated in \eqref{4.8} and \eqref{4.9} hold. Once we obtain $\Phi_+(k,x)$ and $\Phi_-(k,x)$ in their respective $k$-domains
$\overline{\mathcal P^+}$ and $\overline{\mathcal P^-},$ as seen from \eqref{4.1} and \eqref{4.2}, we can recover the four basic solutions
$f(k,x),$ $g(k,x),$ $h^\text{\rm{down}}(k,x),$ and $h^\text{\rm{up}}(k,x)$ in their respective $k$-domains $\overline{\Omega_1},$ $\overline{\Omega_2},$
$\overline{\Omega_3},$ $\overline{\Omega_4}.$ Having those four basic solutions at hand, we can use them
in their respective $k$-domains and we can recover the two potentials $Q(x)$ and $P(x).$ 
For example, we can recover $Q(x)$ and $P(x)$ by using either 
the left Jost solution $f(k,x)$ obtained in $k\in\Omega_1$ or the right Jost solution $g(k,x)$ obtained
in $k\in\Omega_3$ with the help of the large $k$-asymptotics of those solutions. The determination of $Q(x)$ and $P(x)$ from $f(k,x)$ can be achieved 
as follows. It is known \cite{T2024} that we have the asymptotic expansion
\begin{equation}\label{4.12}
 f(k,x) = e^{kx}\left[1 +\ds\frac{u_1(x)}{k} + \ds\frac{u_2(x)}{k^2} + O\left(\ds\frac{1}{k^3}\right)\right],\qquad k\to\infty  \text{\rm{ in }} \overline{\Omega_1},\quad x\in\mathbb R, 
 \end{equation}
 where we have defined
\begin{equation}\label{4.13}
 u_1(x):= \ds\frac{1}{3}\int_x^{\infty} dy\, Q(y),\qquad x\in\mathbb R,
 \end{equation}
 \begin{equation}\label{4.14}
 u_2(x):= -\ds\frac{1}{3}\int_x^{\infty} dy \left[Q'(y)-P(y)\right] + \ds\frac{1}{18}\left[\int_x^{\infty} dy\, Q(y)\right]^2,
 \qquad x\in\mathbb R.
  \end{equation}
With the help of \eqref{4.13} and \eqref{4.14}, the potentials $Q(x)$ and $P(x)$ can be recovered from $u_1(x)$ and $u_2(x)$ as
\begin{equation}\label{4.15}
 Q(x)= -3\,\ds\frac{du_1(x)}{dx}, \qquad x\in\mathbb R,
 \end{equation}
 \begin{equation}\label{4.16}
 P(x)= 3\left[u_1(x)\,\ds\frac{du_1(x)}{dx}- \ds\frac{d^2u_1(x)}{dx^2}-\ds\frac{du_2(x)}{dx}\right], \qquad x\in\mathbb R.
 \end{equation}
Alternatively, we can recover the potentials $Q(x)$ and $P(x)$ from the right Jost solution $g(k,x)$ as follows. It is known \cite{T2024} that
we have the asymptotic expansion
\begin{equation*}
%\label{4.17}
 g(k,x) = e^{kx}\left[1 +\ds\frac{v_1(x)}{k} + \ds\frac{v_2(x)}{k^2} + O\left(\ds\frac{1}{k^3}\right)\right],\qquad k\to\infty  \text{\rm{ in }}\overline{\Omega_4}, 
 \end{equation*}
 where we have defined
\begin{equation}\label{4.18}
 v_1(x):= -\ds\frac{1}{3}\int_{-\infty}^x dy\, Q(y),\qquad x\in\mathbb R,
 \end{equation}
 \begin{equation}\label{4.19}
 v_2(x):= \ds\frac{1}{3}\int_{-\infty}^x dy \left[Q'(y)-P(y)\right] + \ds\frac{1}{18}\left[\int_{-\infty}^x dy\, Q(y)\right]^2, \qquad x\in\mathbb R.
  \end{equation}
With the help of \eqref{4.18} and \eqref{4.19},
we can recover the potentials $Q(x)$ and $P(x)$ from $v_1(x)$ and $v_2(x)$ as
\begin{equation*}
%\label{4.20}
 Q(x)= -3\,\ds\frac{dv_1(x)}{dx}, \qquad x\in\mathbb R,
 \end{equation*}
 \begin{equation*}
%\label{4.21}
 P(x)= 3\left[v_1(x)\,\ds\frac{dv_1(x)}{dx}- \ds\frac{d^2v_1(x)}{dx^2}-\ds\frac{dv_2(x)}{dx}\right], \qquad x\in\mathbb R.
 \end{equation*}

We recall that we suppress the dependence on the parameter $t$ in our notation. The procedure outlined above
holds regardless of whether the potentials $Q$ and $P$ contain the parameter $t.$
The parameter $t$ in the input data set only appears in the time evolution of the bound-state dependency constants.
For example, if the  left transmission coefficient
$T_{\text{\rm{l}}}(k)$ has a simple pole at $k=k_j$ in $\Omega_1^\text{\rm{down}},$
then the corresponding bound-state  dependency constant $D(k_j)$ evolves in
time as in \eqref{3.24}. If the  left transmission coefficient
$T_{\text{\rm{l}}}(k)$ has a simple pole at $k=k_j^\ast$ in $\Omega_1^\text{\rm{up}},$
then the corresponding bound-state  dependency constant $D(k_j^\ast)$ evolves in
time as in the second line of \eqref{3.31}.

\section{Soliton solutions}
\label{section5}

In this section we present some explicit solutions to the system \eqref{1.6} and its special case given in \eqref{1.10}. This is done
by solving the inverse scattering problem for \eqref{1.1} explicitly with the appropriate input data set in the reflectionless case.
We recall that we suppress the appearance of the time parameter $t$ in the arguments of the functions we use.

In the first example below we assume that our input data set contains the left transmission coefficient $T_{\text{\rm{l}}}(k)$ with $\mathbf N$ simple poles occurring at $k=k_j$ 
for $1\le j\le \mathbf N,$ all located in the sector $\arg[k]\in[7\pi/6,4\pi/3).$ Our input data set is given by
$\{k_j,E(k_j)\}_{j=1}^{\mathbf N},$ where $E(k_j)$ is the initial bound-state dependency constant at $k=k_j$ appearing in \eqref{3.24}.
In case the simple poles $k_j$ occur in the sector  $\arg[k]\in(\pi,7\pi/6),$ the example can be modified so that one can instead use the input data set
$\{k_j, U(k_j)\}_{j=1}^{\mathbf N},$ where we recall that $U(k_j)$
is the initial bound-state dependency constant at $k=k_j$
appearing in \eqref{3.18}.

\begin{example}\label{example5.1}
\normalfont
In this example, we construct the $\mathbf N$-soliton solution to \eqref{1.6} by solving
the inverse scattering problem for \eqref{1.1} in the reflectionless case.
Our input data set consists of a transmission coefficient and the 
associated bound-state dependency constants.
We choose the left transmission
coefficient $T_\text{\rm{l}}(k)$ as
\begin{equation}\label{5.1}
T_\text{\rm{l}}(k)=\ds\frac{F(k)}{G(k)}, \qquad k\in\overline{\Omega_1},
\end{equation}
where the scalar quantities $F(k)$ and $G(k)$ are defined as
\begin{equation}\label{5.2}
F(k):=\ds\prod_{j=1}^{\mathbf N} (k+k_j^\ast),
\quad
G(k):=\ds\prod_{j=1}^{\mathbf N} (k-k_j).
\end{equation}
In \eqref{5.2}, $\mathbf N$ is a fixed positive integer
and the complex constants $k_j$ are distinct and all occur in the sector
$\arg[k]\in[7\pi/6,4\pi/3).$ 
From \eqref{3.46} we know that the right transmission coefficient $T_\text{\rm{r}}(k)$ is given by the reciprocal
of $T_\text{\rm{l}}(k),$ and hence we have
\begin{equation}\label{5.3}
T_\text{\rm{r}}(k)=\ds\frac{G(k)}{F(k)}, \qquad k\in\overline{\Omega_3}.
\end{equation}
We remark that \eqref{5.1} and \eqref{5.3} are compatible with the constraint \eqref{4.11}. Our Riemann--Hilbert problem \eqref{4.10}
with the large $k$-asymptotics in \eqref{4.8} and \eqref{4.9} can be solved by multiplying both sides of \eqref{4.10} by
$e^{-kx}\,G(k).$ From \eqref{4.10} we get
\begin{equation}\label{5.4}
e^{-kx}\,G(k)\,\Phi_+(k,x)=e^{-kx}\,G(k)\,\Phi_-(k,x), \qquad k\in \mathcal L.
\end{equation}
The left-hand side of \eqref{5.4} has an analytic extension from $k\in\mathcal L$ to the plus region $\mathcal P^+$ and the right-hand
side of \eqref{5.4} has an analytic extension from $k\in\mathcal L$ to the minus region $\mathcal P^-.$ Using the large $k$-asymptotics given in \eqref{4.8}
and \eqref{4.9}, with the help of a generalization of the Liouville theorem \cite{R1987}, we conclude that each side of \eqref{5.4} is entire and equal
to a monic polynomial in $k$ of degree $\mathbf N.$ We define $V(k)$ as the row vector with $\mathbf N$ components given by
\begin{equation}\label{5.5}
V(k):=\begin{bmatrix}
k^{\mathbf N-1} &k^{\mathbf N-2} & \cdots & k &1
\end{bmatrix},\qquad k\in\mathbb C.
\end{equation}
We use $\mathbf A(x)$ to denote the column vector with $\mathbf N$ components, which is given by
\begin{equation}\label{5.6}
\mathbf A(x):=\begin{bmatrix}
A_{\mathbf N-1}(x)\\
A_{\mathbf N-2}(x)\\
\vdots\\
A_1(x)\\
A_0(x)
\end{bmatrix},\qquad x\in\mathbb R,
\end{equation}
where $A_j(x)$ is a scalar-valued function of $x\in\mathbb R$ for $0\le j\le \mathbf N-1.$ Thus, each side of \eqref{5.4} is equal to the
monic polynomial $[k^{\mathbf N}+V(k)\,\mathbf A(x)],$ and we obtain the solution to the Riemann--Hilbert problem \eqref{4.10} as
\begin{equation}\label{5.7}
\Phi_+(k,x)=e^{kx}\,\ds\frac{k^{\mathbf N}+V(k)\,\mathbf A(x)}{G(k)}, \qquad k\in \overline{\mathcal P^+}, \quad x\in\mathbb R,
\end{equation}
\begin{equation}\label{5.8}
\Phi_-(k,x)=e^{kx}\,\ds\frac{k^{\mathbf N}+V(k)\,\mathbf A(x)}{G(k)}, \qquad k\in \overline{\mathcal P^-}, \quad x\in\mathbb R.
\end{equation}
Comparing \eqref{5.7} with \eqref{4.3}, with the help of \eqref{5.1} we recover the solutions $f(k,x)$ and $m(k,x)$ to \eqref{1.1} as
\begin{equation}\label{5.9}
f(k,x)=e^{kx}\,\ds\frac{k^{\mathbf N}+V(k)\,\mathbf A(x)}{F(k)}, \qquad k\in\overline{\Omega_1}, \quad x\in \mathbb R,
\end{equation}
\begin{equation}\label{5.10}
m(k,x)=e^{kx}\,\ds\frac{k^{\mathbf N}+V(k)\,\mathbf A(x)}{G(k)}, \qquad k\in\overline{\Omega_2}, \quad x\in \mathbb R.
\end{equation}
Similarly, comparing \eqref{5.8} and \eqref{4.4} we recover the solutions $g(k,x)$ and $n(k,x)$ to \eqref{1.1} as
\begin{equation}\label{5.11}
g(k,x)=e^{kx}\,\ds\frac{k^{\mathbf N}+V(k)\,\mathbf A(x)}{G(k)}, \qquad k\in\overline{\Omega_3}, \quad x\in \mathbb R,
\end{equation}
\begin{equation}\label{5.12}
n(k,x)=e^{kx}\,\ds\frac{k^{\mathbf N}+V(k)\,\mathbf A(x)}{G(k)}, \qquad k\in\overline{\Omega_4}, \quad x\in \mathbb R.
\end{equation}
We note that the column vector $\mathbf A(x)$ with $\mathbf N$ components is uniquely determined by using the $\mathbf N$ initial dependency constants
$E(k_j)$ given in our input data set for $1\le j\le \mathbf N.$ From \eqref{3.4} and \eqref{3.24} we get
\begin{equation}\label{5.13}
f(k_j,x)=E(k_j)
\, e^{9(z^2-1)k_j^5t}
\, g(zk_j,x), \qquad 1\le j\le \mathbf N.
\end{equation}
Using \eqref{5.9} and \eqref{5.11} in \eqref{5.13}, we obtain
\begin{equation}\label{5.14}
e^{k_jx}\, \ds\frac{k_j^{\mathbf N}+V(k_j)\,\mathbf A(x)}{F(k_j)}
=E(k_j)
\,e^{9(z^2-1)k_j^5t}
e^{zk_jx}
\,\ds\frac{(zk_j)^{\mathbf N}+V(zk_j)\,\mathbf A(x)}{G(zk_j)}, \qquad 1\le j\le \mathbf N.
\end{equation}
In terms of the initial dependency constants $E(k_j),$ we introduce the modified initial dependency constants
$\gamma(k_j)$ as
\begin{equation}\label{5.15}
\gamma(k_j):=-\ds\frac{F(k_j)\, E(k_j)}{G(zk_j)}, \qquad 1\le j\le \mathbf N.
\end{equation}
We emphasize that each $\gamma(k_j)$ is a complex-valued constant independent of $x$ and $t.$ We also introduce 
the quantities $\chi(k_j)$ by letting
\begin{equation}\label{5.16}
\chi(k_j):=\exp\left((z-1)k_j x+9(z^2-1)k_j^5 t\right), \qquad 1\le j\le \mathbf N.
\end{equation}
We remark that each $\chi(k_j)$ is a function of $x$ and $t.$ Since $x$ and
$t$ are
real-valued parameters, it follows that each $\chi(k_j)$ is real valued if and only if
$k_j=iz\eta_j$ for some positive constant $\eta_j.$ This can be justified as follows. From \eqref{5.16} we see that for $\chi(k_j)$ to be real valued, it
is necessary that $(z-1)k_j$ is real. With the help of \eqref{2.1}, we evaluate the real and imaginary parts of $(z-1)k_j$ and we get
\begin{equation}\label{5.17}
(z-1)k_j=\left(-\ds\frac{3}{2}\,\text{\rm{Re}}[k_j]-\ds\frac{\sqrt{3}}{2}\,\text{\rm{Im}}[k_j]\right)
+i\left(\ds\frac{\sqrt{3}}{2}\,\text{\rm{Re}}[k_j]-\ds\frac{3}{2}\,\text{\rm{Im}}[k_j]\right),
\end{equation}
where we use $\text{\rm{Re}}[k_j]$ and $\text{\rm{Im}}[k_j]$ to denote the real and imaginary parts,
respectively, of the complex
constant $k_j.$ From \eqref{5.17} we see that $(z-1)k_j$ is real if and only if
 $\text{\rm{Re}}[k_j]=\sqrt{3}\,\text{\rm{Im}}[k_j],$ which is equivalent to having $k_j=iz\eta_j$ for a real
 constant $\eta_j.$ Since $k_j$ is in the sector $\arg[k]\in[7\pi/6,4\pi/3),$ we must have $\eta_j>0.$
Then, with $k_j=iz\eta_j,$ from \eqref{5.16} we get
\begin{equation}\label{5.18}
\chi(iz\eta_j)=\exp\left(\sqrt{3}\,\eta_j\,x-9\,\sqrt{3}\,\eta_j^5\,t\right), \qquad 1\le j\le \mathbf N.
\end{equation}
Comparing \eqref{5.16} and \eqref{5.18}, we observe that if the quantity $(z-1)k_j$ is real then the quantity $(z^2-1)k_j^5$ is also real.
Hence, we have shown that each $\chi(k_j)$ is real valued if and only if
$k_j=iz\eta_j$ for some positive constant $\eta_j.$ 
Using \eqref{5.15} and \eqref{5.18}, we write \eqref{5.14} as
\begin{equation}\label{5.19}
k_j^{\mathbf N}+V(k_j)\,\mathbf A(x)=-\gamma(k_j)\,\chi(k_j)\left[(zk_j)^{\mathbf N}+V(zk_j)\,\mathbf A(x)\right],\qquad 1\le j\le \mathbf N.
\end{equation}
We observe that \eqref{5.19} for $1\le j\le \mathbf N$ yields a linear algebraic system of $\mathbf N$ equations with the $\mathbf N$ unknowns $A_l(x)$ for
$0\le l\le \mathbf N-1.$ This can be seen by defining the scalar quantities $m_l(k_j)$ as
\begin{equation}\label{5.20}
m_l(k_j):=k_j^l+(zk_j)^l\,\gamma(k_j)\,\chi(k_j),\qquad 1\le j\le\mathbf N,\quad 0\le l\le \mathbf N.
\end{equation}
We remark that the dependence on $x$ and $t$ in $m_l(k_j)$ appears only through $\chi(k_j).$ Using
\eqref{5.20} we form the $\mathbf N\times \mathbf N$ matrix $\mathbf M(x)$ and the column vector $\mathbf B(x)$ with $\mathbf N$ components as
\begin{equation}\label{5.21}
\mathbf M(x):=\begin{bmatrix}
m_{\mathbf N-1}(k_1) & m_{\mathbf N-2}(k_1) & \cdots & m_1(k_1) & m_0(k_1)\\
m_{\mathbf N-1}(k_2) & m_{\mathbf N-2}(k_2) & \cdots & m_1(k_2) & m_0(k_2)\\
\vdots & \vdots & \ddots & \vdots & \vdots \\
m_{\mathbf N-1}(k_{\mathbf N}) & m_{\mathbf N-2}(k_{\mathbf N}) & \cdots & m_1(k_{\mathbf N}) & m_0(k_{\mathbf N})
\end{bmatrix},
\end{equation}
\begin{equation}\label{5.22}
\mathbf B(x):=\begin{bmatrix}
m_{\mathbf N}(k_1)\\
m_{\mathbf N}(k_2)\\
\vdots\\
m_{\mathbf N}(k_{\mathbf N})
\end{bmatrix}.
\end{equation}
With the help of \eqref{5.21} and \eqref{5.22}, we write the linear algebraic system \eqref{5.19} as
\begin{equation}\label{5.23}
\mathbf M(x)\, \mathbf A(x)=-\mathbf B(x),
\end{equation}
and hence the column vector $\mathbf A(x)$ is obtained as
\begin{equation}\label{5.24}
\mathbf A(x)=-\mathbf M(x)^{-1}\mathbf B(x),
\end{equation}
where we stress that we suppress in our notation the $t$-dependence in both $\mathbf M(x)$ and $\mathbf A(x).$ Alternatively, the
 $\mathbf N$ components $A_j(x)$ with $0\le j\le \mathbf N-1$ of the column vector $\mathbf A(x)$ can be obtained as the ratio of two
determinants, and this is achieved by applying Cramer's rule on the linear algebraic system \eqref{5.23}. For example, we have
\begin{equation}\label{5.25}
A_{\mathbf N-1}(x)=-\ds\frac{\det[\mathbf M_1(x)]}{\det[\mathbf M(x)]},\qquad \mathbf N\ge 1,
\end{equation}
\begin{equation}\label{5.26}
A_{\mathbf N-2}(x)=-\ds\frac{\det[\mathbf M_2(x)]}{\det[\mathbf M(x)]},\qquad \mathbf N\ge 2,
\end{equation}
where we use $\mathbf M_1(x)$ to denote the $\mathbf N\times \mathbf N$ matrix obtained by replacing the first column of $\mathbf M(x)$ with the column
vector $\mathbf B(x),$ and we use $\mathbf M_2(x)$ to denote the $\mathbf N\times \mathbf N$ matrix obtained by replacing the second column of
$\mathbf M(x)$ with $\mathbf B(x).$ Let us remark that, when $\mathbf N=1,$ the column vector $\mathbf A(x)$ contains only one entry, namely, $A_0(x).$
Hence, \eqref{5.26} is valid and needed only for $\mathbf N\ge 2.$
Having obtained the column vector $\mathbf A(x)$ explicitly in terms of the input data set
$\{k_j,E(k_j)\}_{j=1}^{\mathbf N},$ as seen from \eqref{5.9}--\eqref{5.12}, we have the solutions $f(k,x),$ $m(k,x),$
$g(k,x),$ and $n(k,x)$ to \eqref{1.1}, where each of those solutions in their respective $k$-domains is explicitly expressed in
terms of the input data set consisting of the bound-state poles of $T_\text{\rm{l}}(k)$ in $\Omega_1^\text{\rm{down}}$ and the corresponding
bound-state
dependency constants. Having $f(k,x)$ at hand, we can recover the potentials $Q(x)$ and $P(x)$ with the help of
\eqref{4.15} and \eqref{4.16}, respectively. For this we proceed as follows. Comparing \eqref{4.12} with \eqref{5.9} we observe
that
\begin{equation}\label{5.27}
\ds\frac{k^{\mathbf N}+V(x)\,\mathbf A(x)}{F(k)}=1+\ds\frac{u_1(x)}{k}+\ds\frac{u_2(x)}{k^2}+O\left(\ds\frac{1}{k^3}\right), \qquad k\to \infty \text{\rm{ in }} \overline{\Omega_1}.
\end{equation}
With the help of \eqref{5.2}, \eqref{5.5}, and \eqref{5.6}, we expand the left-hand side of \eqref{5.27} in powers of $1/k$ and obtain
\begin{equation}\label{5.28}
\begin{split}
\ds\frac{k^{\mathbf N}+V(x)\,\mathbf A(x)}{F(k)}=&1+\ds\frac{A_{\mathbf N-1}(x)-\Sigma_{\mathbf N}}{k}\\
    &+\ds\frac{A_{\mathbf N-2}(x)-\Sigma_{\mathbf N} A_{\mathbf N-1}(x)+\Pi_{\mathbf N}}{k^2}+O\left(\ds\frac{1}{k^3}\right), \qquad k\to \infty \text{\rm{ in }} \overline{\Omega_1},
\end{split}
\end{equation}
where we have defined
\begin{equation}\label{5.29}
\Sigma_{\mathbf N}:=k_1^\ast+k_2^\ast+\cdots+k_{\mathbf N}^\ast,
\end{equation}
\begin{equation}\label{5.30}
\Pi_{\mathbf N}:=k_1^\ast\left(k_1^\ast+k_2^\ast\cdots+k_{\mathbf N}^\ast \right)+k_2^\ast\left(k_2^\ast+k_3^\ast+\cdots+k_{\mathbf N}^\ast \right)
 +\cdots+k_{\mathbf N}^\ast\left(k_{\mathbf N}^\ast \right).
\end{equation}
Comparing \eqref{5.28} with \eqref{5.27}, we see that
\begin{equation}\label{5.31}
\begin{cases}
u_1(x)=A_{\mathbf N-1}(x)-\Sigma_{\mathbf N},\\
\noalign{\medskip}
u_2(x)=A_{\mathbf N-2}(x)-\Sigma_{\mathbf N} A_{\mathbf N-1}(x)+\Pi_{\mathbf N}.
\end{cases}
\end{equation}
Using \eqref{5.31} in \eqref{4.15} and \eqref{4.16}, we express the potentials $Q(x)$ and $P(x)$ in terms of $A_{\mathbf N-1}(x)$ and
$A_{\mathbf N-2}(x)$ as
\begin{equation}\label{5.32}
\begin{cases}
Q(x)=-3\,\ds\frac{dA_{\mathbf N-1}(x)}{dx},\\
\noalign{\medskip}
P(x)=3\left(A_{\mathbf N-1}(x)\,\ds\frac{dA_{\mathbf N-1}(x)}{dx}-\ds\frac{d^2A_{\mathbf N-1}(x)}{dx^2}-\ds\frac{dA_{\mathbf N-2}(x)}{dx}\right).
\end{cases}
\end{equation}
We remark that \eqref{5.27}--\eqref{5.32} are valid for $\mathbf N\ge 1$ with the understanding that we use $A_{-1}(x)\equiv 0$
on the right-hand sides of \eqref{5.28}, \eqref{5.31}, and \eqref{5.32} when $\mathbf N=1.$
From \eqref{5.31} and \eqref{5.32} we observe that the constant terms $\Sigma_{\mathbf N}$ and $\Pi_{\mathbf N}$ do not appear in \eqref{5.32}.
Using \eqref{5.25} and \eqref{5.26} in \eqref{5.32}, we see that the potentials $Q(x)$ and $P(x)$ are explicitly constructed in terms of
the input data set $\{k_j, E(k_j)\}_{j=1}^{\mathbf N}.$ As seen from the first line of \eqref{5.32}, the potential
$Q(x)$ is determined by $A_{\mathbf N-1}(x)$ only. Similarly, as seen from the second line of \eqref{5.32}, the potential
$P(x)$ is determined by $A_{\mathbf N-1}(x)$ and $A_{\mathbf N-2}(x)$ only.
The remaining entries of $\mathbf A(x)$ are only needed to construct the solutions to
\eqref{1.1}.
Having constructed $Q(x)$ and $P(x)$ in terms of our input data set,
we now have  the explicit solution pair $Q(x)$ and
$P(x)$ to the integrable nonlinear system given in \eqref{1.6}. Thus, we have demonstrated
the construction of the $\mathbf N$-soliton solution to \eqref{1.6}
by using the input data set consisting of the locations of the poles $k_j$ for
$1\le j\le \mathbf N$ as well as the complex-valued initial dependency constants $E(k_j).$ 
We note that the $1$-soliton solution to \eqref{1.6} when $k_1=iz\eta_1$ for some positive $\eta_1$ is obtained 
from \eqref{5.32} as
\begin{equation}\label{5.33}
Q(x)=\ds\frac{9\,\eta_1^2\,\gamma(k_1)\,e^{\sqrt{3}\eta_1(x-9\eta_1^4 t)}
}{\left(1+\gamma(k_1)\,e^{\sqrt{3}
\eta_1(x-9\eta_1^4 t)
}\right)^2},
 \qquad x\in \mathbb R,
\end{equation}
\begin{equation*}
%\label{5.34}
P(x)=\ds\frac{9iz\,\eta_1^3\,\gamma(k_1)\,e^{\sqrt{3}\eta_1(x-9\eta_1^4 t)}\left(z+\gamma(k_1)\,e^{\sqrt{3}
\eta_1(x-9\eta_1^4 t)
}\right)
}{\left(1+\gamma(k_1)\,e^{\sqrt{3}
\eta_1(x-9\eta_1^4 t)
}\right)^3},
 \qquad x\in \mathbb R,
\end{equation*}
where we recall that $z$ is the special constant appearing in \eqref{2.1}.
From \eqref{5.33} we see that
$Q(x)$ becomes real valued and nonsingular when $\gamma(k_1)$ is chosen as
positive. However, in that case $P(x)$ remains complex valued.

\end{example}

In the next example, we construct the $2\mathbf N$-soliton solution to \eqref{1.6}
again by solving the inverse scattering problem for \eqref{1.1} in the reflectionless case with
a particular input data set. One of the reasons for us to present this example is that 
our $2\mathbf N$-soliton solution to \eqref{1.6} yields Hirota's real-valued $\mathbf N$-soliton solution to the Sawada--Kotera
equation by using an appropriate restriction on our input data set.

\begin{example}\label{example5.2}
\normalfont
In this example we choose our
left transmission coefficient
 $T_\text{\rm{l}}(k)$ as
\begin{equation}\label{5.35}
T_\text{\rm{l}}(k)=\ds\frac{\Gamma(k)}{\Gamma(-k)}, \qquad k\in\overline{\Omega_1},
\end{equation}
where the scalar quantity $\Gamma(k)$ is defined as
\begin{equation}\label{5.36}
\Gamma(k):=\ds\prod_{j=1}^{\mathbf N} (k+k_j)(k+k_j^\ast).
\end{equation}
Here, $\mathbf N$ is a fixed positive integer
and the complex constants $k_j$ are distinct and all occur in the sector
$\arg[k]\in[7\pi/6,4\pi/3).$ 
From \eqref{3.46} and \eqref{5.35} we 
get
\begin{equation}\label{5.37}
T_\text{\rm{r}}(k)=\ds\frac{\Gamma(-k)}{\Gamma(k)}, \qquad k\in\overline{\Omega_3},
\end{equation}
and hence \eqref{5.35} and \eqref{5.37} are compatible with \eqref{4.11}. We solve the Riemann--Hilbert problem \eqref{4.10}
with the large $k$-asymptotics in \eqref{4.8} and \eqref{4.9} by multiplying both sides of \eqref{4.10} by
$e^{-kx}\,\Gamma(-k).$ From \eqref{4.10} we get
\begin{equation}\label{5.38}
e^{-kx}\,\Gamma(-k)\,\Phi_+(k,x)=e^{-kx}\,\Gamma(-k)\,\Phi_-(k,x), \qquad k\in \mathcal L.
\end{equation}
The left-hand side of \eqref{5.38} has an analytic extension from $k\in\mathcal L$ to the plus region $\mathcal P^+$ and the right-hand
side of \eqref{5.38} has an analytic extension from $k\in\mathcal L$ to the minus region $\mathcal P^-.$ Using the large $k$-asymptotics given in \eqref{4.8}
and \eqref{4.9}, with the help of the generalized Liouville theorem we conclude that each side of \eqref{5.38} is entire and equal
to a monic polynomial in $k$ of degree $2\mathbf N.$ We define $V(k)$ as the row vector with $2\mathbf N$ components given by
\begin{equation}\label{5.39}
V(k):=\begin{bmatrix}
k^{2\mathbf N-1} &k^{2\mathbf N-2} & \cdots & k &1
\end{bmatrix},\qquad k\in\mathbb C.
\end{equation}
We use $\mathbf A(x)$ to denote the column vector with $2\mathbf N$ components, which is given by
\begin{equation}\label{5.40}
\mathbf A(x):=\begin{bmatrix}
A_{2\mathbf N-1}(x)\\
A_{2\mathbf N-2}(x)\\
\vdots\\
A_1(x)\\
A_0(x)
\end{bmatrix},\qquad x\in\mathbb R,
\end{equation}
where each $A_j(x)$ is a scalar-valued function of $x\in\mathbb R$ for $0\le j\le 2\mathbf N-1.$ Thus, each side of \eqref{5.38} is equal to the
monic polynomial $[k^{2\mathbf N}+V(k)\,\mathbf A(x)],$ and we obtain the solution to the Riemann--Hilbert problem \eqref{4.10} as
\begin{equation}\label{5.41}
\Phi_+(k,x)=e^{kx}\,\ds\frac{k^{2\mathbf N}+V(k)\,\mathbf A(x)}{\Gamma(-k)}, \qquad k\in \overline{\mathcal P^+}, \quad x\in\mathbb R,
\end{equation}
\begin{equation}\label{5.42}
\Phi_-(k,x)=e^{kx}\,\ds\frac{k^{2\mathbf N}+V(k)\,\mathbf A(x)}{\Gamma(-k)}, \qquad k\in \overline{\mathcal P^-}, \quad x\in\mathbb R.
\end{equation}
Having $\Phi_+(k,x)$ expressed as the right-hand side of \eqref{5.41}, we compare \eqref{5.41} with \eqref{4.3}. That comparison
shows that the solutions $f(k,x)$ and $m(k,x)$ to \eqref{1.1} are expressed as
%
%
%
%Comparing \eqref{5.41} with \eqref{4.3}, with the help of \eqref{5.35}, we recover the solutions $f(k,x)$ and $m(k,x)$ to \eqref{1.1} as
\begin{equation}\label{5.43}
f(k,x)=e^{kx}\,\ds\frac{k^{2\mathbf N}+V(k)\,\mathbf A(x)}{\Gamma(k)}, \qquad k\in\overline{\Omega_1}, \quad x\in \mathbb R,
\end{equation}
\begin{equation}\label{5.44}
m(k,x)=e^{kx}\,\ds\frac{k^{2\mathbf N}+V(k)\,\mathbf A(x)}{\Gamma(-k)}, \qquad k\in\overline{\Omega_2}, \quad x\in \mathbb R.
\end{equation}
Similarly, comparing \eqref{5.42} and \eqref{4.4}, we recover the solutions $g(k,x)$ and $n(k,x)$ to \eqref{1.1} as
\begin{equation}\label{5.45}
g(k,x)=e^{kx}\,\ds\frac{k^{2\mathbf N}+V(k)\,\mathbf A(x)}{\Gamma(-k)}, \qquad k\in\overline{\Omega_3}, \quad x\in \mathbb R,
\end{equation}
\begin{equation}\label{5.46}
n(k,x)=e^{kx}\,\ds\frac{k^{2\mathbf N}+V(k)\,\mathbf A(x)}{\Gamma(-k)}, \qquad k\in\overline{\Omega_4}, \quad x\in \mathbb R.
\end{equation}
We note that the column vector $\mathbf A(x)$ with $2\mathbf N$ components is uniquely determined by using the $2\mathbf N$ initial dependency constants
$E(k_j)$ and $E(k_j^\ast)$ given in our input data set for $1\le j\le \mathbf N.$ From \eqref{3.4} and \eqref{3.24} we get
\begin{equation}\label{5.47}
f(k_j,x)=E(k_j)\, e^{9(z^2-1)k_j^5t}\, g(zk_j,x), \qquad 1\le j\le \mathbf N,
\end{equation}
and from the second lines of \eqref{3.31} and \eqref{3.38} we have
\begin{equation}\label{5.48}
f(k_j^\ast,x)=E(k_j^\ast)\, e^{9(z-1)(k_j^\ast)^5t}\, g(z^2k_j^\ast,x), \qquad 1\le j\le \mathbf N.
\end{equation}
Using \eqref{5.43} and \eqref{5.45} in \eqref{5.47}, we obtain
\begin{equation}\label{5.49}
e^{k_jx}\, \ds\frac{k_j^{2\mathbf N}+V(k_j)\,\mathbf A(x)}{\Gamma(k_j)}
=E(k_j)\,\,e^{9(z^2-1)k_j^5t} e^{zk_jx}\,\ds\frac{(zk_j)^{2\mathbf N}+V(zk_j)\,\mathbf A(x)}{\Gamma(-zk_j)}, \qquad 1\le j\le \mathbf N,
\end{equation}
and using \eqref{5.43} and \eqref{5.45} in \eqref{5.48}, we get
\begin{equation}\label{5.50}
\begin{split}
e^{k_j^\ast x}\,&\ds\frac{(k_j^\ast)^{2\mathbf N}+V(k_j^\ast)\,\mathbf A(x)}{\Gamma(k_j^\ast)}\\
&\phantom{xxxxx}=E(k_j^\ast)\,e^{9(z-1)(k_j^\ast)^5t}\,e^{z^2k_j^\ast x}\,
\ds\frac{(z^2k_j^\ast)^{2\mathbf N}+V(z^2k_j^\ast)
\,\mathbf A(x)}{\Gamma(-z^2k_j^\ast)}, \qquad 1\le j\le \mathbf N.
\end{split}
\end{equation}
In terms of the initial dependency constants $E(k_j)$ and $E(k_j^\ast),$ we introduce the modified initial dependency constants
$\gamma(k_j)$ and $\gamma(k_j^\ast)$ as
\begin{equation}\label{5.51}
\gamma(k_j):=-\ds\frac{\Gamma(k_j)\, E(k_j)}{\Gamma(-zk_j)}, \qquad 1\le j\le \mathbf N,
\end{equation}
\begin{equation}\label{5.52}
\gamma(k_j^\ast):=-\ds\frac{\Gamma(k_j^\ast)\, E(k_j^\ast)}{\Gamma(-z^2k_j^\ast)}, \qquad 1\le j\le \mathbf N.
\end{equation}
Since the dependency constants $E(k_j)$ and $E(k_j^\ast)$ are nonzero, the modified dependency constants
$\gamma(k_j)$ and $\gamma(k_j^\ast)$ are also nonzero.
We emphasize that $\gamma(k_j)$ and $\gamma(k_j^\ast)$ are complex-valued constants independent of $x$ and $t.$ We also introduce 
the quantities $\chi(k_j)$ and $\chi(k_j^\ast)$ by letting
\begin{equation}\label{5.53}
\chi(k_j):=\exp\left((z-1)k_j x+9(z^2-1)k_j^5 t\right), \qquad 1\le j\le \mathbf N,
\end{equation}
\begin{equation}\label{5.54}
\chi(k_j^\ast):=\exp\left((z^2-1)k_j^\ast x+9(z-1)(k_j^\ast)^5 t\right), \qquad 1\le j\le \mathbf N.
\end{equation}
We remark that $\chi(k_j)$ and $\chi(k_j^\ast)$ are functions of $x$ and $t.$ Since $x$ and
$t$ are
real-valued parameters, it follows that each of $\chi(k_j)$ and $\chi(k_j^\ast)$ is real valued if and only if
$k_j=iz\eta_j$ for some positive constant $\eta_j.$ This can be justified as follows.
From the same argument used in \eqref{5.16}--\eqref{5.18}, we already know that
$\chi(k_j)$ is real valued if and only if
$k_j=iz\eta_j$ for some positive constant $\eta_j.$ 
Then, with $k_j^\ast=-iz^2\eta_j,$ from \eqref{5.54} we get
\begin{equation}\label{5.55}
\chi(-iz^2\eta_j)=\exp\left(\sqrt{3}\,\eta_j\,x-9\,\sqrt{3}\,\eta_j^5\,t\right), \qquad 1\le j\le \mathbf N.
\end{equation}
Hence, we have shown that each of $\chi(k_j)$ and $\chi(k_j^\ast)$ is real valued if and only if
$k_j=iz\eta_j$ for some positive constant $\eta_j.$ 
With the help of \eqref{2.1}, from \eqref{5.53} and \eqref{5.54} we observe that
\begin{equation}\label{5.56}
\chi(k_j^\ast)=\chi(k_j)^\ast, \qquad 1\le j\le \mathbf N.
\end{equation}
From \eqref{5.36} we get
\begin{equation}\label{5.57}
\Gamma(k_j^\ast)=\Gamma(k_j)^\ast, \qquad 1\le j\le \mathbf N,
\end{equation}
\begin{equation}\label{5.58}
\Gamma(-z^2k_j^\ast)=\Gamma(-zk_j)^\ast, \qquad 1\le j\le \mathbf N.
\end{equation}
In general, we do not have $E(k_j^\ast)=E(k_j)^\ast$ for $1\le j\le \mathbf N,$ but as indicated in the second equality of the second line of \eqref{3.32}, we
have $E(k_j^\ast)=E(k_j)^\ast$ for $1\le j\le \mathbf N$ when the potentials $Q$ and $P$ are real valued. 
Using \eqref{5.51}--\eqref{5.54}, we write \eqref{5.49} and \eqref{5.50}, respectively, for $1\le j\le \mathbf N$ as
\begin{equation}\label{5.59}
\begin{cases}
k_j^{2\mathbf N}+V(k_j)\,\mathbf A(x)=-\gamma(k_j)\,\chi(k_j)\left[(zk_j)^{2\mathbf N}+V(zk_j)\,\mathbf A(x)\right],\\
\noalign{\medskip}
(k_j^\ast)^{2\mathbf N}+V(k_j^\ast)\,\mathbf A(x)=-\gamma(k_j^\ast)\,\chi(k_j^\ast)\left[(z^2k_j^\ast)^{2\mathbf N}+V(z^2k_j^\ast)\,\mathbf A(x)\right].
\end{cases}
\end{equation}
We observe that \eqref{5.59} for $1\le j\le \mathbf N$ yields a linear algebraic system of $2\mathbf N$ equations with the $2\mathbf N$ unknowns $A_l(x)$ for
$0\le l\le 2\mathbf N-1.$ This can be seen by defining the scalar quantities $m_l(k_j)$ and $m_l(k_j^\ast)$ for $1\le j\le \mathbf N$ and
$0\le l\le 2\mathbf N,$ where we have let
\begin{equation}\label{5.60}
\begin{cases}
m_l(k_j):=k_j^l+(zk_j)^l\,\gamma(k_j)\,\chi(k_j),\\
\noalign{\medskip}
m_l(k_j^\ast):=(k_j^\ast)^l+(z^2k_j^\ast)^l\,\gamma(k_j^\ast)\,\chi(k_j^\ast).
\end{cases}
\end{equation}
We remark that the dependence on $x$ and $t$ in $m_l(k_j)$ and $m_l(k_j^\ast)$ appears only through $\chi(k_j)$ and $\chi(k_j^\ast),$ respectively. Using
\eqref{5.60} we form the $2\mathbf N\times 2\mathbf N$ matrix $\mathbf M(x)$ and the column vector $\mathbf B(x)$ with $2\mathbf N$ components as
\begin{equation}\label{5.61}
\mathbf M(x):=\begin{bmatrix}
m_{2\mathbf N-1}(k_1) & m_{2\mathbf N-2}(k_1) & \cdots & m_1(k_1) & m_0(k_1)\\
m_{2\mathbf N-1}(k_1^\ast) & m_{2\mathbf N-2}(k_1^\ast) & \cdots & m_1(k_1^\ast) & m_0(k_1^\ast)\\
\vdots & \vdots & \ddots & \vdots & \vdots \\
m_{2\mathbf N-1}(k_{\mathbf N}) & m_{2\mathbf N-2}(k_{\mathbf N}) & \cdots & m_1(k_{\mathbf N}) & m_0(k_{\mathbf N})\\
m_{2\mathbf N-1}(k_{\mathbf N}^\ast) & m_{2\mathbf N-2}(k_{\mathbf N}^\ast) & \cdots & m_1(k_{\mathbf N}^\ast) & m_0(k_{\mathbf N}^\ast)
\end{bmatrix},
\end{equation}
\begin{equation}\label{5.62}
\mathbf B(x):=\begin{bmatrix}
m_{2\mathbf N}(k_1)\\
m_{2\mathbf N}(k_1^\ast)\\
\vdots\\
m_{2\mathbf N}(k_{\mathbf N})\\
m_{2\mathbf N}(k_{\mathbf N}^\ast)
\end{bmatrix}.
\end{equation}
With the help of \eqref{5.61} and \eqref{5.62}, we write the linear algebraic system \eqref{5.59} as
\begin{equation}\label{5.63}
\mathbf M(x)\, \mathbf A(x)=-\mathbf B(x),
\end{equation}
and hence the column vector $\mathbf A(x)$ is obtained as

\begin{equation}\label{5.64}
\mathbf A(x)=-\mathbf M(x)^{-1}\mathbf B(x),
\end{equation}
where we emphasize that we suppress in our notation the $t$-dependence in both $\mathbf M(x)$ and $\mathbf A(x).$ Alternatively, by
using Cramer's rule on \eqref{5.63}, the
 $2\mathbf N$ components $A_j(x)$ with $0\le j\le 2\mathbf N-1$ of the column vector $\mathbf A(x)$ can be obtained as the ratio of two
determinants. For example, we have
\begin{equation}\label{5.65}
A_{2\mathbf N-1}(x)=-\ds\frac{\det[\mathbf M_1(x)]}{\det[\mathbf M(x)]},\qquad \mathbf N\ge 1,
\end{equation}
\begin{equation}\label{5.66}
A_{2\mathbf N-2}(x)=-\ds\frac{\det[\mathbf M_2(x)]}{\det[\mathbf M(x)]},\qquad \mathbf N\ge 1,
\end{equation}
where we use $\mathbf M_1(x)$ to denote the $2\mathbf N\times 2\mathbf N$ matrix obtained by replacing the first column of $\mathbf M(x)$ with the column
vector $\mathbf B(x),$ and we use $\mathbf M_2(x)$ to denote the $2\mathbf N\times 2\mathbf N$ matrix obtained by replacing the second column of
$\mathbf M(x)$ with $\mathbf B(x).$ Having obtained the column vector $\mathbf A(x)$ as in \eqref{5.64} explicitly in terms of the input data set
$\{k_j, k_j^\ast, E(k_j), E(k_j^\ast)\}_{j=1}^{\mathbf N},$ as seen from \eqref{5.43}--\eqref{5.46}, we have the solutions $f(k,x),$ $m(k,x),$
$g(k,x),$ and $n(k,x)$ to \eqref{1.1}, where each of those solutions in their respective $k$-domains is explicitly expressed in
terms of the input data set consisting of the bound-state poles of $T_\text{\rm{l}}(k)$ in $\Omega_1^\text{\rm{down}}$ and the corresponding
bound-state
dependency constants. Having $f(k,x)$ at hand, we can recover the potentials $Q(x)$ and $P(x)$ with the help of
\eqref{4.15} and \eqref{4.16}, respectively. For this we proceed as follows. Comparing \eqref{4.12} with \eqref{5.43} we observe
that
\begin{equation}\label{5.67}
\ds\frac{k^{2\mathbf N}+V(x)\,\mathbf A(x)}{\Gamma(k)}=1+\ds\frac{u_1(x)}{k}+\ds\frac{u_2(x)}{k^2}+O\left(\ds\frac{1}{k^3}\right), \qquad k\to \infty \text{\rm{ in }} \overline{\Omega_1}.
\end{equation}
With the help of \eqref{5.36}, \eqref{5.39}, and \eqref{5.40}, we expand the left-hand side of \eqref{5.67} in powers of $1/k$ and obtain
\begin{equation}\label{5.68}
\begin{split}
\ds\frac{k^{2\mathbf N}+V(x)\,\mathbf A(x)}{\Gamma(k)}=&1+\ds\frac{A_{2\mathbf N-1}(x)-\Sigma_{\mathbf N}}{k}\\
    &+\ds\frac{A_{2\mathbf N-2}(x)-\Sigma_{\mathbf N} A_{2\mathbf N-1}(x)+\Pi_{\mathbf N}}{k^2}+O\left(\ds\frac{1}{k^3}\right), \qquad k\to \infty \text{\rm{ in }} \overline{\Omega_1},
\end{split}
\end{equation}
where we have defined
\begin{equation}\label{5.69}
\Sigma_{\mathbf N}:=k_1+k_1^\ast+k_2+k_2^\ast+\cdots+k_{\mathbf N}+k_{\mathbf N}^\ast,
\end{equation}
\begin{equation*}
%\label{5.70}
\begin{split}
\Pi_{\mathbf N}:=&k_1\left(k_1+k_1^\ast+\cdots+k_{\mathbf N}+k_{\mathbf N}^\ast \right)+k_1^\ast\left(k_1^\ast+k_2+\cdots+k_{\mathbf N}+k_{\mathbf N}^\ast \right)\\
       &+k_2\left(k_2+k_2^\ast+\cdots+k_{\mathbf N}+k_{\mathbf N}^\ast \right)+\cdots
       +k_{\mathbf N}\left(k_{\mathbf N}+k_{\mathbf N}^\ast\right)+ k_{\mathbf N}^\ast\left(k_{\mathbf N}^\ast\right),
\end{split}
\end{equation*}
which are the analogs of \eqref{5.29} and \eqref{5.30}, respectively.
Comparing \eqref{5.68} with \eqref{5.67}, we see that
\begin{equation}\label{5.71}
\begin{cases}
u_1(x)=A_{2\mathbf N-1}(x)-\Sigma_{\mathbf N},\\
\noalign{\medskip}
u_2(x)=A_{2\mathbf N-2}(x)-\Sigma_{\mathbf N} A_{2\mathbf N-1}(x)+\Pi_{\mathbf N}.
\end{cases}
\end{equation}
Using \eqref{5.71} in \eqref{4.15} and \eqref{4.16}, we express the potentials $Q(x)$ and $P(x)$ in terms of $A_{2\mathbf N-1}(x)$ and
$A_{2\mathbf N-2}(x)$ as
\begin{equation}\label{5.72}
\begin{cases}
Q(x)=-3\,\ds\frac{dA_{2\mathbf N-1}(x)}{dx},\\
\noalign{\medskip}
P(x)=3\left(A_{2\mathbf N-1}(x)\,\ds\frac{dA_{2\mathbf N-1}(x)}{dx}-\ds\frac{d^2A_{2\mathbf N-1}(x)}{dx^2}-\ds\frac{dA_{2\mathbf N-2}(x)}{dx}\right).
\end{cases}
\end{equation}
From \eqref{5.71} and \eqref{5.72} we observe that the constant terms $\Sigma_{\mathbf N}$ and $\Pi_{\mathbf N}$ do not appear in \eqref{5.72}.
Using \eqref{5.65} and \eqref{5.66} in \eqref{5.72}, we see that the potentials $Q(x)$ and $P(x)$ are explicitly constructed in terms of
the input data set $\{k_j, k_j^\ast, E(k_j), E(k_j^\ast)\}_{j=1}^{\mathbf N}.$ As seen from the first line of \eqref{5.72}, the potential
$Q(x)$ is determined by $A_{2\mathbf N-1}(x)$ only. Similarly, as seen from the second line of \eqref{5.72}, the potential
$P(x)$ is determined by $A_{2\mathbf N-1}(x)$ and $A_{2\mathbf N-2}(x)$ only.
The remaining entries of $\mathbf A(x)$ are only needed to construct the solutions to
\eqref{1.1}.
Having constructed $Q(x)$ and $P(x)$ in terms of our input data set,
we now have  the explicit solution pair $Q(x)$ and
$P(x)$ to the integrable nonlinear system given in \eqref{1.6}. Thus, we have demonstrated
the construction of the $2\mathbf N$-soliton solution to \eqref{1.6}
by using the input data set consisting of the locations of the poles $k_j$ and $k_j^\ast$ for
$1\le j\le \mathbf N$ as well as the complex-valued initial dependency constants $E(k_j)$ and $E(k_j^\ast),$ 

\end{example}

In the next example, by using the appropriate restrictions on the locations of the bound-state $k$-values and the
appropriate restrictions on the corresponding bound-state dependency constants,
we show that the $2\mathbf N$-soliton solution to \eqref{1.6} constructed in
Example~\ref{example5.2} yields the real-valued $\mathbf N$-soliton solution to the Sawada--Kotera equation \eqref{1.10}.

\begin{example}\label{example5.3}
\normalfont

From Example~\ref{example5.2}  we know that for the potentials $Q(x)$ and  $P(x)$ to be real valued, it is necessary that we 
 choose the complex constants $k_j$ as distinct and satisfying
\begin{equation}\label{5.73}
k_j=iz\eta_j, \qquad  1\le j\le \mathbf N,
\end{equation}
with positive $\eta_j$-values. Without loss of generality, we assume that we have $0<\eta_1<\eta_2<\dots<\eta_\mathbf N.$ In that case, as we see from \eqref{5.35},
the left transmission coefficient $T_\text{\rm{l}}(k)$ is given by
\begin{equation}
\label{5.74}
T_\text{\rm{l}}(k)=\ds\prod_{j=1}^{\mathbf N} \left(\ds\frac{k+iz\eta_j}{k-iz\eta_j}\,\ds\frac{k-iz^2\eta_j}{k+iz^2\eta_j}\right).
\end{equation}
Using \eqref{5.73} in \eqref{5.53} we obtain
\begin{equation}\label{5.75}
\chi(k_j)=e^{\sqrt{3}\eta_j(x-9\eta_j^4 t)}, \qquad 1\le j\le \mathbf N,
\end{equation}
and from \eqref{5.56} and \eqref{5.75} we see that
\begin{equation}\label{5.76}
\chi(k_j^\ast)=\chi(k_j), \qquad 1\le j\le \mathbf N.
\end{equation}
We have the modified initial dependency constants $\gamma(k_j)$ and $\gamma(k_j^\ast)$ appearing in \eqref{5.51} and \eqref{5.52},
respectively.
We introduce the real constants $r_j$ and $s_j$ as the real and imaginary parts of $\gamma(k_j),$ i.e. we let
\begin{equation}\label{5.77}
\gamma(k_j)=r_j+is_j, \qquad 1\le j\le \mathbf N.
\end{equation}
Since $\gamma(k_j)$ is nonzero, we know that $r_j$ and $s_j$ cannot vanish simultaneously.
In order to have the real-valued potentials $Q$ and $P,$ as seen from \eqref{5.51}, \eqref{5.52}, \eqref{5.57}, \eqref{5.58}, and
the second equality of \eqref{3.32}, we have
\begin{equation*}
%\label{5.78}
\gamma(k_j^\ast)=r_j-is_j, \qquad 1\le j\le \mathbf N.
\end{equation*}
To obtain the $\mathbf N$-soliton solution to \eqref{1.10} with the constraint $P(x)\equiv 0,$ from the second line of \eqref{5.72} we see that
the real constants $r_j$ and $s_j$ must be restricted. Similarly, in order to obtain the $\mathbf N$-soliton solution to \eqref{1.10} with the
constraint $P(x)\equiv Q_x(x),$ from the second line of \eqref{5.72} we observe that the real constants $r_j$ and $s_j$ must be restricted. 
Those restrictions can be evaluated explicitly, and they allow us to obtain the
real-valued $\mathbf N$-soliton solution $Q(x)$ to 
the Sawada--Kotera equation \eqref{1.10}. As seen from the argument presented in the steps involving
\eqref{1.10}--\eqref{1.14}, the real-valued $\mathbf N$-soliton solution to
\eqref{1.10} can be obtained by using the inverse scattering theory
either for \eqref{1.1} with real-valued
$Q(x)$ and $P(x)\equiv 0$ or for
\eqref{1.14} with that real-valued $Q(x).$
As indicated in Section~\ref{section1}, the transmission coefficients for
\eqref{1.1} with $P(x)\equiv 0$ and the transmission coefficients for
\eqref{1.14} coincide. Thus, only the bound-state dependency constants
for \eqref{1.1} with $P(x)\equiv 0$ and the bound-states dependency constants for
\eqref{1.14} differ from each other. In our example, we specify
the corresponding dependency constants in both cases. 
When we have \eqref{5.73}, the determinant of the matrix $\mathbf M_1(x)$ appearing in 
\eqref{5.65} as well as the determinant of the matrix $\mathbf M(x)$ appearing in \eqref{5.61}
are related to each other via
\begin{equation}\label{5.79}
\det[\mathbf M_1(x)]=\Sigma_{\mathbf N}\, \det[\mathbf M(x)]+\ds\frac{d}{dx}\left( \det[\mathbf M(x)]\right),\qquad 
\mathbf N\ge 1,
\end{equation}
where $\Sigma_{\mathbf N}$ is the scalar quantity defined in \eqref{5.69}.
A proof of \eqref{5.79} can be given by using mathematical induction.
Using \eqref{5.65} and \eqref{5.79} in the first line of \eqref{5.72}, we obtain the potential $Q(x)$ explicitly in terms of
the determinant of the matrix $\mathbf M(x)$ as
\begin{equation}\label{5.80}
Q(x)=3 \ds\frac{d}{dx}\left[\ds\frac{1}{\det[\mathbf M(x)]}\,
\ds\frac{d}{dx}\left( \det[\mathbf M(x)]\right)
\right],
\qquad 
\mathbf N\ge 1.
\end{equation}
The determinant of the matrix $\mathbf M(x)$ in \eqref{5.61} can be further simplified when we have \eqref{5.73}. From
\eqref{5.75} and \eqref{5.76} we get
\begin{equation}\label{5.81}
\ds\lim_{x\to-\infty} \chi(k_j)=0,\quad
\ds\lim_{x\to-\infty} \chi(k_j^\ast)=0,\qquad 1\le j\le \mathbf N.
\end{equation}
With the help of \eqref{5.60}, \eqref{5.61}, and \eqref{5.81}, we observe that, as $x\to-\infty,$ the matrix
$\mathbf M(x)$ has a simple limit as a Vandermonde matrix \cite{GV2013}, which is given as
\begin{equation}\label{5.82}
\ds\lim_{x\to-\infty} \mathbf M(x)=\begin{bmatrix}
k_1^{2\mathbf N-1} & k_1^{2\mathbf N-2}& \cdots &k_1 &1\\
\noalign{\medskip}
(k_1^\ast)^{2\mathbf N-1}&(k_1^\ast)^{2\mathbf N-2} & \cdots & k_1^\ast & 1\\
\noalign{\medskip}
\vdots & \vdots & \ddots & \vdots & \vdots \\
k_{\mathbf N} ^{2\mathbf N-1}& k_{\mathbf N} ^{2\mathbf N-2}& \cdots & k_{\mathbf N}& 1\\
\noalign{\medskip}
(k_{\mathbf N}^\ast) ^{2\mathbf N-1}& (k_{\mathbf N}^\ast) ^{2\mathbf N-2}& \cdots & k_{\mathbf N}^\ast & 1
\end{bmatrix},
\end{equation}
and hence the determinant of the matrix on the right-hand side of \eqref{5.82} can be expressed in a simple manner as a Vandermonde
determinant \cite{GV2013} in terms of the elements in the set $\{k_j,k_j^\ast\}_{j=1}^{\mathbf N}.$
From \eqref{5.55}, \eqref{5.56}, \eqref{5.60}, \eqref{5.61} it follows that $\mathbf M(x)$ is a continuous function of
$x$ in $\mathbb R.$ Hence, we have
\begin{equation*}
%\label{5.83}
\det\left[\ds\lim_{x\to-\infty} \mathbf M(x)\right]=
\ds\lim_{x\to-\infty} \det[\mathbf M(x)].
\end{equation*}
With the help of \eqref{5.61} and \eqref{5.82}, we define the scalar quantity $\Delta(\mathbf N,x)$ as
\begin{equation}\label{5.84}
\Delta(\mathbf N,x):=\ds\sqrt{\ds\frac{\det[\mathbf M(x)]}{
\ds\lim_{x\to-\infty} \det[\mathbf M(x) ]}}.
\end{equation}
Let us use $\chi_j$ to denote the quantity $\chi(k_j)$ when $k=iz\eta_j.$ From \eqref{5.75} we get
\begin{equation*}
%\label{5.85}
\chi_j:=e^{\sqrt{3}\eta_j(x-9\eta_j^4t)},\qquad 1\le j\le \mathbf N.
\end{equation*}
%One can prove, by using mathematical induction, that 
%$\Delta(\mathbf N,x)$ defined in \eqref{5.84} is a particular
%polynomial in the variables $\chi_j$ for $1\le j\le \mathbf N$ of degree $\mathbf N.$
From \eqref{5.61} and \eqref{5.82} we observe that the quantity inside the square root on the right-hand side of \eqref{5.84} 
is a 
polynomial of degree $2\mathbf N$ in the variables $\chi_j$ for $1\le j\le \mathbf N,$ and in fact the 
constant term in that polynomial is $1.$
The coefficients in that polynomial are uniquely determined by our input data set given by
$\{k_j,E(k_j)\}_{j=1}^{\mathbf N}$ or equivalently in terms of the elements of the set
$\{k_j,\gamma(k_j)\}_{j=1}^{\mathbf N},$ 
where we recall that the modified initial
bound-state dependency constant $\gamma(k_j)$ is related to
the initial bound-state dependency constant $E(k_j)$ as in \eqref{5.51}.
Since $\gamma(k_j)$ is expressed in terms of its real and imaginary parts as in \eqref{5.77}, we 
observe that the coefficients in the aforementioned polynomial
%$\Delta(\mathbf N,x)$ 
are explicitly expressed in terms of
the elements of the set $\{\eta_j,r_j,s_j\}_{j=1}^{\mathbf N}.$ 
Moreover, each of the two restrictions $P(x)\equiv 0$ and $P(x)\equiv Q_x(x)$ in \eqref{1.9}
allows us to express each $s_j$ proportional to
$r_j$ with the coefficient determined by the elements of the set $\{\eta_j\}
_{j=1}^{\mathbf N}.$ 
One can prove, by using mathematical induction, that
those expressions for $s_j$
make the aforementioned polynomial of degree $2\mathbf N$ a perfect square of a
polynomial of degree $\mathbf N,$ which is the polynomial
$\Delta(\mathbf N,x)$ appearing on the left-hand side of \eqref{5.84}.
For each integer $j,$ the coefficient of $\chi_j$
in $\Delta(\mathbf N,x)$
is a linear combination of $r_j$ and $s_j.$
Replacing $s_j$ by its equivalent expressed in terms of
$r_j$ and $\{\eta_j\}
_{j=1}^{\mathbf N},$ we express $\Delta(\mathbf N,x)$
in terms of the elements of the set
$\{\eta_j,r_j\}
_{j=1}^{\mathbf N}.$
The positivity of the coefficient of each $\chi_j$ in $\Delta(\mathbf N,x)$ for $1\le j\le \mathbf N$ guarantees that the resulting
potentials $Q(x)$ and $P(x)$ do not have any singularities. 
We list the first few of the values of $\Delta(\mathbf N,x)$ so that the reader can see the general pattern.

\begin{enumerate}
    \item[$\bullet$] When $\mathbf N=1,$ we have
\begin{equation}\label{5.86}
\Delta(1,x)=1+w_1^{(1)}\chi_1,\quad w_1^{(1)}:=r_1-\sqrt{3}\,s_1.
\end{equation}

 \item[$\bullet$] When $\mathbf N=2,$ we have
\begin{equation}\label{5.87}
\Delta(2,x)=1+w_1^{(2)}\chi_1+w_2^{(2)}\chi_2+w^{(2)}_{12}\chi_1\, \chi_2,
\end{equation}
where we let
\begin{equation}\label{5.88}
w_1^{(2)}:=\ds\frac{(\eta_1+\eta_2)^3[r_1\,(\eta_1^2-\eta_2^2)+\sqrt{3}\,s_1\,(\eta_1^2+\eta_2^2)]}
{(\eta_1^2-\eta_2^2)(\eta_1^3-\eta_2^3)},
\end{equation}
\begin{equation}\label{5.89}
w_2^{(2)}:=\ds\frac{(\eta_1+\eta_2)^3[r_2\,(\eta_1^2-\eta_2^2)-\sqrt{3}\,s_2\,(\eta_1^2+\eta_2^2)]}
{(\eta_1^2-\eta_2^2)(\eta_1^3-\eta_2^3)},
\end{equation}
\begin{equation}\label{5.90}
w_{12}^{(2)}:=\ds\frac{(\eta_1+\eta_2)[a_1 r_1 r_2+a_2 s_1 s_2+a_3 (r_2 s_1-r_1 s_2)}
{(\eta_1^2-\eta_2^2)(\eta_1^3-\eta_2^3)},
\end{equation}
with the constants $a_1,$ $a_2,$ $a_3$ expressed in terms of $\eta_1$ and $\eta_2$ as
\begin{equation*}
%\label{5.91}
a_1:=\eta_1^6-4\eta_1^5 \eta_2-13\eta_1^4 \eta_2^2+20\eta_1^3 \eta_2^3-13\eta_1^2 \eta_2^4-
4\eta_1\eta_2^5+\eta_2^6,
\end{equation*}
\begin{equation*}
%\label{5.92}
a_2:=-3\left(\eta_1^6-4\eta_1^5 \eta_2-\eta_1^4 \eta_2^2-4\eta_1^3 \eta_2^3-\eta_1^2 \eta_2^4-
4\eta_1\eta_2^5+\eta_2^6\right),
\end{equation*}
\begin{equation*}
%\label{5.93}
a_3:=\sqrt{3}\,(\eta_1^4-\eta_2^4)(\eta_1^2-4\eta_1\eta_2+\eta_2^2).
\end{equation*}
We remark that there are various symmetries in \eqref{5.88}--\eqref{5.90} coming from \eqref{5.73}.

 \item[$\bullet$] When $\mathbf N=3,$ we have
\begin{equation}\label{5.94}
\begin{split}
\Delta(3,x)=&1+w_1^{(3)}\chi_1+w_2^{(3)}\chi_2+w_3^{(3)}\chi_3\\
&+w^{(3)}_{12}\chi_1\, \chi_2+w^{(3)}_{13}\chi_1\, \chi_3
+w^{(3)}_{23}\chi_2\,\chi_3+w^{(3)}_{123}\chi_1\, \chi_2\,\chi_3
,\quad
\end{split}
\end{equation}
with the coefficients explicitly expressed in terms of $\eta_1,$ $\eta_2,$ $\eta_3,$ $r_1,$ $r_2,$ $r_3,$ $s_1,$ $s_2,$ and $s_3.$
Since the expressions are lengthy, we do not display those coefficients here. Using the symbolic software Mathematica we can evaluate and
 explicitly display those coefficients.

 \item[$\bullet$] When $\mathbf N=4,$ we have
\begin{equation}\label{5.95}
\begin{split}
\Delta(4,x)=&1+w_1^{(4)}\chi_1+w_2^{(4)}\chi_2+w_3^{(4)}\chi_3+w_4^{(4)}\chi_4
+w^{(4)}_{12}\chi_1\, \chi_2+w^{(4)}_{13}\chi_1\, \chi_3
+w^{(4)}_{14}\chi_1 \,\chi_4\\
&+w^{(4)}_{23}\chi_2\, \chi_3
+w^{(4)}_{24}\chi_2\, \chi_4+w^{(4)}_{34}\chi_3\, \chi_4
+w^{(4)}_{123}\chi_1\, \chi_2\,\chi_3+w^{(4)}_{124}\chi_1\, \chi_2\,\chi_4\\
&+w^{(4)}_{134}\chi_1\, \chi_3\,\chi_4+w^{(4)}_{234}\chi_2\, \chi_3\,\chi_4+w^{(4)}_{1234}\chi_1\, \chi_2\,\chi_3\,\chi_4,
\end{split}
\end{equation}
with the coefficients explicitly expressed in terms of $\eta_1,$ $\eta_2,$ $\eta_3,$ $\eta_4,$ $r_1,$ $r_2,$ $r_3,$ $r_4,$ $s_1,$ $s_2,$ $s_3,$ and $s_4.$
Since the expressions are lengthy, we do not list those coefficients. With the help of Mathematica we can evaluate and display
those explicitly expressed coefficients.
\end{enumerate}

Using \eqref{5.84} in \eqref{5.80} we express the potential $Q(x)$ in terms of the scalar quantity $\Delta(\mathbf N,x)$ as
\begin{equation}\label{5.96}
Q(x)=6 \,\ds\frac{d}{dx}\left[\ds\frac{\Delta'(\mathbf N,x)}{\Delta(\mathbf N,x)}\right],
\qquad 
\mathbf N\ge 1,
\end{equation}
where $\Delta'(\mathbf N,x)$ denotes the $x$-derivative of $\Delta(\mathbf N,x).$
In the following list, we provide the restrictions on the constants $r_j$ and $s_j$ appearing in the coefficients
in \eqref{5.86}, \eqref{5.87}, \eqref{5.94}, and \eqref{5.95} so that
the quantity $Q(x)$ in \eqref{5.96} yields the real-valued $\mathbf N$-soliton solution to
\eqref{1.10} for each of the two cases
corresponding to \eqref{1.1} with
$P(x)\equiv 0$ and to \eqref{1.14}, respectively.

\begin{enumerate}

\item[$\bullet$] When $\mathbf N=1,$ the quantity $Q(x)$ given in \eqref{5.96} yields a $1$-soliton solution to the Sawada--Kotera equation, with $P(x)\equiv 0,$ if we
use in \eqref{5.86} the restriction
\begin{equation}\label{5.97}
s_1=\sqrt{3}\,r_1.
\end{equation}
In that case, the coefficient $w_1^{(1)}$ is evaluated as $-2r_1$ by using \eqref{5.97} in 
the second equality of \eqref{5.86}, and we get the explicit expression for $Q(x)$ given by
\begin{equation}\label{5.98}
Q(x)=-\ds\frac{36  r_1 \eta_1^2\,e^{\sqrt{3} \eta_1 (x-9\eta_1^4 t)}}{\left(1- 2r_1\, e^{\sqrt{3}\eta_1(x-9 \eta_1^4 t)}\right)^2},\qquad x\in\mathbb R,
\end{equation}
where we recall that $\eta_1>0.$ The choice $r_1<0$ ensures that $Q(x)$ in \eqref{5.98} does not have any singularities. 
The explicit expression for $Q(x)$ is equivalent to that obtained by Hirota's bilinear method \cite{H1989}, but our own method clearly indicates the physical
origins of the two constants $\eta_1$ and $r_1$ appearing in \eqref{5.98} and relates them to the bound-state information
for \eqref{1.1}. On the other hand, in Hirota's method those two constants are introduced algebraically without any
reference to any physical quantities. In fact, for any choice of those two constants, the expression in \eqref{5.98} satisfies
\eqref{1.10} because those two constants originate from the input data set used to solve the inverse problem for
\eqref{1.1}. The quantity $Q(x)$ in \eqref{5.98} is a complex-valued solution to \eqref{1.10}.
If $\eta_1$ is chosen as zero, we get the trivial solution to \eqref{1.10}.
If $\eta_1$ is restricted to nonnegative real values, we still get a solution to \eqref{1.10}, where the solution
is complex valued if the constant $r_1$ is nonreal, the solution is the trivial solution if $r_1=0,$ and 
the solution has a singularity if $r_1>0.$
Let us also compare \eqref{5.98} with the quantity $Q(x)$ given in \eqref{5.33}. We recall that
$Q(x)$ in \eqref{5.33} is related to the transmission coefficient
$T_\text{\rm{l}}(k)$ in \eqref{5.1} having $\mathbf N$ simple poles.
On the other hand, the quantity $Q(x)$ in \eqref{5.98} 
 is related to the transmission coefficient
$T_\text{\rm{l}}(k)$ in \eqref{5.35} having $2\mathbf N$ simple poles.
It is impossible to choose
the complex-valued dependency constant $\gamma(k_1)$ appearing in \eqref{5.33} so that
the quantity $Q(x)$ in \eqref{5.33} and the quantity $Q(x)$ in \eqref{5.98} coincide. This shows that the
$1$-soliton solution to \eqref{1.10} is obtained as a special case of a $2$-soliton solution to \eqref{1.6}, but not as
a special case of a $1$-soliton solution to \eqref{1.6}.

\item[$\bullet$] When $\mathbf N=1,$ the quantity in \eqref{5.96} again yields a $1$-soliton solution 
to the Sawada--Kotera equation, with $P(x)\equiv Q_x(x),$ if we use the restriction
\begin{equation}\label{5.99}
s_1=0.
\end{equation}
In that case,  the coefficient $w_1^{(1)}$ is evaluated as $r_1$ by using \eqref{5.99} in 
the second equality of \eqref{5.86}, and we have the explicit expressions for $Q(x)$ given by
\begin{equation}\label{5.100}
Q(x)=\ds\frac{18 r_1 \eta_1^2\,e^{\sqrt{3} \eta_1 (x-9\eta_1^4 t)}}{\left(1+r_1\,e^{\sqrt{3} \eta_1(x-9 \eta_1^4 t)}\right)^2},\qquad x\in\mathbb R,
\end{equation}
where $\eta_1>0.$ The choice $r_1>0$ ensures that $Q(x)$ appearing in \eqref{5.100}
does not have any singularities. 
Again, our method clearly indicates the 
physical origins of the two constants $\eta_1$ and $r_1$ appearing in \eqref{5.100} and relates them to the bound-state information
for \eqref{1.14}. 
By comparing \eqref{5.100} with the quantity $Q(x)$ in \eqref{5.33}, we see that it is impossible to choose
the complex-valued dependency constant $\gamma(k_1)$ appearing in \eqref{5.33} so that
the quantity $Q(x)$ in \eqref{5.33} and the quantity $Q(x)$ in \eqref{5.100} coincide. 
The quantity
$Q(x)$ in \eqref{5.33} is related to the transmission coefficient
$T_\text{\rm{l}}(k)$ in \eqref{5.1} with $\mathbf N$ simple poles, and 
the quantity $Q(x)$ in \eqref{5.100} 
 is related to the transmission coefficient
$T_\text{\rm{l}}(k)$ in \eqref{5.35} with $2\mathbf N$ simple poles.
It is impossible to choose
the complex-valued dependency constant $\gamma(k_1)$ appearing in \eqref{5.33} so that
the quantity $Q(x)$ in \eqref{5.33} and the quantity $Q(x)$ in \eqref{5.100} coincide. Thus, the
$1$-soliton solution to \eqref{1.10} is obtained as a special case of a $2$-soliton solution to \eqref{1.6}, but not as
a special case of a $1$-soliton solution to \eqref{1.6}.
%
%
%
%Again, the
%$1$-soliton solution to \eqref{1.10} is obtained as a special case of a $2$-soliton solution to \eqref{1.6}, but not as
%a special case of a $1$-soliton solution to \eqref{1.6}.

\item[$\bullet$] For $\mathbf N=2,$ the quantity $Q(x)$ in \eqref{5.96} yields a $2$-soliton solution to the Sawada--Kotera equation, with $P(x)\equiv 0,$ if we
use in \eqref{5.87}--\eqref{5.90} the restrictions
\begin{equation*}
%\label{5.101}
s_1=\left(\ds\frac{\sqrt{3}\, \eta_2^2}{2 \eta_1^2 + \eta_2^2}\right)r_1, \quad s_2=\left(\ds\frac{\sqrt{3}\, \eta_1^2}{\eta_1^2+2 \eta_2^2 }\right)r_2.
\end{equation*}
In that case, the coefficients in \eqref{5.87} are given by
\begin{equation}\label{5.102}
w_1^{(2)}=\ds\frac{2 r_1(\eta_1+\eta_2)\,(\eta_1^3-\eta_2^3)}
{(\eta_1-\eta_2)^2(2\eta_1^2+\eta_2^2)},
\end{equation}
\begin{equation}\label{5.103}
w_2^{(2)}=-\ds\frac{2r_2(\eta_1+\eta_2)\,(\eta_1^3-\eta_2^3)}
{(\eta_1-\eta_2)^2(\eta_1^2+2\eta_2^2)},
\end{equation}
\begin{equation}\label{5.104}
w_{12}^{(2)}=-\ds\frac{4 r_1 r_2 (\eta_1^6-\eta_2^6)}
{(\eta_1^2-\eta_2^2)(2\eta_1^2+\eta_2^2)(\eta_1^2+2\eta_2^2)}.
\end{equation}
We remark that the symmetry induced by \eqref{5.73} allows us to determine \eqref{5.103} from \eqref{5.102} and vice versa.
The right-hand side of \eqref{5.104} has the corresponding symmetry when the subscripts $1$ and $2$ are interchanged.

\item[$\bullet$] When $\mathbf N=2,$ the quantity $Q(x)$ in \eqref{5.96} yields a $2$-soliton solution to the 
Sawada--Kotera equation, with $P(x)\equiv Q_x(x),$ if we
use in \eqref{5.87}--\eqref{5.90} the restrictions
\begin{equation}\label{5.105}
s_1=-\left(\ds\frac{\sqrt{3}\, \eta_1^2}{\eta_1^2+2 \eta_2^2}\right)r_1, \quad s_2=-\left(\ds\frac{\sqrt{3}\, \eta_2^2}{2 \eta_1^2 + \eta_2^2}\right)r_2.
\end{equation}
In that case, the coefficients in \eqref{5.87} are given by
\begin{equation}\label{5.106}
w_1^{(2)}=-\ds\frac{2 r_1(\eta_1+\eta_2)\,(\eta_1^3-\eta_2^3)}
{(\eta_1-\eta_2)^2(\eta_1^2+2\eta_2^2)},
\end{equation}
\begin{equation}\label{5.107}
w_2^{(2)}=\ds\frac{2r_2(\eta_1+\eta_2)\,(\eta_1^3-\eta_2^3)}
{(\eta_1-\eta_2)^2(2\eta_1^2+\eta_2^2)},
\end{equation}
\begin{equation}\label{5.108}
w_{12}^{(2)}=-\ds\frac{4 r_1 r_2 (\eta_1^6-\eta_2^6)}
{(\eta_1^2-\eta_2^2)(2\eta_1^2+\eta_2^2)(\eta_1^2+2\eta_2^2)}.
\end{equation}
We observe various symmetries in \eqref{5.105}--\eqref{5.108} induced by \eqref{5.73}.

\item[$\bullet$] When $\mathbf N=3,$ the quantity $Q(x)$ in \eqref{5.96} yields a $3$-soliton solution to the Sawada--Kotera equation, with $P(x)\equiv 0,$ if
the coefficients in \eqref{5.94} are restricted as
\begin{equation}\label{5.109}
s_1=\left(\ds\frac{\sqrt{3}(-\eta_1^4 + \eta_2^2 \,\eta_3^2)}{\eta_1^4 +\eta_2^2\, \eta_3^2 + 
2 \eta_1^2 (\eta_2^2 + \eta_3^2)} \right)r_1, 
\end{equation}
\begin{equation}\label{5.110}
s_2=\left(\ds\frac{\sqrt{3}(-\eta_2^4 + \eta_1^2\, \eta_3^2)}{\eta_2^4 +\eta_1^2\, \eta_3^2 + 2 \eta_2^2 (\eta_1^2 + \eta_3^2)} \right)r_2,
\end{equation}
\begin{equation}\label{5.111}
s_3=\left(\ds\frac{\sqrt{3}(-\eta_3^4 + \eta_2^2\, \eta_1^2)}{\eta_3^4 +\eta_2^2\, \eta_1^2 + 2 \eta_3^2 (\eta_2^2 + \eta_1^2)} \right)r_3.
\end{equation}
In that case, the coefficients in \eqref{5.94} for $\chi_1,$ $\chi_2,$ and $\chi_3$ 
are given by
\begin{equation}\label{5.112}
w_1^{(3)}=-\ds\frac{2 r_1(\eta_1+\eta_2)(\eta_1+\eta_3)\,(\eta_1^3-\eta_2^3)(\eta_1^3-\eta_3^3)}
{(\eta_1-\eta_2)^2(\eta_1-\eta_3)^2
\left(\eta_1^4 +\eta_2^2\, \eta_3^2 + 
2 \eta_1^2 (\eta_2^2 + \eta_3^2)\right)},
\end{equation}
\begin{equation}\label{5.113}
w_2^{(3)}=\ds\frac{2r_2(\eta_1+\eta_2)
(\eta_2+\eta_3)
\,(\eta_1^3-\eta_2^3)(\eta_2^3-\eta_2^3)}
{(\eta_1-\eta_2)^2(\eta_2-\eta_3)^2
\left(\eta_2^4 +\eta_1^2\, \eta_3^2 + 2 \eta_2^2 (\eta_1^2 + \eta_3^2)\right)},
\end{equation}
\begin{equation}\label{5.114}
w_3^{(3)}=-\ds\frac{2r_3 (\eta_1+\eta_3)
(\eta_2+\eta_3)
\,(\eta_1^3-\eta_3^3)(\eta_2^3-\eta_3^3)}
{(\eta_1-\eta_3)^2(\eta_2-\eta_3)^2
\left(\eta_3^4 +\eta_2^2\, \eta_1^2 + 2 \eta_3^2 (\eta_2^2 + \eta_1^2)\right)}.
\end{equation}
Using Mathematica,
one can display all the coefficients appearing in \eqref{5.94} explicitly.
We do not display the explicit expressions for the remaining coefficients here because they are lengthy.
We observe various symmetries in \eqref{5.109}--\eqref{5.114} induced by \eqref{5.73}.

\item[$\bullet$] When $\mathbf N=3,$ the quantity $Q(x)$ in \eqref{5.96} yields a $3$-soliton solution to the 
Sawada--Kotera equation, with $P(x)\equiv Q_x(x),$ if we
impose on the expressions for the coefficients in \eqref{5.94} the restrictions
\begin{equation}\label{5.115}
s_1=\left(\ds\frac{\sqrt{3}\,\eta_1^2 (\eta_1^2 + \eta_2^2 + \eta_3^2)}{\eta_1^4-2 \eta_2^2 \,\eta_3^2-\eta_1^2
(\eta_2^2 + \eta_3^2)}\right)r_1, 
\end{equation}
\begin{equation}\label{5.116}
s_2=\left(\ds\frac{\sqrt{3}\,\eta_2^2 (\eta_1^2 + \eta_2^2 + \eta_3^2)}{\eta_2^4-2 \eta_1^2 \,\eta_3^2-\eta_2^2(\eta_1^2 + \eta_3^2)}\right)r_2,
\end{equation}
\begin{equation}\label{5.117}
s_3=\left(\ds\frac{\sqrt{3}\,\eta_3^2 (\eta_1^2 + \eta_2^2 + \eta_3^2)}{\eta_3^4-2 \eta_2^2\, \eta_1^2-\eta_3^2
(\eta_2^2 + \eta_1^2)}\right)r_3.
\end{equation}
In that case the coefficients in \eqref{5.94} for $\chi_1,$ $\chi_2,$ and $\chi_3$ 
are given by
\begin{equation}\label{5.118}
w_1^{(3)}=-\ds\frac{2 r_1(\eta_1+\eta_2)(\eta_1+\eta_3)\,(\eta_1^3-\eta_2^3)(\eta_1^3-\eta_3^3)}
{(\eta_1-\eta_2)^2(\eta_1-\eta_3)^2
(\eta_1^4-\eta_1^2\eta_2^2-\eta_1^2\eta_3^2-2\eta_2^2\eta_3^2)},
\end{equation}
\begin{equation}\label{5.119}
w_2^{(3)}=\ds\frac{2r_2(\eta_1+\eta_2)
(\eta_2+\eta_3)
\,(\eta_1^3-\eta_2^3)(\eta_2^3-\eta_3^3)}
{(\eta_1-\eta_2)^2(\eta_2-\eta_3)^2
(\eta_2^4-\eta_1^2\eta_2^2-\eta_2^2\eta_3^2-2\eta_1^2\eta_3^2)},
\end{equation}
\begin{equation}\label{5.120}
w_3^{(3)}=-\ds\frac{2r_3 (\eta_1+\eta_3)
(\eta_2+\eta_3)
\,(\eta_1^3-\eta_3^3)(\eta_2^3-\eta_3^3)}
{(\eta_1-\eta_3)^2(\eta_2-\eta_3)^2
(\eta_3^4-\eta_1^2\eta_3^2-\eta_2^2\eta_3^2-2\eta_1^2\eta_2^2)}.
\end{equation}
Using Mathematica,
one can display all the coefficients appearing in \eqref{5.94} explicitly.
We do not display the explicit expressions for the remaining coefficients here because they are lengthy.
The choice given in \eqref{5.73} induces various symmetries in \eqref{5.115}--\eqref{5.120}.

\item[$\bullet$] When $\mathbf N=4,$ the quantity $Q(x)$ in \eqref{5.96} yields a $4$-soliton solution to the Sawada--Kotera equation, with $P(x)\equiv 0,$ if the expressions for the coefficients in \eqref{5.95} satisfy the restrictions
\begin{equation}\label{5.121}
s_1=\left(\ds\frac{\sqrt{3}\, \left(\eta_1^6 - \eta_2^2\,\eta_3^2\,\eta_4^2 + \eta_1^4\left(\eta_2^2 + \eta_3^2 + \eta_4^2\right)\right)}
{\eta_1^6 - \eta_2^2\,\eta_3^2\,\eta_4^2 - \eta_1^4 \left(\eta_2^2 + \eta_3^2 + \eta_4^2\right) - 
 2 \eta_1^2 \left(\eta_3^2 \,\eta_4^2 + \eta_2^2 \,\eta_3^2 + \eta_2^2 \,\eta_4^2\right)}\right)r_1,
\end{equation}
\begin{equation}\label{5.122}
s_2=\left(\ds\frac{\sqrt{3}\, \left(\eta_2^6 - \eta_1^2\,\eta_3^2\,\eta_4^2 + \eta_2^4\left(\eta_1^2 + \eta_3^2 + \eta_4^2\right)\right)}
{\eta_2^6 - \eta_1^2\,\eta_3^2\,\eta_4^2 - \eta_2^4 \left(\eta_1^2 + \eta_3^2 + \eta_4^2\right) - 
 2 \eta_2^2 \left(\eta_3^2 \,\eta_4^2 + \eta_1^2 \,\eta_3^2 + \eta_1^2 \,\eta_4^2\right)}\right)r_2,
\end{equation}
\begin{equation}\label{5.123}
s_3=\left(\ds\frac{\sqrt{3}\, \left(\eta_3^6 - \eta_2^2\,\eta_1^2\,\eta_4^2 + \eta_3^4\left(\eta_2^2 + \eta_1^2 + \eta_4^2\right)\right)}
{\eta_3^6 - \eta_2^2\,\eta_1^2\,\eta_4^2 - \eta_3^4 \left(\eta_2^2 + \eta_1^2 + \eta_4^2\right) - 
 2 \eta_3^2 \left(\eta_1^2 \,\eta_4^2 + \eta_2^2 \,\eta_1^2 + \eta_2^2 \,\eta_4^2\right)}\right)r_3,
\end{equation}
\begin{equation}\label{5.124}
s_4=\left(\ds\frac{\sqrt{3}\, \left(\eta_4^6 - \eta_2^2\,\eta_3^2\,\eta_1^2 + \eta_4^4\left(\eta_2^2 + \eta_3^2 + \eta_1^2\right)\right)}
{\eta_4^6 - \eta_2^2\,\eta_3^2\,\eta_1^2 - \eta_4^4 \left(\eta_2^2 + \eta_3^2 + \eta_1^2\right) - 
 2 \eta_4^2 \left(\eta_3^2 \,\eta_1^2 + \eta_2^2 \,\eta_3^2 + \eta_2^2 \,\eta_1^2\right)}\right)r_4.
\end{equation}
With the help of Mathematica,
one can display all the coefficients appearing in \eqref{5.95} explicitly.
We do not display those explicit expressions here because they are lengthy.
We remark that the symmetry induced by \eqref{5.73} enables us to obtain all four equalities in
\eqref{5.121}--\eqref{5.124} from any one of them.

\item[$\bullet$] When $\mathbf N=4,$ the quantity $Q(x)$ in \eqref{5.96} yields a $4$-soliton solution to the 
Sawada--Kotera equation, with $P(x)\equiv Q_x(x),$ if we
use  in the expressions for the coefficients in \eqref{5.95} the restrictions
\begin{equation}\label{5.125}
s_1=\left(\ds\frac{\sqrt{3}\, \eta_1^2 \left(\eta_3^2\,\eta_4^2 + \eta_2^2 (\eta_3^2 +\eta_4^2) + \eta_1^2 (\eta_2^2 +
 \eta_3^2 + \eta_4^2)\right)}{2 \eta_1^6-2 \eta_2^2\, \eta_3^2\, \eta_4^2 + \eta_1^4 (\eta_2^2 + \eta_3^2+\eta_4^2) - \eta_1^2
\left(\eta_3^2\, \eta_4^2 + \eta_2^2 (\eta_3^2 + \eta_4^2)\right)}\right)r_1,
\end{equation}
\begin{equation}\label{5.126}
s_2=\left(\ds\frac{\sqrt{3}\, \eta_2^2 \left(\eta_3^2\,\eta_4^2 + \eta_1^2 (\eta_3^2 +\eta_4^2) + \eta_2^2 (\eta_1^2 + \eta_3^2 +
\eta_4^2)\right)}{2 \eta_2^6-2 \eta_1^2\, \eta_3^2\, \eta_4^2 + \eta_2^4 (\eta_1^2 + \eta_3^2+\eta_4^2) - \eta_2^2
\left(\eta_3^2\, \eta_4^2 + \eta_1^2 (\eta_3^2 + \eta_4^2)\right)}\right)r_2,
\end{equation}
\begin{equation}\label{5.127}
s_3=\left(\ds\frac{\sqrt{3}\, \eta_3^2 \left(\eta_1^2\,\eta_4^2 + \eta_2^2 (\eta_1^2 +\eta_4^2) + \eta_3^2 (\eta_2^2 + \eta_1^2 +
\eta_4^2)\right)}{2 \eta_3^6-2 \eta_2^2\, \eta_1^2\, \eta_4^2 + \eta_3^4 (\eta_2^2 + \eta_1^2+\eta_4^2) - \eta_3^2
\left(\eta_1^2\, \eta_4^2 + \eta_2^2 (\eta_1^2 + \eta_4^2)\right)}\right)r_3,
\end{equation}
\begin{equation}\label{5.128}
s_4=\left(\ds\frac{\sqrt{3}\, \eta_4^2 \left(\eta_3^2\,\eta_1^2 + \eta_2^2 (\eta_3^2 +\eta_1^2) + \eta_4^2 (\eta_2^2 + \eta_3^2 +
\eta_1^2)\right)}{2 \eta_4^6-2 \eta_2^2\, \eta_3^2\, \eta_1^2 + \eta_4^4 (\eta_2^2 + \eta_3^2+\eta_1^2) - \eta_4^2
\left(\eta_3^2\, \eta_1^2 + \eta_2^2 (\eta_3^2 + \eta_1^2)\right)}\right)r_4.
\end{equation}
Using Mathematica,
one can display all the coefficients appearing in \eqref{5.95} explicitly.
We do not display those explicit expressions here because they are lengthy.
The symmetry induced by \eqref{5.73} enables us to obtain \eqref{5.125}--\eqref{5.128} from any one of those equalities.

We remark that the explicit expressions for the coefficients in $\Delta(\mathbf N,x)$
agree with the coefficients evaluated by using
Hirota's bilinear method \cite{H1989}.
Our method clearly indicates from where those coefficients come, whereas in Hirota's method
those coefficients are introduced algebraically in an ad hoc manner without
providing any physical insight.

\end{enumerate}

\end{example}

In the next example we present some snapshots for certain specific $\mathbf N$-soliton solutions to the 
Sawada--Kotera equation
\eqref{1.10} for $\mathbf N=1,2,3,4.$
As seen from \eqref{5.96}, each soliton solution is uniquely determined by specifying
the scalar quantity $\Delta(\mathbf N,x).$ We know from Example~\ref{example5.3} that
each $\Delta(\mathbf N,x)$ is uniquely determined by the set $\{\eta_j,r_j\}_{j=1}^{\mathbf N},$ 
where $\eta_j$ is the real-valued constant identifying the location of the bound state at $k=k_j$
and $r_j$ is the real-valued constant used to identify the corresponding bound-state
dependency constant.

\begin{example}\label{example5.4}
\normalfont

In Figure~\ref{figure5.1} we present four snapshots for 
the $1$-soliton solution to the 
Sawada--Kotera equation
\eqref{1.10} by using
$\eta_1=1$ and $r_1=-1$ in \eqref{5.98}.
The right-hand side of \eqref{5.98} contains
$x$ and $t$ in the combination $x-9\eta_1^4t.$ Thus, the $1$-soliton solution $Q(x)$ in \eqref{5.98}
is a solitary wave moving to the right with speed $9\eta_1^4.$ Consequently, in this example
the soliton moves to the right with speed $9.$
In Figure~\ref{figure5.2} we present four snapshots for 
the $2$-soliton solution to \eqref{1.10} by using the four real-valued
parameters $\eta_1,$ $\eta_2,$ $r_1,$ $r_2$ with
$(\eta_1,\eta_2)=(1/\sqrt{3},\sqrt{3})$ and
$(r_1,r_2)=(-1,1).$ We observe that the faster moving taller soliton catches the shorter soliton, and after they interact nonlinearly the
taller soliton moves away from the shorter soliton.
In Figure~\ref{figure5.3} we present four snapshots for 
the $3$-soliton solution to \eqref{1.10} by using the six real-valued
parameters $\eta_1,$ $\eta_2,$ $\eta_3,$ $r_1,$ $r_2,$ $r_3$ with
$(\eta_1,\eta_2,\eta_3)=(1,3/2,2)$ and
$(r_1,r_2,r_3)=(-1,1,-1).$ 
We observe that the fastest moving tallest soliton catches the two shorter solitons, and after they interact nonlinearly the
tallest soliton moves away from the shorter solitons.
Figure~\ref{figure5.4} shows four snapshots for 
the $4$-soliton solution to \eqref{1.10} by using the eight real-valued
parameters $\eta_1,$ $\eta_2,$ $\eta_3,$ $\eta_4,$ $r_1,$ $r_2,$ $r_3,$ $r_4$ with
$(\eta_1,\eta_2,\eta_3,\eta_4)=(1,1.1,1.2,1.3)$ and
$(r_1,r_2,r_3,r_4)=(-1,1,-1,1).$ 
We observe that at $t=-1$ the four solitons are aligned in such a way that 
the shortest soliton is in the front and 
the tallest soliton is behind the other three. Then, they start interacting nonlinearly and the taller solitons pass
the shorter ones. At $t=1$ the four solitons are well separated from each other
with the tallest soliton is in the front and the shortest soliton is behind the rest.
Since the amplitude of each soliton is large, in the figure we do not show the whole
amplitudes.
% so that we can demonstrate the 
%nonlinear interactions as well as the alignments of the four solitons before and after those nonlinear interactions.

\begin{figure}[!ht]
     \centering
         \includegraphics[width=1.45in]{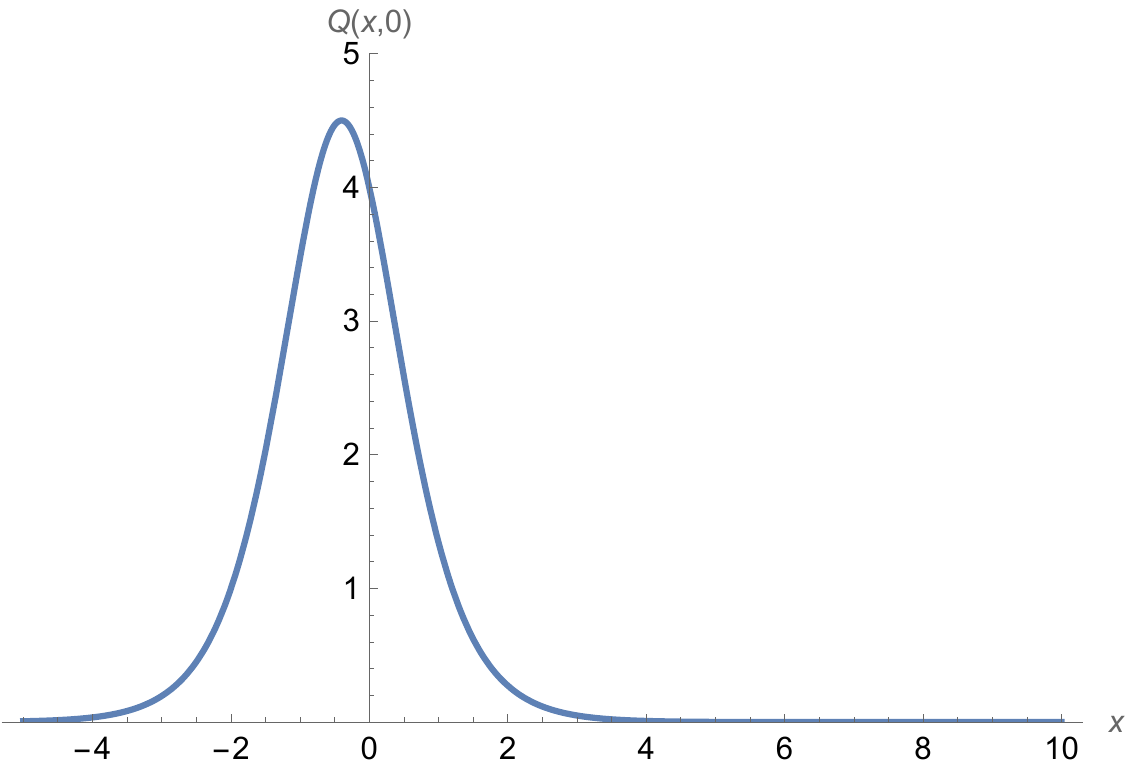}      \hskip .1in
         \includegraphics[width=1.45in]{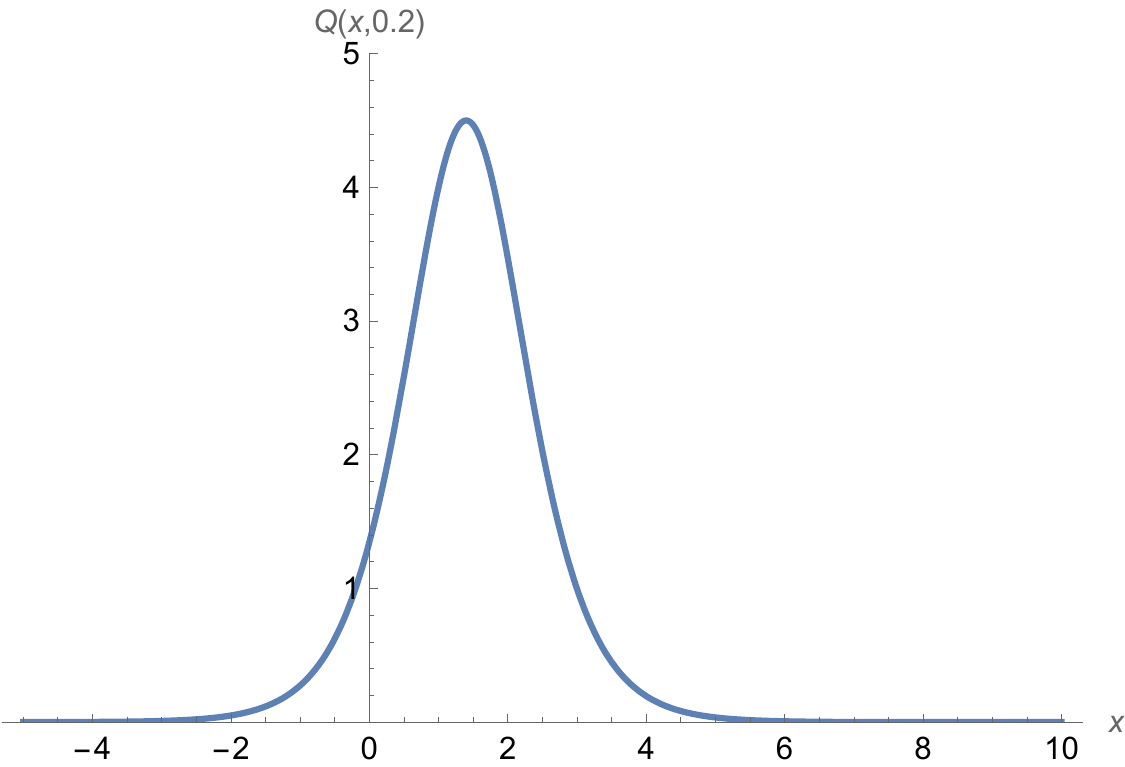} \hskip .1in
                           \includegraphics[width=1.45in]{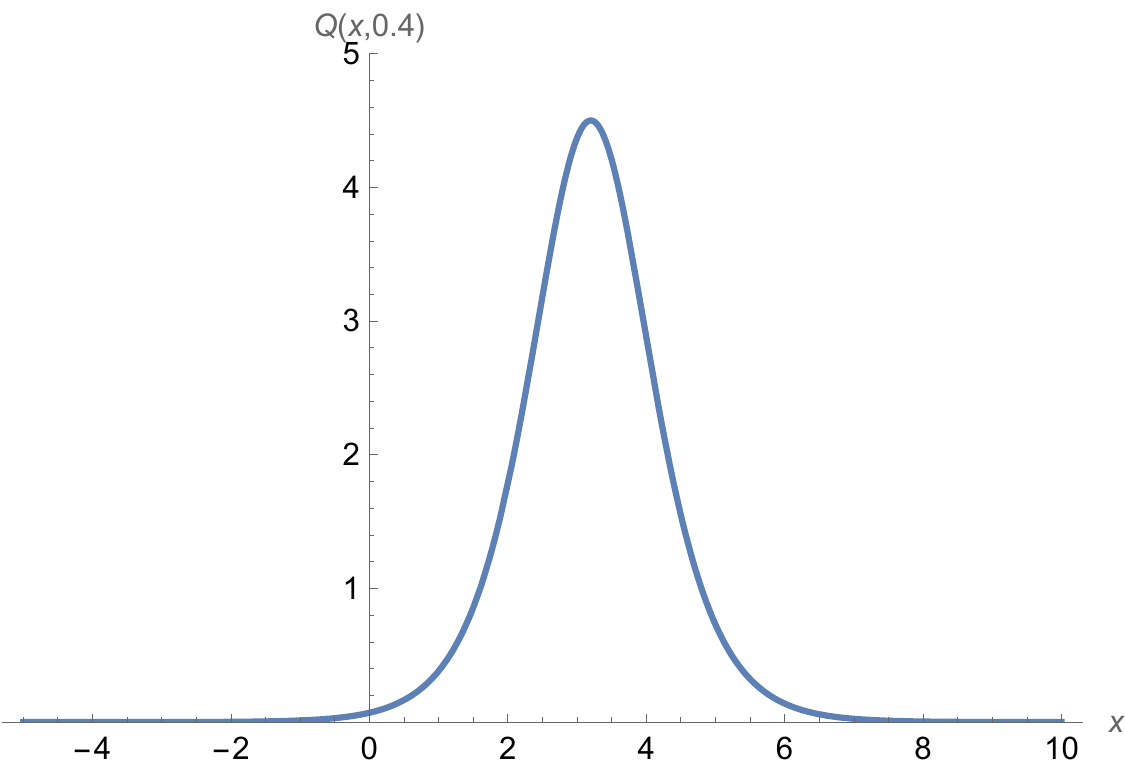} \hskip .1in
                  \includegraphics[width=1.45in]{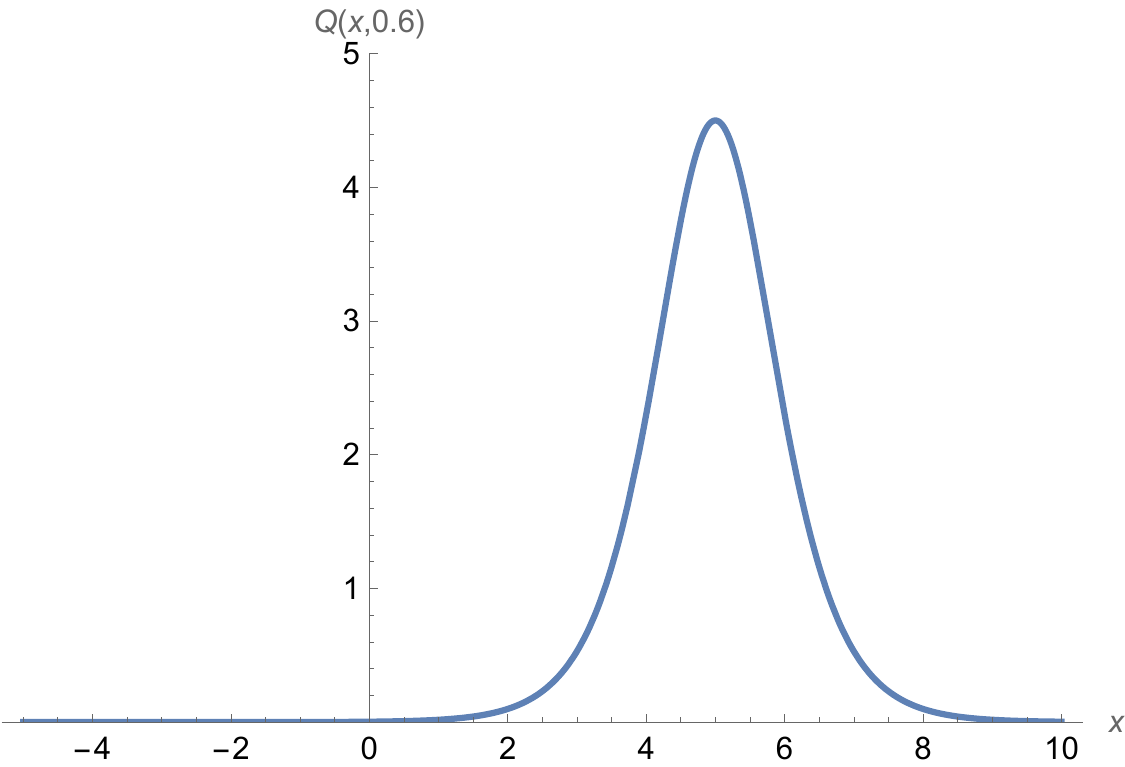}
\caption{The snapshots for the $1$-soliton solution $Q(x,t)$ to
\eqref{1.10} 
%with $(\eta_1,r_1)=(1,-1)$ 
at $t=0,$ $t=0.2,$ $t=0.4,$ and $t=0.6,$ respectively. }
\label{figure5.1}
\end{figure}

\begin{figure}[!ht]
     \centering
         \includegraphics[width=1.45in]{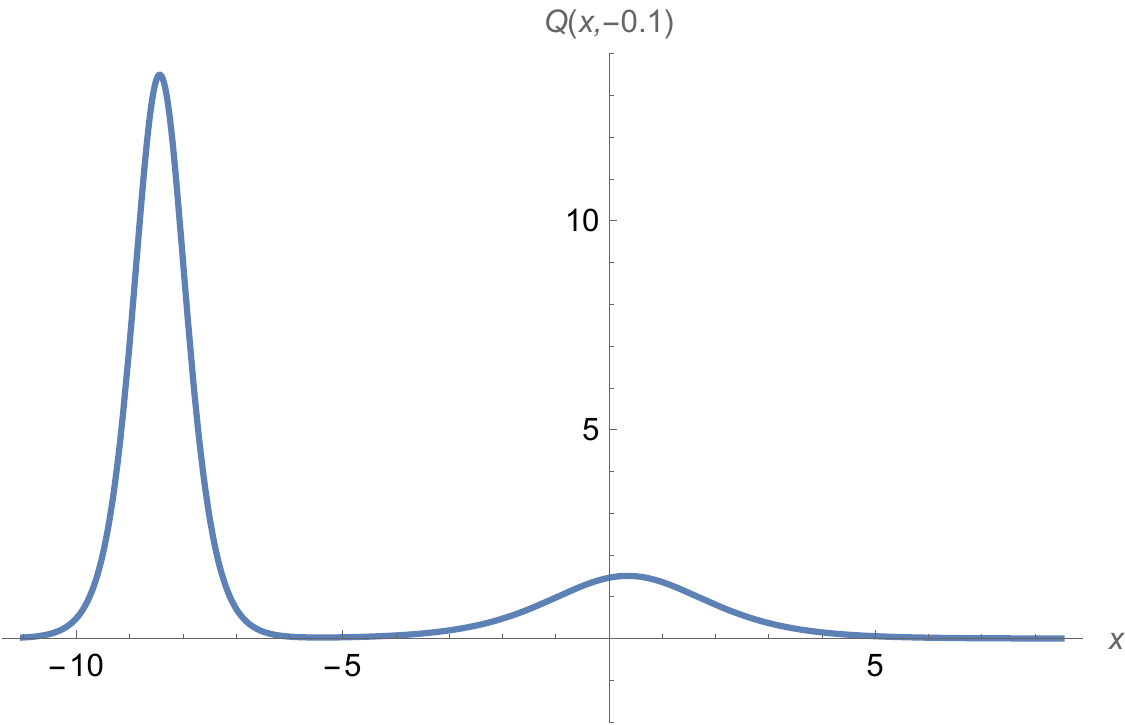}      \hskip .1in
         \includegraphics[width=1.45in]{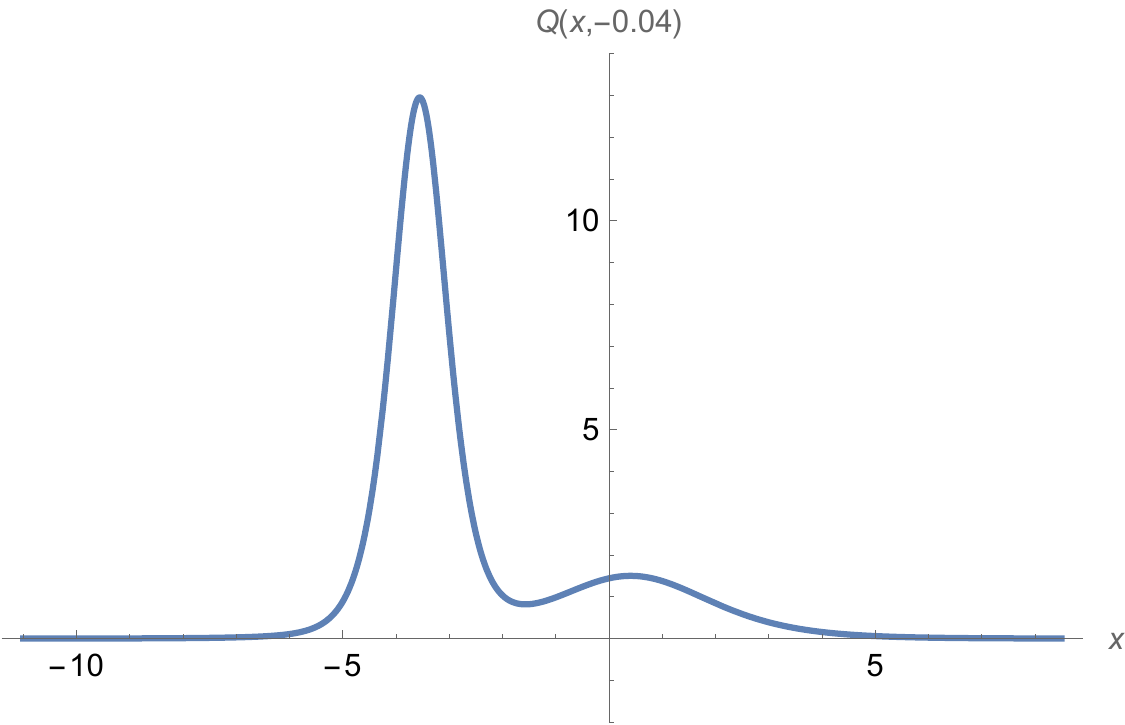} \hskip .1in
                  \includegraphics[width=1.45in]{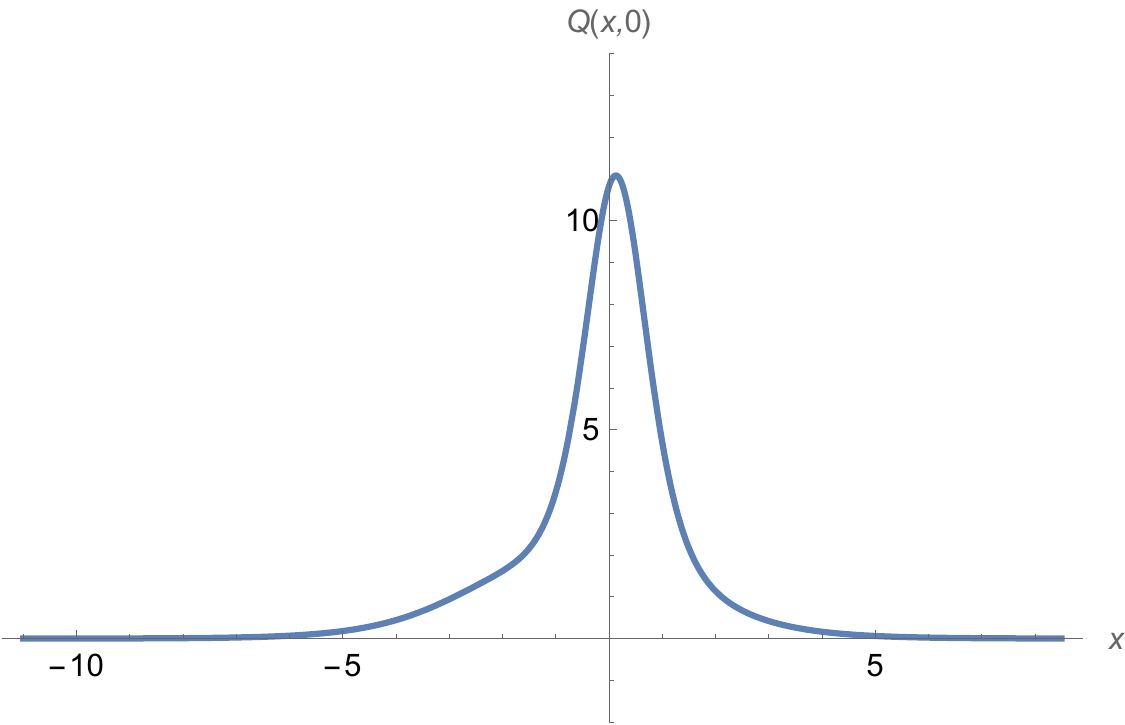} \hskip .1in
                  \includegraphics[width=1.45in]{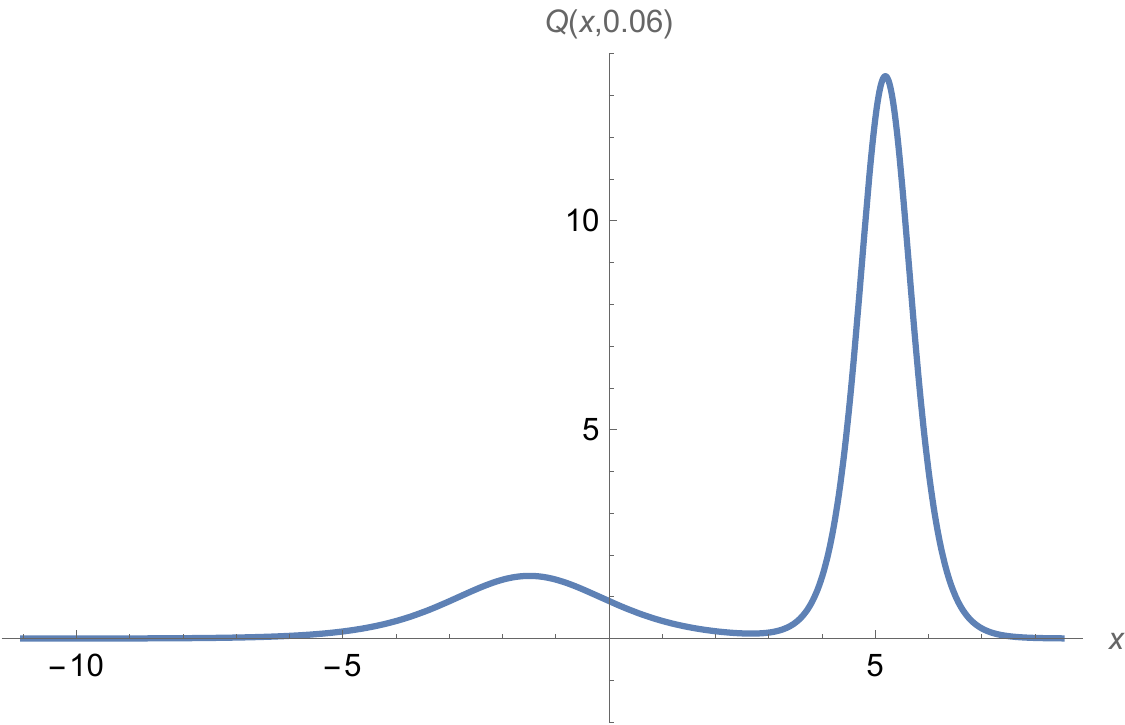} 
\caption{The snapshots for the $2$-soliton solution $Q(x,t)$ to
\eqref{1.10} 
%with $(\eta_1,\eta_2)=(1/\sqrt{3},\sqrt{3})$ and $(r_1,r_2)=(-1,1)$ 
at $t=-0.1,$ $t=-0.04,$ $t=0,$ and $t=0.06,$ respectively.}
\label{figure5.2}
\end{figure}

\begin{figure}[!ht]
     \centering
         \includegraphics[width=1.45in]{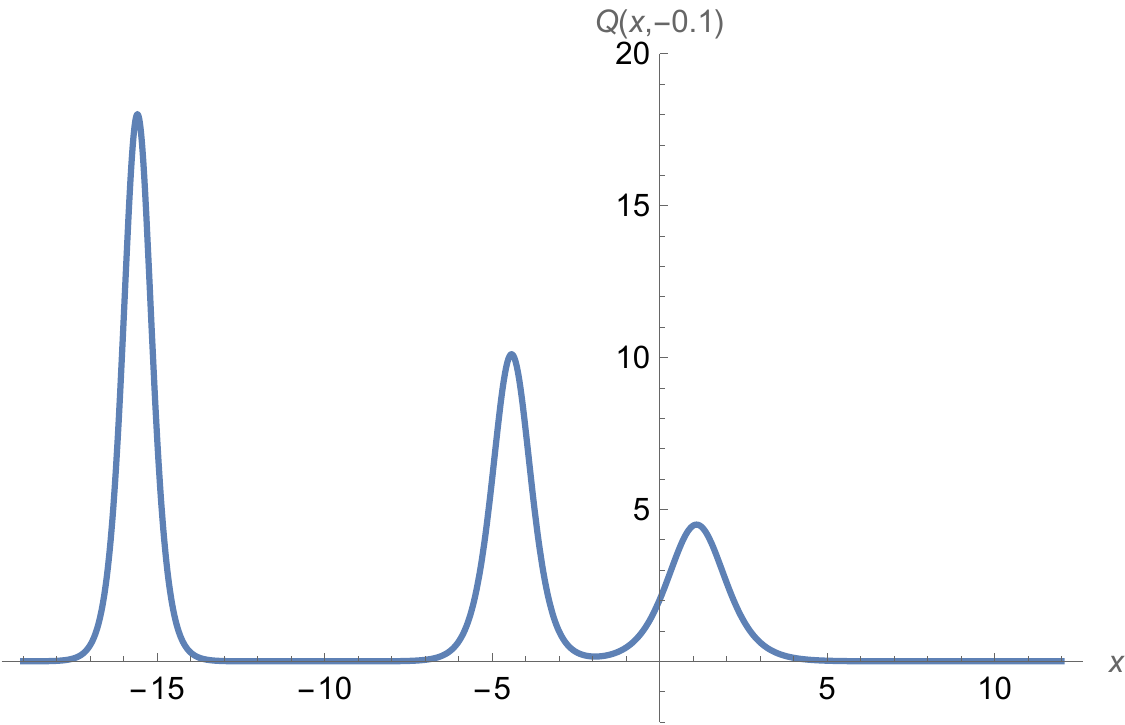}      \hskip .1in
         \includegraphics[width=1.45in]{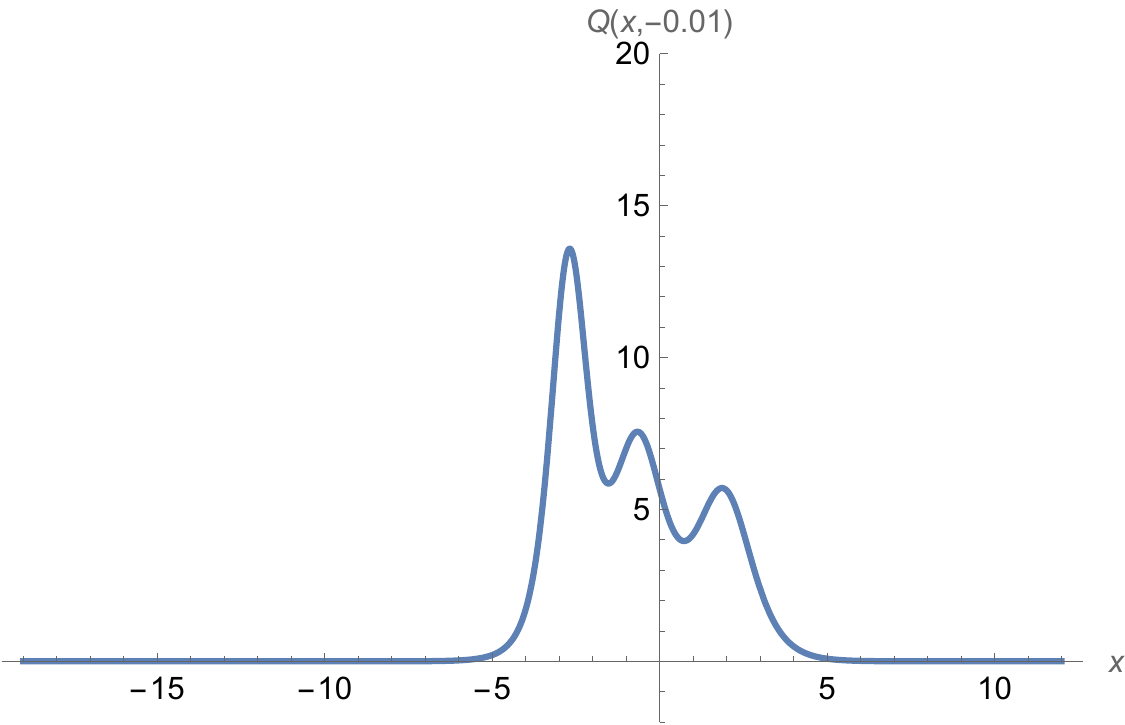} \hskip .1in
                  \includegraphics[width=1.45in]{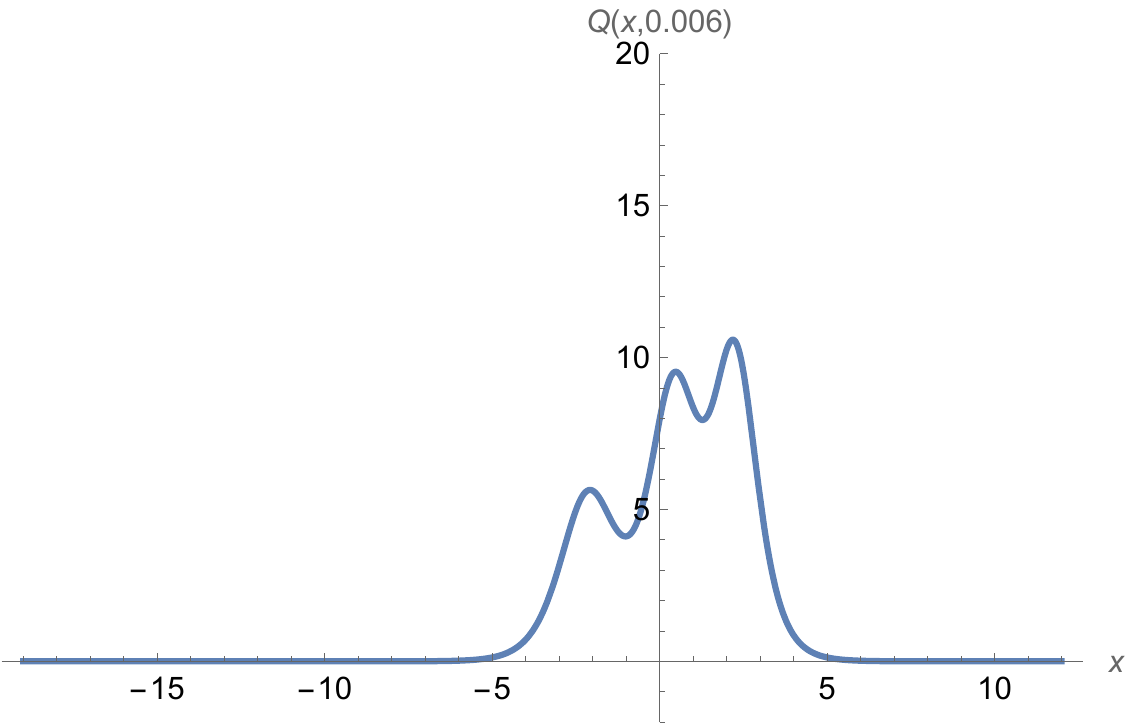} \hskip .1in
                  \includegraphics[width=1.45in]{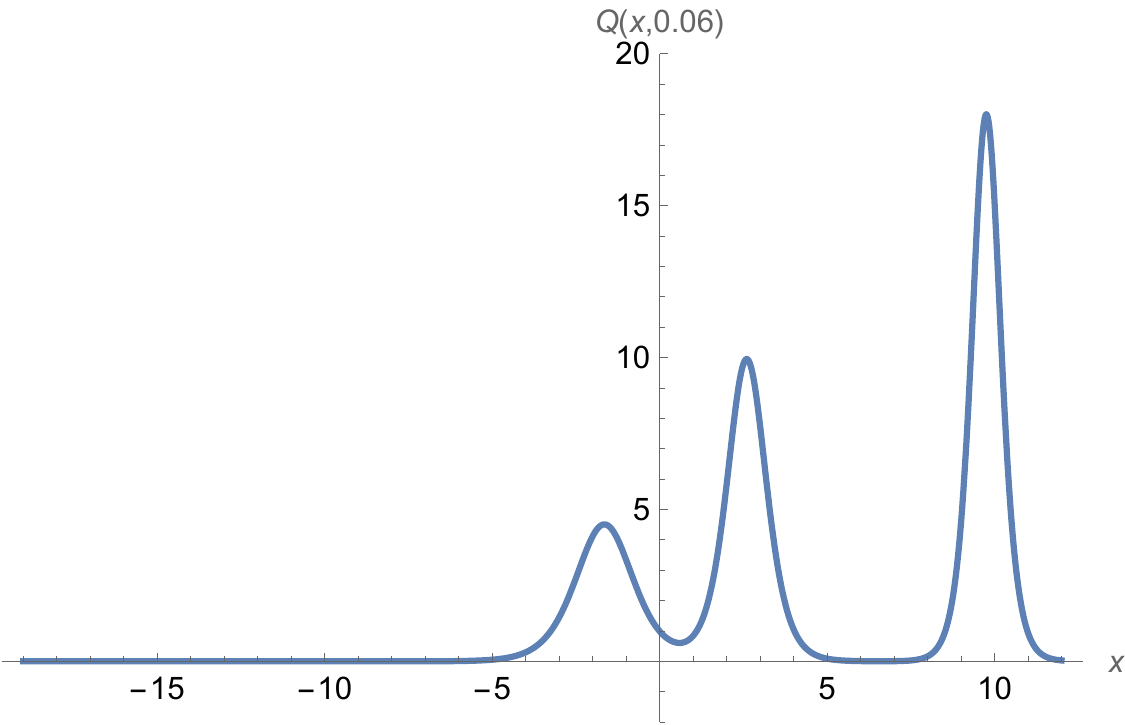} 
\caption{The snapshots for the $3$-soliton solution $Q(x,t)$ to
\eqref{1.10} 
%with
%$(\eta_1,\eta_2,\eta_3)=(1,3/2,2)$ and
%$(r_1,r_2,r_3)=(-1,1,-1)$ 
at $t=-0.1,$ $t=-0.01,$ $t=0.006,$ and $t=0.06,$ respectively.}
\label{figure5.3}
\end{figure}

\begin{figure}[!ht]
     \centering
         \includegraphics[width=1.45in]{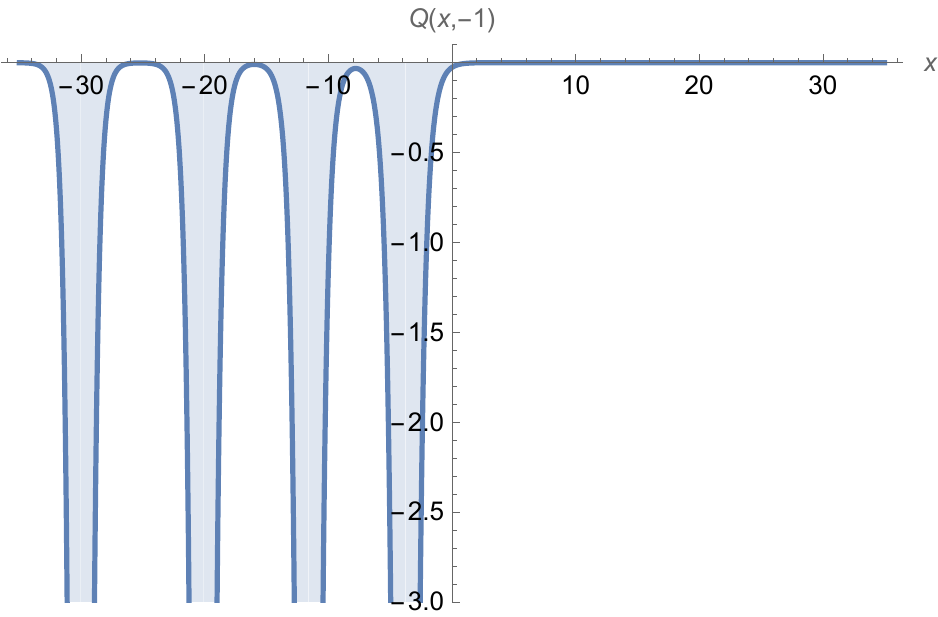}      \hskip .1in
         \includegraphics[width=1.45in]{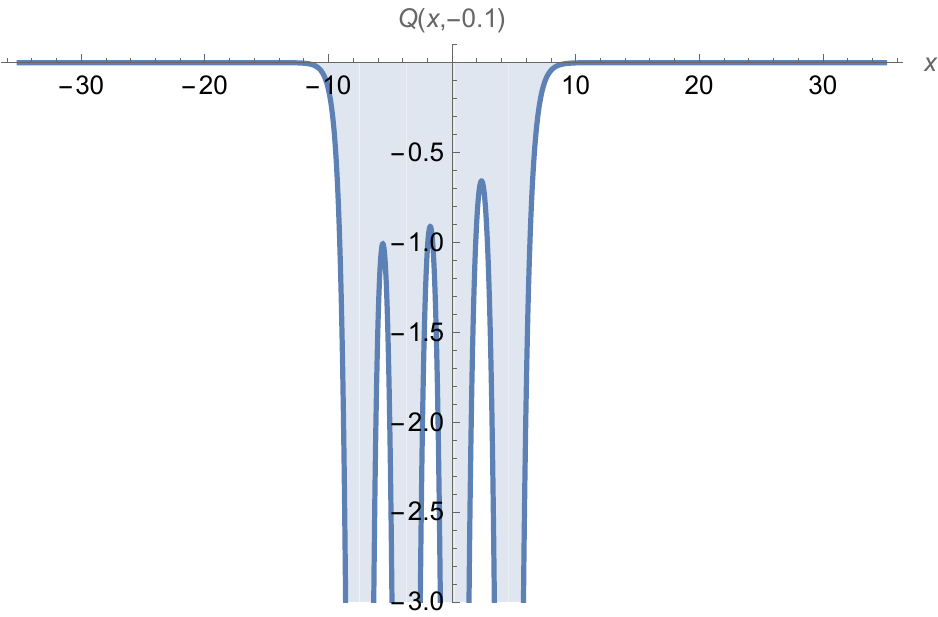} \hskip .1in
                  \includegraphics[width=1.45in]{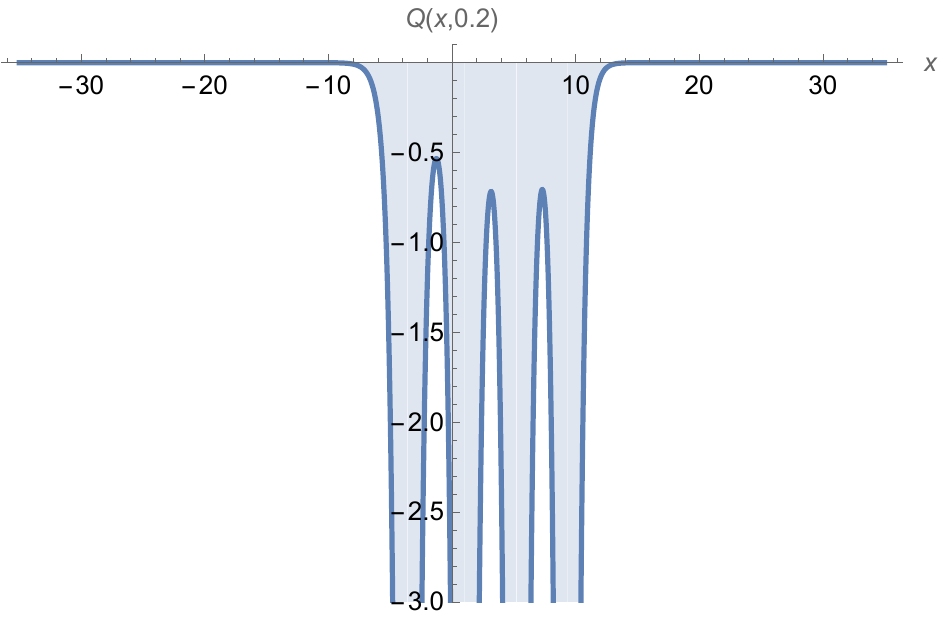} \hskip .1in
                  \includegraphics[width=1.45in]{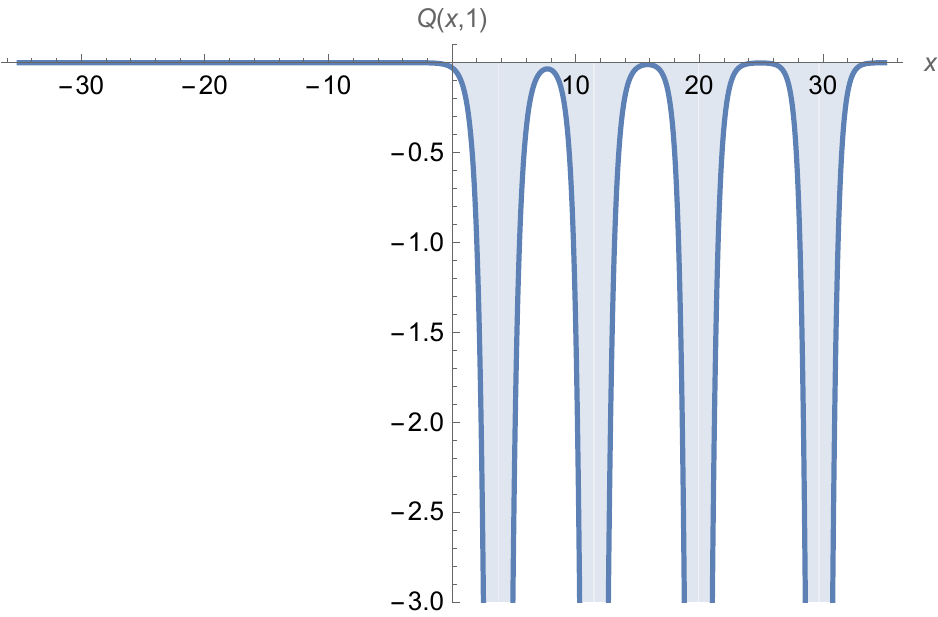} 
\caption{The snapshots for the $4$-soliton solution $Q(x,t)$ to the Sawada--Kotera equation
\eqref{1.10} 
%with
%$(\eta_1,\eta_2,\eta_3,\eta_4)=(1,1.1,1.2,1.3)$ and
%$(r_1,r_2,r_3,r_4)=(-1,1,-1,1)$ 
at $t=-1,$ $t=-0.1,$ $t=0.2,$ and $t=1,$ respectively.}
\label{figure5.4}
\end{figure}

\end{example}

\section{A modified bad Boussinesq equation}
\label{section6}

The method presented
in this paper to obtain
explicit solutions to the relevant integrable evolution equations
is based on the solution to the inverse scattering
problem for \eqref{1.1} in the reflectionless case. Hence, our method is
applicable to obtain explicit solutions to 
other integrable nonlinear partial differential equations
associated with the third-order linear ordinary differential equation \eqref{1.1}. 
In this section, we illustrate the use of our method to obtain explicit solutions to a modified version
\cite{DTT1982} of
the bad Boussinesq equation \cite{B1872}.

Deift, Tomei, and Trubowitz analyzed \cite{DTT1982} the inverse scattering transform for 
a modified version of the integrable bad Boussinesq system \cite{AC1991,AS1981,B1872,DEGM1982,DJ1989,H1973,NMPZ1984}, which is given by
\begin{equation}
\label{6.1}
\begin{cases}
q_t=-3p_x,\\
p_t=-q_{xxx}+8q q_x.
\end{cases}
\end{equation}
The integrable system \eqref{6.1} is obtained from the Lax equation \eqref{1.4}
by using the Lax pair $(L,A)$ given by
\begin{equation}
\label{6.2}
L=iD^3-i(qD+Dq)+p,
\end{equation}
\begin{equation}
\label{6.3}
A=i(3D^2-4q).
\end{equation}
We recall that we suppress the $t$-dependence in the potentials $q$ and $p$ in our notation, and we write
$q(x)$ and $p(x)$ instead of $q(x,t)$ and $p(x,t).$  We also remark that the complex constant $i$ is used in the definition of
the linear operator $L$ in \eqref{6.2}, whereas the operator $L$ in \eqref{1.2} does not contain that complex constant.
The linear operator $L$ appearing in \eqref{6.2} is selfadjoint when $q$ and $p$ are real valued. Using \eqref{6.2} and letting $L\phi=ik^3\phi,$ we obtain the
third-order differential equation given by
\begin{equation}
\label{6.4}
\phi'''(x)-2q(x)\phi'(x)-[q'(x)+ip(x)]\phi(x)=k^3\phi, \qquad x\in\mathbb R,
\end{equation}
which is a special case of \eqref{1.1}. By eliminating $p$ in \eqref{6.1} we obtain the integrable modified bad Boussinesq equation given by
\begin{equation}
\label{6.5}
q_{tt}+12(q^2)_{xx}-3\,q_{xxxx}=0, \qquad x\in\mathbb R.
\end{equation}
By comparing $L\phi=ik^3\phi$ with $L\phi=\lambda\phi,$ we see that the spectral parameter $\lambda$ used in this section is not equal to
$k^3$ but to $ik^3,$ i.e. we have $\lambda:=ik^3.$

By defining the quantity $\tilde q(x,t)$ as
\begin{equation}
\label{6.6}
\tilde q(x,t):=4\,q(x,t/\sqrt{3})+\ds\frac{1}{2},
\end{equation}
where $q(x,t)$ refers to the quantity $q$ appearing in \eqref{6.5}, 
one can directly verify that
$\tilde q(x,t)$ satisfies the bad Boussinesq equation
\begin{equation}
\label{6.7}
\tilde q_{tt}-\tilde q_{xx}+(\tilde q^2)_{xx}-\tilde q_{xxxx}=0, \qquad x\in\mathbb R.
\end{equation}
The nonlinear equation \eqref{6.7} can be obtained from the Lax compatibility equation \eqref{1.4}
by using the Lax pair $(L,A),$ where we let
\begin{equation*}
%\label{6.8}
L=D^3+\left(\ds\frac{1}{4}-\ds\frac{1}{2}\,\tilde q\right)D-\ds\frac{1}{4}\,\tilde q_x+\ds\frac{i}{4\sqrt{3}}\,\tilde\phi,
\end{equation*}
\begin{equation*}
%\label{6.9}
A=\sqrt{3}\,i\,D^2-\frac{i}{\sqrt{3}}\,\tilde q.
\end{equation*}
This yields the integrable nonlinear system
\begin{equation*}
%\label{6.10}
\begin{cases}
\tilde q_t=\tilde\phi_x,\\
\tilde\phi_t=\tilde q_x-(\tilde q^2)_x+\tilde q_{xxx},
\end{cases}
\end{equation*}
from which we obtain \eqref{6.7} by eliminating $\tilde\phi.$

Even though \eqref{6.5} and \eqref{6.7} are related to each other as in \eqref{6.6}, soliton solutions
to \eqref{6.5} are not related to solitons solutions to \eqref{6.7} via \eqref{6.6}. This is seen from \eqref{6.6}
by recalling that solitons solutions decay exponentially as $x\to\pm\infty$ at any fixed $t.$
We refer the reader to \cite{CL2022} and the references therein for the elaboration on this issue
and why the development of an inverse scattering transform approach and the asymptotic analysis of
solutions to \eqref{6.7} have many challenges.
%Thus, even though the absence of the term involving $q_{xx}$ in
%\eqref{6.5} gives the incorrect impression that \eqref{6.5} is a special case of the bad Boussinesq equation
%given in \eqref{6.7}, the transformation in \eqref{6.6} shows that \eqref{6.5} is indeed
%equivalent to \eqref{6.7}.
We refer the reader also to \cite{CL2023a,CL2023b,CL2024,CLW2023} for analysis of various aspects of
the bad Boussinesq equation \eqref{6.7}, the properties of the solution to its initial-value problem, and
the asymptotic behavior of such a solution in the presence of solitons.

With the goal of analyzing \eqref{6.5}, Deift, Tomei, and Trubowitz \cite{DTT1982} presented a formulation of the direct and inverse scattering
problem associated with \eqref{6.4}. As in \eqref{2.14} and \eqref{2.15}, they used two transmission coefficients and four reflection coefficients
in their formulation of the direct scattering problem. However, to obtain a proper formulation of a Riemann--Hilbert problem as in the case of the inverse
scattering problem for the full-line Schr\"odinger equation \cite{AK2001,CS1989,DT1979,F1967,L1987,M2011}, they
assumed \cite{DTT1982} that the two transmission coefficients are
identically equal to 1 for all $k$-values and two of the reflection coefficients are identically zero for all $k$-values. Those assumptions
allowed them to formulate a proper Riemann--Hilbert problem on the complex $k$-plane and also to derive a system of linear integral equations
as in the case of the full-line Schr\"odinger equation. However, the severe restriction on the transmission coefficients prevented their method to yield
any explicit solutions to \eqref{6.5} because such solutions are mainly related to the bound states of \eqref{6.4}, which occur at the
$k$-values corresponding to the poles of the transmission coefficients.

We now demonstrate how the solution to the inverse scattering problem for \eqref{1.1} in the reflectionless case yields explicit
solutions to the modified bad Boussinesq equation \eqref{6.5}.
The construction of the basic solutions $f(k,x),$ $m(k,x),$ $g(k,x),$ and $n(k,x)$ for \eqref{1.1} is the same as the construction
of the corresponding basic solutions to \eqref{6.4}. The only difference comes from the time evolution of those solutions. Using
the linear operator $A$ given in \eqref{6.3} instead of the linear operator $A$ given in \eqref{1.3}, we determine the time evolution
of the basic solutions to \eqref{6.4} as
\begin{equation}
\label{6.11}
\ds\frac{\partial f(k,x)}{\partial t}-A\,f(k,x)=-3ik^2 f(k,x), \qquad x\in\mathbb R,
\end{equation}
\begin{equation}
\label{6.12}
\ds\frac{\partial g(k,x)}{\partial t}-A\,g(k,x)=-3ik^2 g(k,x), \qquad x\in\mathbb R,
\end{equation}
\begin{equation*}
%\label{6.13}
\ds\frac{\partial m(k,x)}{\partial t}-A\,m(k,x)=-3ik^2 m(k,x), \qquad x\in\mathbb R,
\end{equation*}
\begin{equation*}
%\label{6.14}
\ds\frac{\partial n(k,x)}{\partial t}-A\,n(k,x)=-3ik^2 n(k,x), \qquad x\in\mathbb R.
\end{equation*}
We remark that \eqref{6.11} and \eqref{6.12} are the counterparts of \eqref{2.46} and \eqref{2.47}, respectively. The scattering
coefficients $T_\text{\rm{l}}(k),$ $T_\text{\rm{r}}(k),$ $L(k),$ $R(k),$ $M(k),$ and $N(k)$ for \eqref{6.4} are defined exactly the
same way as in \eqref{2.14} and \eqref{2.15} they are defined for \eqref{1.1}. The time evolution of the scattering coefficients for \eqref{6.4} is respectively given by
\begin{equation*}
%\label{6.15}
\begin{cases}
T_{\text{\rm{l}}}(k)\mapsto T_{\text{\rm{l}}}(k),\qquad k\in\overline{\Omega_1},\\
\noalign{\medskip}
T_{\text{\rm{r}}}(k)\mapsto T_{\text{\rm{l}}}(k),\qquad k\in\overline{\Omega_3},\\
\noalign{\medskip}
L(k)\mapsto L(k)\, \exp\left(3i(z^2-1)k^2 t\right),\qquad k\in\mathcal L_1,\\
\noalign{\medskip}
M(k)\mapsto M(k)\, \exp\left(3i(z-1)k^2 t\right),\qquad k\in\mathcal L_2,\\
\noalign{\medskip}
R(k)\mapsto R(k)\, \exp\left(3i(z^2-1)k^2 t\right),\qquad k\in\mathcal L_3,\\
\noalign{\medskip}
N(k)\mapsto N(k)\, \exp\left(3i(z-1)k^2 t\right),\qquad k\in\mathcal L_4,
\end{cases}
\end{equation*}
which is the counterpart of \eqref{2.48}.

The analysis of the bound states for \eqref{6.4} is similar to the analysis of the bound states for \eqref{1.1} presented in
Section~\ref{section3}. The analysis described in \eqref{3.2}--\eqref{3.10} holds verbatim. The respective time evolution of the dependency
constants $W(k_j)$ and $D(k_j)$ appearing in \eqref{3.3} and \eqref{3.4} is given by
\begin{equation*}
%\label{6.16}
W(k_j)=U(k_j)\, e^{3i(z-1)k_j^2 t}, \qquad \arg[k_j]\in(\pi,7\pi/6),
\end{equation*}
\begin{equation}
\label{6.17}
D(k_j)=E(k_j)\, e^{3i(z^2-1)k_j^2 t}, \qquad \arg[k_j]\in[7\pi/6,4\pi/3),
\end{equation}
which are the analogs of \eqref{3.18} and \eqref{3.24}, respectively.

In the next example, we illustrate the $\mathbf N$-soliton solution to \eqref{6.4}. The method used is the analog of the method described in
Section~\ref{section5} for \eqref{1.1}.

\begin{example}\label{example6.1}
\normalfont

As in \eqref{5.1}--\eqref{5.3} we let the left and right transmission coefficients $T_\text{\rm{l}}(k)$ and $T_\text{\rm{r}}(k)$ be given by
\begin{equation}
\label{6.18}
T_\text{\rm{l}}(k)=\ds\frac{F(k)}{G(k)}, \quad T_\text{\rm{r}}(k)=\ds\frac{G(k)}{F(k)},
\end{equation}
where we have let
\begin{equation}\label{6.19}
F(k):=\ds\prod_{j=1}^{\mathbf N} (k+k_j^\ast),
\quad
G(k):=\ds\prod_{j=1}^{\mathbf N} (k-k_j).
\end{equation}
The choice in \eqref{6.18} for the left and right scattering coefficients is compatible with \eqref{3.46}.
We remark that the poles $k_j$ of $T_\text{\rm{l}}(k)$ for $1\le j\le \mathbf N,$ i.e. the zeros of $G(k),$ are all located in $\Omega_1^\text{\rm{down}}.$
By proceeding as in Example~\ref{example5.1}, we recover the basic solutions to \eqref{6.4} as
\begin{equation}\label{6.20}
f(k,x)=e^{kx}\,\ds\frac{k^{\mathbf N}+V(k)\,\mathbf A(x)}{F(k)}, \qquad k\in\overline{\Omega_1}, \quad x\in \mathbb R,
\end{equation}
\begin{equation}\label{6.21}
m(k,x)=e^{kx}\,\ds\frac{k^{\mathbf N}+V(k)\,\mathbf A(x)}{G(k)}, \qquad k\in\overline{\Omega_2}, \quad x\in \mathbb R,
\end{equation}
\begin{equation}\label{6.22}
g(k,x)=e^{kx}\,\ds\frac{k^{\mathbf N}+V(k)\,\mathbf A(x)}{G(k)}, \qquad k\in\overline{\Omega_3}, \quad x\in \mathbb R,
\end{equation}
\begin{equation}\label{6.23}
n(k,x)=e^{kx}\,\ds\frac{k^{\mathbf N}+V(k)\,\mathbf A(x)}{G(k)}, \qquad k\in\overline{\Omega_4}, \quad x\in \mathbb R,
\end{equation}
which are the analogs of \eqref{5.9}--\eqref{5.12}, respectively. We note that $V(k)$ appearing in \eqref{6.20}--\eqref{6.23} is the
row vector with $\mathbf N$-components defined in \eqref{5.5}.
The quantity $\mathbf A(x)$ appearing in \eqref{6.20}--\eqref{6.23} is the column vector with
$\mathbf N$-components, and it is given by
\begin{equation}\label{6.24}
\mathbf A(x):=\begin{bmatrix}
A_{\mathbf N-1}(x)\\
A_{\mathbf N-2}(x)\\
\vdots\\
A_1(x)\\
A_0(x)
\end{bmatrix},\qquad x\in\mathbb R,
\end{equation}
which is analog of \eqref{5.6}. For our illustration purpose, without loss of generality, we assume that the $k$-values $k_j$ for
$1\le j\le \mathbf N$ are all in the sector $\arg[k]\in[7\pi/6,4\pi/3).$ With the help of \eqref{3.4} and \eqref{6.17}, for $1\le j\le \mathbf N$ we have
\begin{equation}\label{6.25}
f(k_j,x)=E(k_j)\,e^{3i(z^2-1)k_j^2 t}\, g(zk_j,x), \qquad \arg[k_j]\in[7\pi/6,4\pi/3), \quad x\in\mathbb R,
\end{equation}
which is the analog of \eqref{5.13}. Using \eqref{6.20} and \eqref{6.22} in \eqref{6.25}, we obtain
\begin{equation}\label{6.26}
e^{k_jx}\,\ds\frac{k_j^{\mathbf N}+V(k_j)\,\mathbf A(x)}{F(k_j)}=E(k_j)\,e^{3i(z^2-1)k_j^2 t}\,e^{zk_jx}\,\ds\frac{(zk_j)^{\mathbf N}+V(zk_j)\,\mathbf A(x)}{G(zk_j)}, \qquad 1\le j\le \mathbf N,
\end{equation}
which is the analog of \eqref{5.14}.
In terms of the initial dependency constants $E(k_j),$ as in \eqref{5.15} we introduce the modified initial dependency constants $\gamma(k_j)$ by letting
\begin{equation}\label{6.27}
\gamma(k_j):=-\ds\frac{F(k_j) E(k_j)}{G(zk_j)}, \qquad 1\le j\le \mathbf N.
\end{equation}
We emphasize that each $\gamma(k_j)$ in \eqref{6.27} is a complex-valued constant independent of $x$ and $t.$
We also introduce the scalar quantity $\chi(k_j)$ as
\begin{equation}\label{6.28}
\chi(k_j):=\exp\left((z-1)k_j x+3i(z^2-1)k_j^2 t\right), \qquad 1\le j\le \mathbf N,
\end{equation}
which is the analog of \eqref{5.16}.
We remark that each $\chi(k_j)$ is a function of $x$ and $t.$ Furthermore, each $\chi(k_j)$ is real valued if and only if
$k_j=iz\eta_j$ for some positive $\eta_j,$ and the proof for this is similar to the proof utilizing \eqref{5.17}.
This shows that each bound-state eigenvalue at $k=k_j$ for the selfadjoint operator $L$ appearing in \eqref{6.2}
occurs when we have $\lambda=\eta_j^3,$ which confirms that all the eigenvalues of the selfadjoint operator
$L$ in \eqref{6.2} are real.
Using \eqref{6.27} and \eqref{6.28} in \eqref{6.26}, we obtain
\begin{equation}\label{6.29}
k_j^{\mathbf N}+V(k_j)\,\mathbf A(x)=-\gamma(k_j)\,\chi(k_j)\left[(zk_j)^{\mathbf N}+V(zk_j)\,\mathbf A(x)\right],
\end{equation}
which is the analog of \eqref{5.19}. We observe that \eqref{6.29} yields a linear system of $\mathbf N$ equations with the $\mathbf N$ unknowns
$A_l(x)$ for $0\le l\le \mathbf N-1.$ This is seen by defining $m_l(k_j)$ as in \eqref{5.20}, i.e. 
\begin{equation}\label{6.30}
m_l(k_j):=k_j^l+(zk_j)^l\,\gamma(k_j)\,\chi(k_j),\qquad 1\le j\le\mathbf N,\quad 0\le l\le \mathbf N.
\end{equation}
Using \eqref{6.30} we form the $\mathbf N\times \mathbf N$ matrix $\mathbf M(x)$ and the column vector $\mathbf B(x)$ with $\mathbf N$ components as
\begin{equation}\label{6.31}
\mathbf M(x):=\begin{bmatrix}
m_{\mathbf N-1}(k_1) & m_{\mathbf N-2}(k_1) & \cdots & m_1(k_1) & m_0(k_1)\\
m_{\mathbf N-1}(k_2) & m_{\mathbf N-2}(k_2) & \cdots & m_1(k_2) & m_0(k_2)\\
\vdots & \vdots & \ddots & \vdots & \vdots \\
m_{\mathbf N-1}(k_{\mathbf N}) & m_{\mathbf N-2}(k_{\mathbf N}) & \cdots & m_1(k_{\mathbf N}) & m_0(k_{\mathbf N})
\end{bmatrix},
\end{equation}
\begin{equation}\label{6.32}
\mathbf B(x):=\begin{bmatrix}
m_{\mathbf N}(k_1)\\
m_{\mathbf N}(k_2)\\
\vdots\\
m_{\mathbf N}(k_{\mathbf N})
\end{bmatrix},
\end{equation}
which are the analogs of \eqref{5.21} and \eqref{5.22}, respectively. Using \eqref{6.31} and \eqref{6.32}, we write the linear system
in \eqref{6.29} as in \eqref{5.23} and recover the column vector $\mathbf A(x)$ as in \eqref{5.24}. Alternatively, the $\mathbf N$ components
$A_j(x)$ with $0\le j\le \mathbf N-1$ of the column vector $\mathbf A(x)$ can be obtained as the ratio of two determinants by
using Cramer's rule on
\eqref{5.23}. In particular, for the components $A_{\mathbf N-1}(x)$ and $A_{\mathbf N-2}(x)$ we have the analogs of \eqref{5.25} and \eqref{5.26}, respectively.
Having obtained the column vector $\mathbf A(x)$ explicitly in terms of the input data set $\{k_j, E(k_j)\}_{j=1}^{\mathbf N},$ as seen from
\eqref{6.20}--\eqref{6.23}, we obtain the basic solutions $f(k,x),$ $m(k,x),$ $g(k,x),$ and $n(k,x)$ to \eqref{6.4}, where each of
those solutions in their respective $k$-domains is explicitly expressed in terms of the input data set consisting of the poles of
$T_\text{\rm{l}}(k)$ and the corresponding dependency constants. To recover the potential $q$ we can use the analog of
\eqref{4.13}, which is given by
\begin{equation}\label{6.33}
q(x)=\ds\frac{3}{2}\,\ds\frac{du_1(x)}{dx},
\end{equation}
where $u_1(x)$ is the term appearing in \eqref{4.12}. We remark that \eqref{6.33} is obtained by comparing the coefficients of the
first derivative in \eqref{1.1} and in \eqref{6.4}, i.e.
\begin{equation}\label{6.34}
Q(x)=-2\,q(x), \qquad t=0, \quad x\in\mathbb R.
\end{equation}
Even though \eqref{6.34} holds at $t=0,$ the time evolution of $Q(x)$ and the time evolution of $q(x)$ are different. Hence, \eqref{6.34} does
not hold when $t\ne 0.$ The quantity $u_1(x)$ appearing in \eqref{6.33} is obtained by using the large $k$-asymptotics of
\eqref{6.20} given by
\begin{equation}\label{6.35}
\ds\frac{k^{\mathbf N}+V(k)\,\mathbf A(x)}{F(k)}=1+\ds\frac{u_1(x)}{k}+O\left(\ds\frac{1}{k^2}\right), \qquad k\to \infty \text{\rm{ in }} \overline{\Omega_1},
\end{equation}
which is the analog of \eqref{5.28}.
With the help of \eqref{5.5}, the first equality of \eqref{6.19}, \eqref{6.24}, and \eqref{6.35}, we have
\begin{equation}\label{6.36}
u_1(x)=A_{\mathbf N-1}(x)-(k_1^\ast+\cdots+k_{\mathbf N}^\ast),
\end{equation}
which is the analog of the first equality in \eqref{5.31}.
Hence, comparing \eqref{6.33} and \eqref{6.36}, we see that we can recover $q(x)$ from $A_{\mathbf N-1}(x)$ as
\begin{equation}\label{6.37}
q(x)=\ds\frac{3}{2}\ds\frac{dA_{\mathbf N-1}(x)}{dx}.
\end{equation}
We remark that the potential $q(x)$ given in \eqref{6.37} is in general complex valued even though it satisfies the modified bad Boussinesq
equation \eqref{6.5}. In order to make $q(x)$ real valued, we need to use the appropriate restrictions on the
locations of the $k_j$-values as well as the choice of the initial dependency constants $E(k_j).$
The restriction $k_j=iz\eta_j$ with $\eta_j>0$ is the necessary and sufficient restriction on the location of $k_j.$ 
We obtain the appropriate restriction on the initial dependency constants $E(k_j)$ or equivalently on the modified initial dependency constants $\gamma(k_j)$
as follows. We require that $u_1(x)$ appearing in \eqref{6.33} and \eqref{6.35} be real valued, or equivalently we require that $A_{\mathbf N-1}(x)$ 
appearing in \eqref{6.36} and \eqref{6.37} be real valued. Those initial dependency constants appear in the $\mathbf N\times \mathbf N$ matrix $\mathbf M(x)$
and the column vector $\mathbf B(x)$ in \eqref{6.32}. The value of $A_{\mathbf N-1}(x)$ is recovered from $\mathbf M(x)$ and 
$\mathbf B(x)$ by using the analog of \eqref{5.25}. Hence, the requirement that $A_{\mathbf N-1}(x)$ be real valued yields 
the appropriate restrictions on the initial dependency constants
$E(k_j)$ or on the modified initial dependency constants $\gamma(k_j)$ for $1\le j\le \mathbf N.$    
For example, when $\mathbf N=1$ in \eqref{6.19} and
$k_1=iz\eta_1$ with $\eta_1>0,$ by choosing the modified initial dependency constant $\gamma(k_1)$ as positive, we obtain
\begin{equation}\label{6.38}
q(x)=-\ds\frac{9\,\gamma(k_1)\,\eta_1^2\,e^{\sqrt{3}\eta_1(x+3\eta_1 t)}
}{2\left[1+\gamma(k_1)\,e^{\sqrt{3}
\eta_1(x+3\eta_1 t)
}\right]^2},
 \qquad x\in \mathbb R,
\end{equation}
corresponding to the left transmission coefficient
\begin{equation*}
%\label{6.39}
T_\text{\rm{l}}(k)=\ds\frac{k-iz^2\eta_1}{k-iz\eta_1}.
\end{equation*}
Using \eqref{6.38} in the first line of \eqref{6.1}, we obtain the real-valued potential $p(x)$ as
\begin{equation}\label{6.40}
p(x)=\ds\frac{9\,\gamma(k_1)\,\eta_1^3
\,e^{\sqrt{3}\eta_1(x+3\eta_1 t)}
}{2\left[1+\gamma(k_1)\,e^{\sqrt{3}\eta_1(x+3\eta_1 t)}\right]^2},
 \qquad x\in \mathbb R.
\end{equation}  
\end{example}
  
The modified bad Boussinesq equation \eqref{6.5} remains unchanged if we replace $x$ by $-x$ or if we replace $t$ by $-t.$
Hence, from any solution to \eqref{6.5} we get other solutions to \eqref{6.5} by using the replacement $x\mapsto -x$ or $t\mapsto -t.$
In particular, such transformations applied to \eqref{6.38} and \eqref{6.40} yield solutions where the soliton moves to the right
with the speed $3\eta_1,$ whereas the soliton in \eqref{6.38} and \eqref{6.40} moves to the left with the speed $3\eta_1.$

In the next example we present the real-valued $2$-soliton solution $q(x,t)$ to \eqref{6.5} containing four arbitrary
real parameters $\eta_1,$ $\eta_2,$ $b_1,$ and $b_2.$ 
%The corresponding solution $\tilde q(x,t)$ to \eqref{6.7} is obtained
%via the transformation in \eqref{6.6}, and hence it also contains those four arbitrary real parameters. 
The method used in this example is the same as the general method that we use and is based on solving the inverse scattering problem for \eqref{1.1}. The method yields a complex-valued solution to \eqref{6.5} with
the four complex-valued parameters $k_1,$ $k_2,$ $\gamma(k_1),$ $\gamma(k_2),$
where the first two parameters identify the location of the bound states in the complex $k$-plane and
the latter two parameters correspond to the modified initial bound-state dependency constants. Since the solution $q(x,t)$
to \eqref{6.5} is expected to be real valued, we then restrict the locations of the complex parameters $k_1$ and $k_2$ so that
they each are identified by one of the real parameters $\eta_1$ and $\eta_2,$ respectively. Similarly, we also restrict the
values of the complex-valued parameters $\gamma(k_1)$ and $\gamma(k_2)$
so that
they each are identified by one of the real parameters $b_1$ and $b_2,$ respectively.
In our example, we only present the final result by omitting the details of the derivation.

\begin{example}\label{example6.2}
\normalfont

We choose our transmission coefficients as in \eqref{6.18} where $\mathbf N=2$ is used in \eqref{6.19}.
The choice in \eqref{6.18} for the left and right scattering coefficients is compatible with \eqref{3.46}.
In order to obtain a real-valued $2$-soliton solution, we choose $k_1$ and $k_2$ as
\begin{equation*}
%\label{6.41}
k_1=iz\eta_1,\quad k_2=iz\eta_2,
\end{equation*}  
where we have $0<\eta_1<\eta_2$
and we recall that $z$ is the special constant defined in \eqref{2.1}.
We choose our complex-valued modified initial dependency constants appearing in \eqref{6.27} as
\begin{equation}\label{6.42}
\gamma(k_1)=\ds\frac{b_1}{\eta_2-z\eta_1},\quad
\gamma(k_2)=\ds\frac{b_2}{\eta_1-z\eta_2},
\end{equation}  
where we have $b_1>0$ and $b_2<0$ as a result of our choice $0<\eta_1<\eta_2.$
We notice the symmetry in \eqref{6.42} so that knowing $\gamma(k_1)$ we are also able to specify
$\gamma(k_2)$ by symmetry.
Using \eqref{6.37} with $\mathbf N=2,$ we obtain the real-valued $2$-soliton solution to \eqref{6.5} as
\begin{equation}
\label{6.43}
q(x)=-\ds\frac{9 \left[{\text{\rm{num}}}_1+{\text{\rm{num}}}_2\right]
}{2\,\left(\eta_2^3-\eta_1^3 \right)\left[{\text{\rm{den}}}\right]^2},
 \qquad x\in \mathbb R,
\end{equation}
where we have defined
\begin{equation*}
%\label{6.44}
{\text{\rm{num}}}_1:=
\left(b_1\eta_1^2\chi_1-b_2\eta_2^2\chi_2\right)\ds\frac{\eta_2^3-\eta_1^3}{\eta_2-\eta_1},
\end{equation*}
\begin{equation*}
%\label{6.45}
{\text{\rm{num}}}_2:=
-b_1 b_2 \chi_1\chi_2 \left[2(\eta_2^3-\eta_1^3)+\eta_1\eta_2(\eta_2-\eta_1)+
b_1 \eta_2^2 \chi_1-b_2 \eta_1^2\chi_2\right],
\end{equation*}
\begin{equation*}
%\label{6.46}
{\text{\rm{den}}}:=1+\ds\frac{b_1 \chi_1}{\eta_2-\eta_1}-\ds\frac{b_2 \chi_2}{\eta_2-\eta_1}-
\ds\frac{b_1 b_2 (\eta_2-\eta_1) \chi_1\chi_2}
{\eta_2^3-\eta_1^3},
\end{equation*}
with $\chi_1$ and $\chi_2$ defined as
\begin{equation*}
%\label{6.47}
\chi_1:=e^{\sqrt{3}\eta_1(x+3\eta_1 t)}
,\quad
\chi_2:=e^{\sqrt{3}\eta_2(x+3\eta_2 t)}.
\end{equation*}

\begin{figure}[!ht]
     \centering
         \includegraphics[width=1.45in]{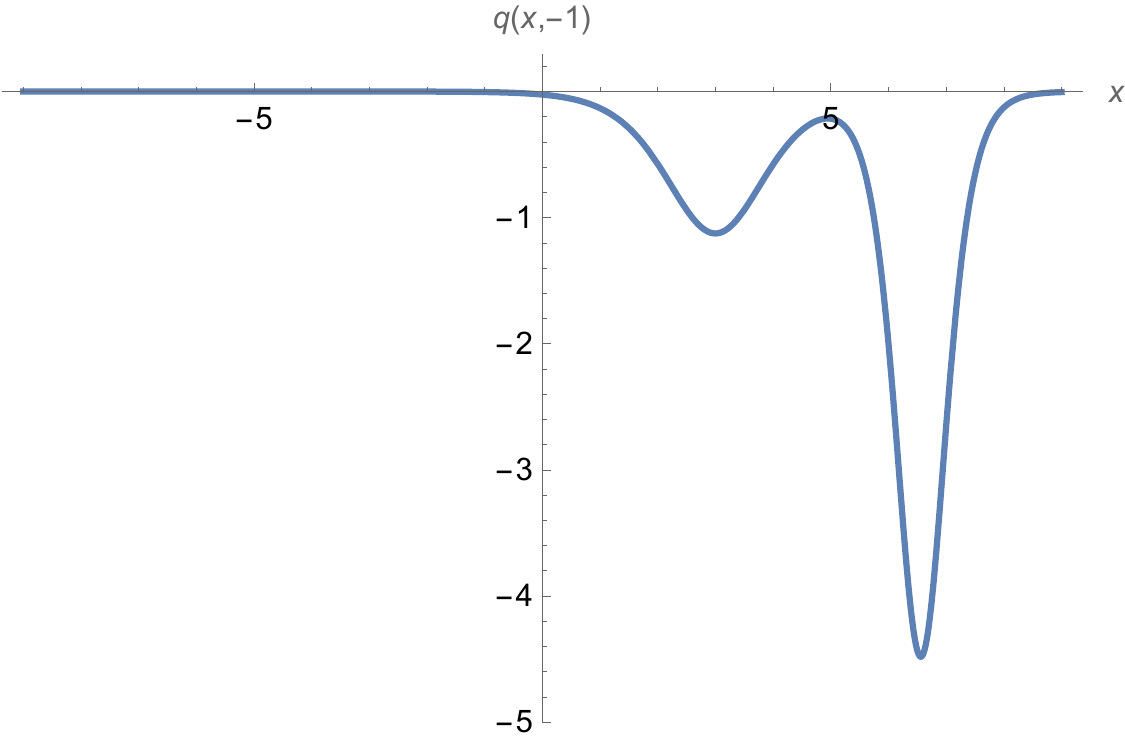}      \hskip .1in
         \includegraphics[width=1.45in]{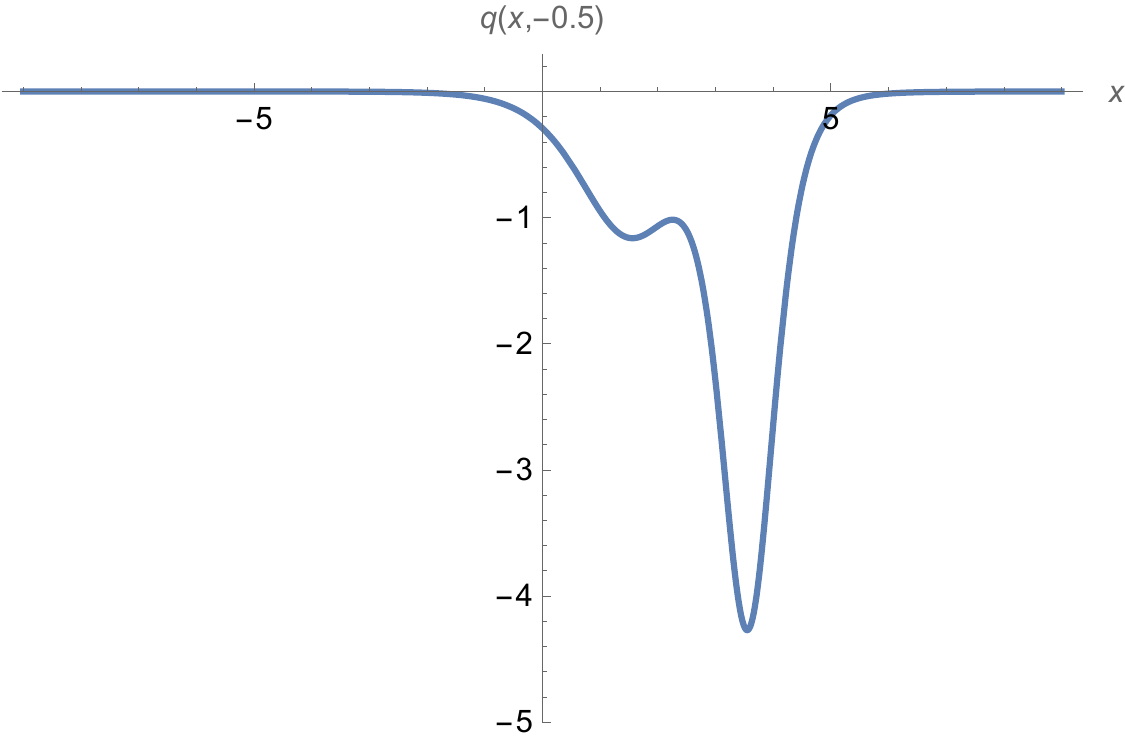} \hskip .1in
                  \includegraphics[width=1.45in]{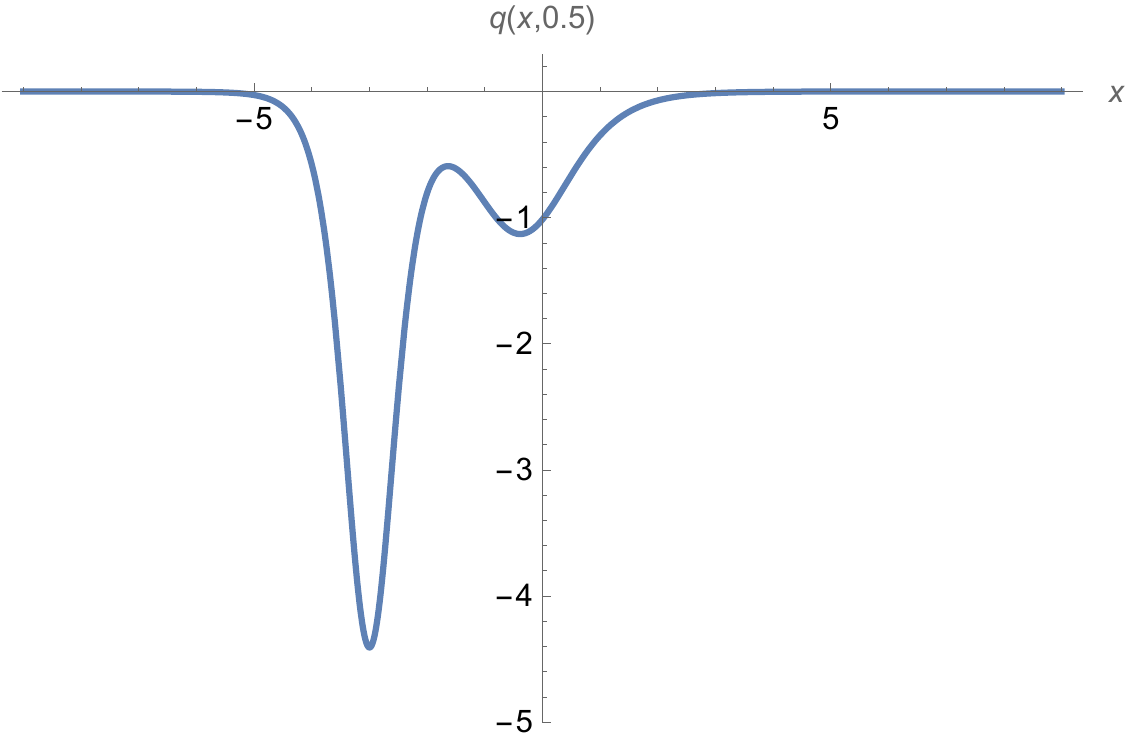} \hskip .1in
                  \includegraphics[width=1.45in]{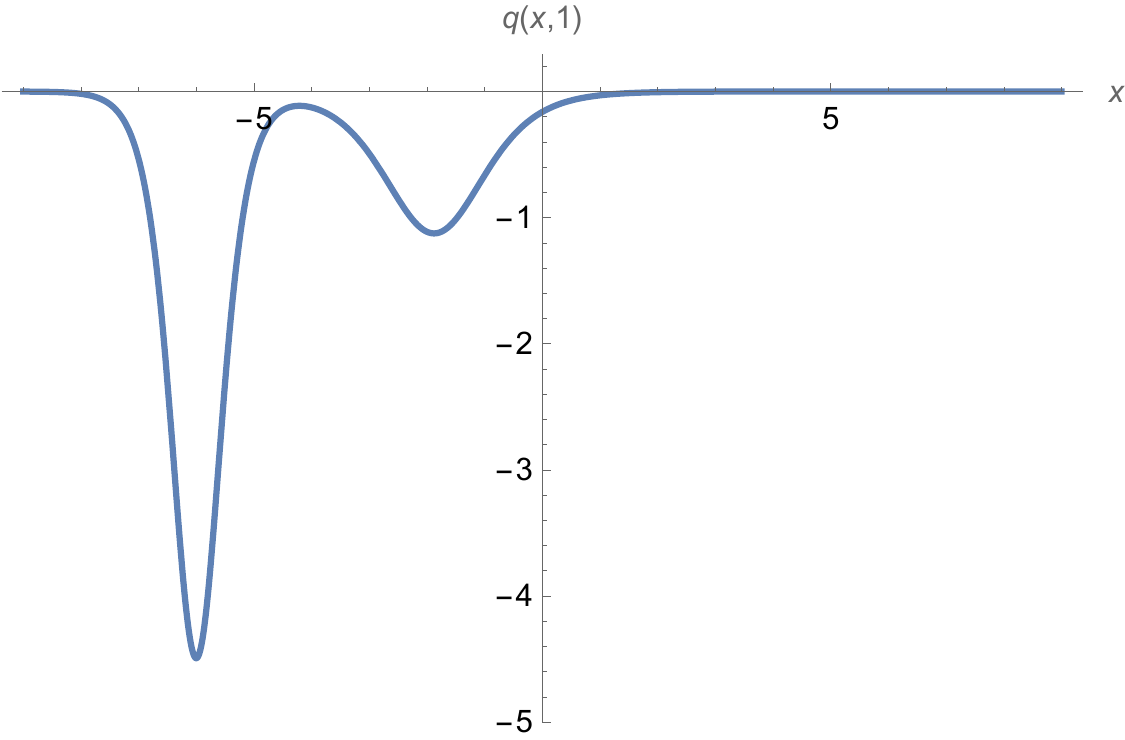} 
\caption{The snapshots for the $2$-soliton solution $q(x,t)$ given in \eqref{6.49}, which is the solution to the modified bad Boussinesq equation
\eqref{6.5} 
%with
%$(\eta_1,\eta_2,\eta_3,\eta_4)=(1,1.1,1.2,1.3)$ and
%$(r_1,r_2,r_3,r_4)=(-1,1,-1,1)$ 
at $t=-1,$ $t=-0.5,$ $t=0.5,$ and $t=1,$ respectively.}
\label{figure6.1}
\end{figure}

\noindent For example, by choosing
\begin{equation*}
%\label{6.48}
\eta_1=1,\quad \eta_2=2,\quad b_1=1,\quad b_2=-1,
\end{equation*}
from \eqref{6.43} we obtain the particular solution $q(x)$ to \eqref{6.5} given by
\begin{equation}
\label{6.49}
q(x)=-\ds\frac{9\left( 7\chi_1+28\,\chi_2+16\,\chi_1\chi_2+4\,\chi_1^2\chi_2+\chi_1\chi_2^2\right)}
{14\left(1+\chi_1+\chi_2+\ds\frac{1}{7}\,\chi_1\chi_2\right)^2},
\end{equation}
with the quantities $\chi_1$ and $\chi_2$ given by
\begin{equation*}
%\label{6.50}
\chi_1:=e^{\sqrt{3}(x+3t)}
,\quad
\chi_2:=e^{2\sqrt{3}(x+6t)}.
\end{equation*}
In Figure~\ref{figure6.1} we present four snapshots for the $2$-soliton solution $q(x)$ given in 
\eqref{6.49}. We observe that both solitons move from the right to the left, the faster moving taller soliton catches the shorter soliton, and after they interact nonlinearly
the taller soliton moves away from the shorter soliton.

\end{example}

\end{document}